\documentclass[twocolumn]{aastex701}

\usepackage[T1]{fontenc}

\shorttitle{}
\shortauthors{Hasegawa et al.}
\graphicspath{{./}{figures/}}
\begin{document}

\title{Formation and evolution pathways of planets. I. Comparison between theory and observations}

\author[0000-0002-9017-3663]{Yasuhiro Hasegawa}
\affiliation{Jet Propulsion Laboratory, California Institute of Technology, Pasadena, CA 91109, USA}
\email[show]{yasuhiro.hasegawa@jpl.nasa.gov}

\author{Renyu Hu}
\affiliation{Jet Propulsion Laboratory, California Institute of Technology, Pasadena, CA 91109, USA}
\affiliation{Department of Astronomy \& Astrophysics, The Pennsylvania State University, University Park, PA 16802, USA}
\affiliation{Center for Exoplanets and Habitable Worlds, The Pennsylvania State University, University Park, PA 16802, USA}
\affiliation{Institute for Computational and Data Science, The Pennsylvania State University, University Park, PA 16802, USA}
\email{}

\author{Amine Bouzerzour}
\affiliation{Jet Propulsion Laboratory, California Institute of Technology, Pasadena, CA 91109, USA}
\affiliation{Institut sup\'{e}rieur de l’a\'{e}ronautique et de l’espace, Universit\'{e} de Toulouse, F-31400 Toulouse, France}
\email{}



\begin{abstract}
Discoveries of numerous exoplanets by various methods enable detailed characterization including bulk density.
Formation and evolution pathways of planets can thus be probed in the mass-radius and mass-density diagrams.
We develop a framework to identify dominant processes shaping parameter space in these diagrams by integrating previous studies.
These include interior structure models, gas accretion/retention recipes, and photoevaporative and collisional mass losses.
We find that the distribution of planets in the diagrams is diversified by two evolution processes: photoevaporative and collisional mass losses,
and the properties of planets before experiencing these processes are consistent with predictions of standard core accretion.
In particular, collisional mass growth and loss move planets to the parameter space, which is otherwise occupied by water-dominated (i.e., nearly pure water) planets,
gathering non-necessity of invoking such planets.
A potentially high abundance of water-rich planets are possible with the ice-to-rock ratio capped at $1/3$, similar to solar system comets.
We propose a new classification scheme and apply to observed exoplanets.
The classification scheme recovers four canonical planet types widely used in the literature
and is expended to eight classes in total due to evolution processes.
We divide formation and evolution pathways into four stages (core formation, gas accretion, collisional mass growth and loss, and photoevaporation) 
and trace how planets populate in the mass-radius and mass-density diagrams with time.
We apply the framework to habitable zone planets and discuss possible predictions.
This work emphasizes the importance of precise mass and radius measurements, especially for small-sized, potentially habitable planets.
\end{abstract}



\section{Introduction} \label{sec:intro}

Revealing the origin of planets is one fundamental task for astronomy and planetary science.
Rapid progress has recently been made,
thanks to the discovery of exoplanets \citep[e.g.,][]{1995Natur.378..355M,2010Sci...327..977B,2021ARA&A..59..291Z}.
The population of confirmed exoplanets now amounts to more than 6,000 and constantly challenges the present view of planet formation.

A high abundance of observed exoplanets have opened a new window of statistically examining formation and evolution pathways of planets. 
One standard means used in the commnuity is the so-called population synthesis calculation 
\citep[e.g.,][]{2004ApJ...604..388I,2009A&A...501.1139M,2013ApJ...778...78H,2021A&A...656A..69E}.
In the calculation, mass growth and orbital evolution of planets forming in protoplanetary disks are traced 
in the semimajor-axis-mass or orbital period-mass diagrams.
The resulting trajectories in these diagrams are referred to as evolutionary tracks \citep{2012ApJ...760..117H} or growth tracks \citep{2021SciA....7..444J} 
and directly reflect where planets accrete disk gas and solid at what rates.
Computing the tracks is thus important for determining the composition of planets \citep[e.g.,][]{2011ApJ...743L..16O,2014ApJ...794L..12M,2019A&A...632A..63C}.
Verifying the validity of computed tracks is not straightforward, however.
This is not only because of uncertainties about the efficiency of planet growth and migration, 
but also due to poor constraints on the composition of accreted materials.

Invention of various observational techniques to discover exoplanets made it possible to characterize their properties in detail 
\citep[e.g.,][]{2007ARA&A..45..397U,2015ARA&A..53..409W};
radial velocity constrains planet mass, and transit measures planet radius.
The combination of these two quantities enabled estimating the bulk density of planets.
The mass-radius and mass-density diagrams thus become complementary but invaluable tools 
to alternatively examine the formation and evolution pathway of planets \citep[e.g.,][]{2019PNAS..116.9723Z,2019ApJ...881..117S,2022Sci...377.1211L}.

Access to multi-dimensional parameter space to examine the origin of planets stimulates 
improvement and expansion of theoretical models in various ways and leads to new recognition and better understanding of underlying processes.
For instance, {\it Kepler} observations coupled with follow-up characterization of host stellar properties result in the discovery of a radius valley \citep{2017AJ....154..109F},
highlighting the importance of envelope mass loss \citep[e.g.,][]{2012ApJ...761...59L,2012MNRAS.425.2931O}.
As another example, discoveries of high bulk density planets \citep[e.g.,][]{2018NatAs...2..393S,2020A&A...641A..92T,2021Sci...374.1271L}
point out the possibility of mantle stripping by catastrophic collisions \citep[e.g.,][]{2010ApJ...712L..73M,2019NatAs...3..416B,2024MNRAS.529.2577D},
an analog to the formation of Mercury in the solar system.

In this work, we synthesize these recent efforts made in the literature and draw a unified picture of how planets form and evolve with time.
We particularly focus on the mass-radius and mass-density diagrams and investigate what parameter space is filled out by which process(es).
We show below that formation and evolution processes cover different parameter spaces in these diagrams;
the distribution of exoplanets in the diagrams becomes diversified by evolution processes.
Importantly, the limitation of the power of radiation-driven envelope mass loss in turn points out the significance of collisional mass growth and loss.
Exoplanets, which are otherwise viewed as water-dominated (i.e., nearly pure water) planets, likely possess very tenuous envelopes as the result of less energetic collisions,
while highly energetic collisions remove (some) mantle of planets and account for iron-rich planets.
It is possible that water-rich planets that have composition similar to solar system comets are abundant in the observed population.

Direct links between underlying processes and the corresponding parameter space in the mass-radius and mass-density diagrams enable careful classification of planets.
We find that the canonical classification scheme (e.g., super-Mercuries, super-Earths, sub-Neptunes, and gas giants) widely used in the literature can be expanded.
In particular, both super-Earths and sub-Neptunes are divided into three sub-classes,
and the total of eight classes (super-Mercuries, bare cores, collisionally sculpted cores, photoevaporated cores, 
collisionally sculpted sub-giants, photoevaporated sub-giants, vapor-rich sub-giants, and gas giants) is proposed based on the current properties of exoplanets.
Formation and evolution processes responsible for these properties are also identified.

The plan of this paper is as follows.
Section \ref{sec:current} introduces a framework to constrain the current and primordial properties of planets.
The framework is applied to exoplanets currently observed and classifies them.
The resulting classification is consistent with the one widely used in the literature.
Section \ref{sec:post} describes post-formation processes.
We especially focuses on radiation-driven envelope mass loss and collisional mass growth and loss.
These processes enrich the distribution of planets in the mass-radius and mass-density diagrams, and so do classification categories.
Section \ref{sec:test} summarizes theoretical predictions deduced from our analyses.
These include constraints on the onset of rapid gas accretion, a new classification scheme of exoplanets, 
predictions of formation and evolution pathways in the mass-radius and mass-density diagrams, and new insights into planets in habitable zones.
This work is developed from the results of previous studies, where simplification and assumptions are used.
Section \ref{sec:disc} discusses the resulting uncertainties that may affect our finding and relevant caveats.
Section \ref{sec:sum} is devoted to the summary and conclusions.

\begin{figure*}
\begin{minipage}{17cm}
%\begin{figure}%[!ht]
\begin{center}
\includegraphics[width=8.4cm]{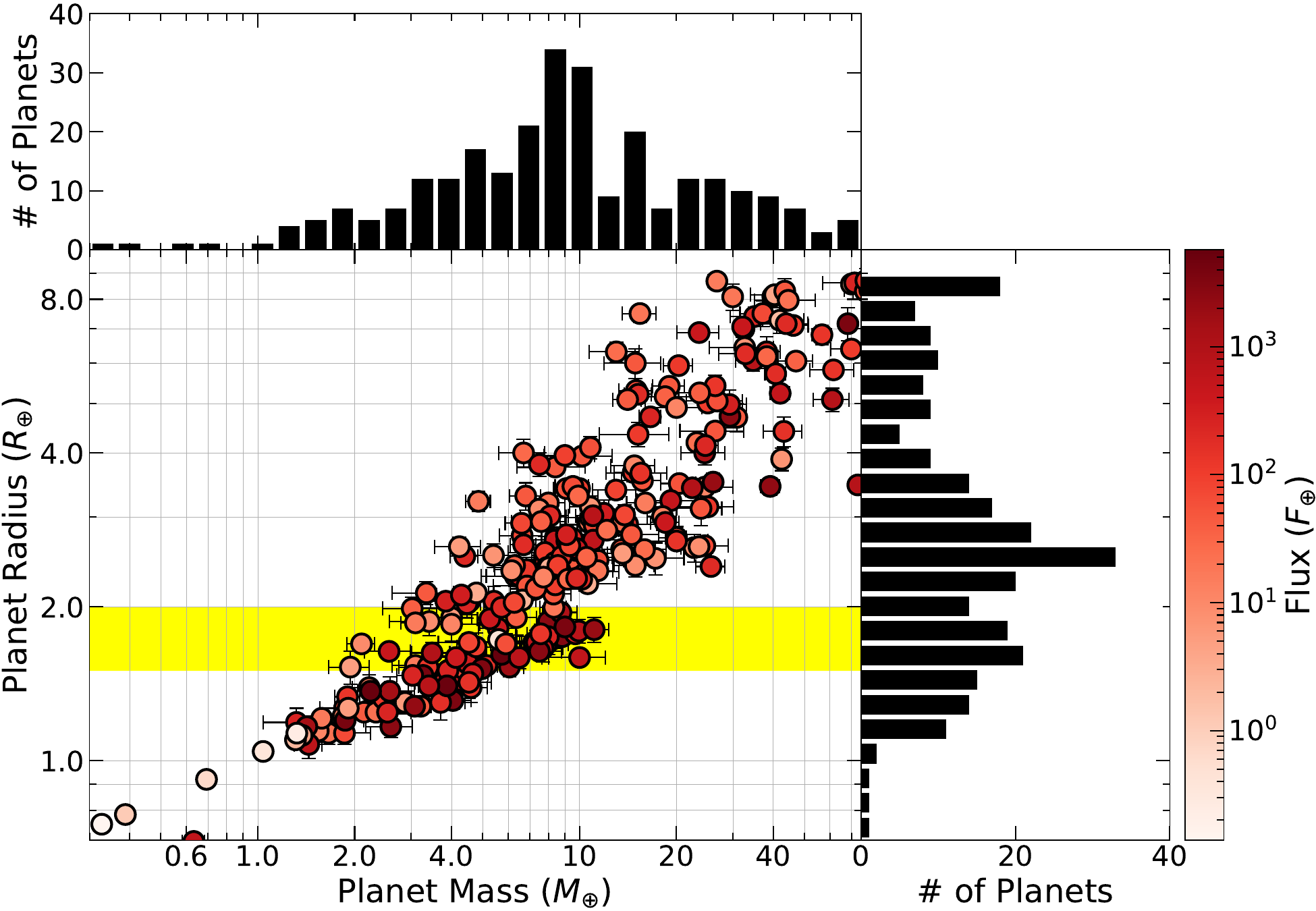}
\includegraphics[width=8.4cm]{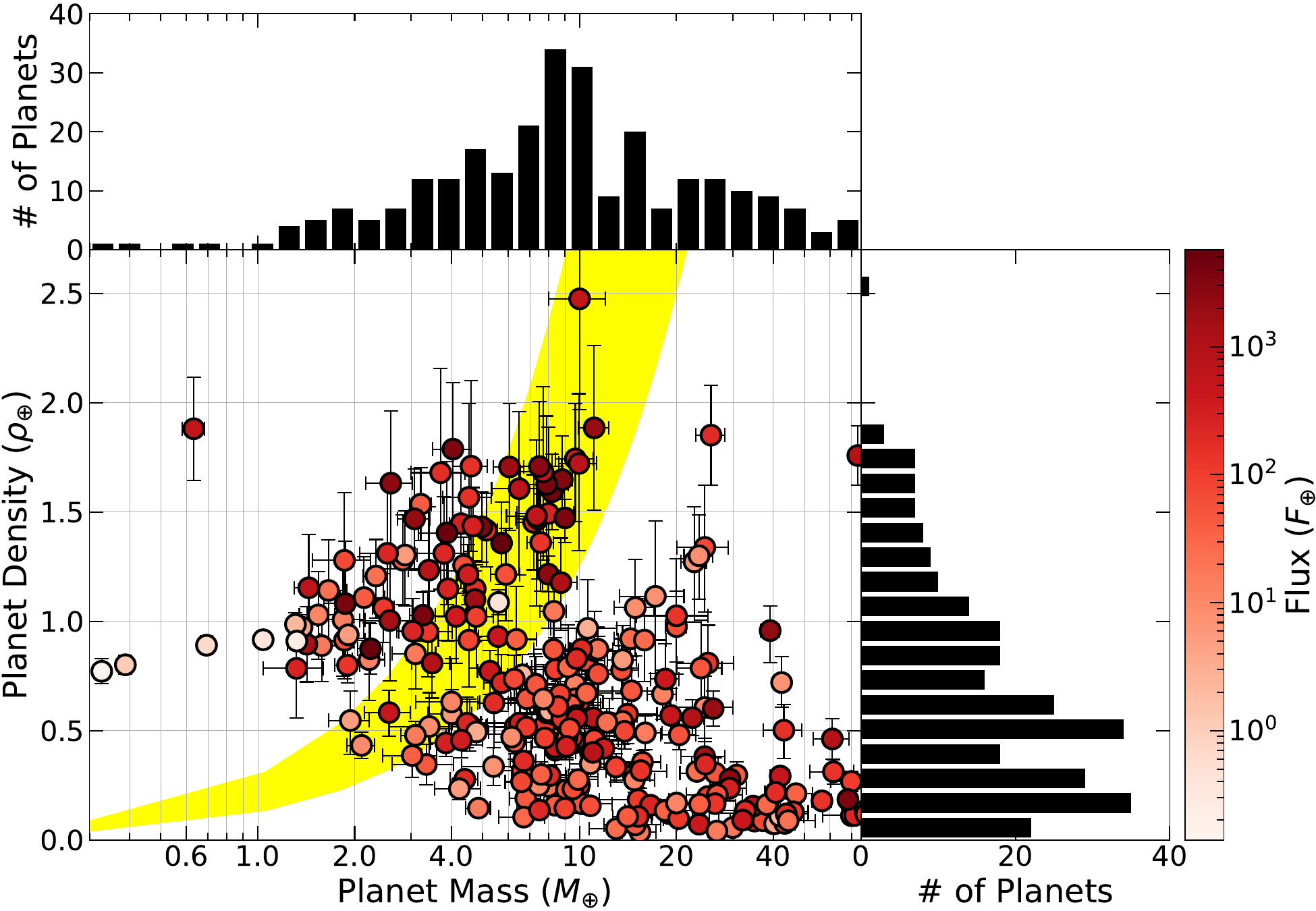}
\caption{Exoplanet samples used in this work.
The left panel represents the mass-radius diagram, and the right one denotes the mass-density diagram.
They are adopted from the PlanetS catalog and 
have accurate measurements of mass (relatively measurement uncertainties of 25 \% or lower) and radius (relatively measurement uncertainties of 8 \% or lower)
as shown by the error bar.
Insolation flux planets receive currently corresponds to the color of the dot with the color bar.
Two histograms are plotted along the x- and y-axes.
For reference, the radius valley ($R_{\rm p} =1.5-2 R_{\oplus}$) is represented by the yellow shaded region on both the panels.}
\label{fig1}
\end{center}
%\end{figure}
\end{minipage}
\end{figure*}

Figure \ref{fig1} shows exoplanet samples used in this work.
These data are adopted from the PlanetS catalog\footnote{\url{https://dace.unige.ch/exoplanets}} \citep{2024A&A...688A..59P}.
The catalog extracts exoplanets from the NASA Exoplanet Archive\footnote{\url{https://exoplanetarchive.ipac.caltech.edu}}
and includes those who have accurate measurements of mass (relatively measurement uncertainties of 25 \% or lower) 
and radius (relatively measurement uncertainties of 8 \% or lower). 
While error bars are included in Figure \ref{fig1},
they will be omitted in the following figure.
This is partly due to the purpose of clear presentation and partly because error propagation calculations are not straightforward for some computations conducted in this work.

\section{Current and primordial properties of planets} \label{sec:current}

We here describe how the current and primordial properties of planets are determined for given values of planet mass and radius.
We make use of various mass-radius relations and gas accretion and retention receipts to characterize observed exoplanets.

\subsection{Mass budget}

We begin with the mass budget of planets.
For planets that are composed of planetary cores and surrounding gaseous envelopes,
their total mass ($M_{\rm p}$) is expressed as
\begin{equation}
\label{eq:Mp}
M_{\rm p} = M_{\rm core} + M_{\rm env},
\end{equation}
where $M_{\rm core}$ is the core mass, and $M_{\rm env}$ is the envelope mass.
The radius ($R_{\rm p}$) of such planets is then written as
\begin{equation}
\label{eq:Rp}
R_{\rm p} = R_{\rm core} + R_{\rm env},
\end{equation}
where $R_{\rm core}$ is the core radius, and $R_{\rm env}$ is the radius due to the presence of gaseous envelopes.

Compositionally, the planet mass is decomposed into
\begin{equation}
\label{eq:Mp_comp}
M_{\rm p} = M_{XY} + M_{Z},
\end{equation}
where $M_{XY}$ and $M_{Z}$ are the mass of hydrogen and helium and that of the so-called metals (heavier than these two elements), respectively.
The composition of $M_{Z}$ is further specified by re-writing it into
\begin{eqnarray}
M_{Z} & = & M_{\rm ref} + M_{\rm ice} \\ \nonumber
          & = & M_{\rm Fe} + M_{\rm Si} + M_{\rm ice} 
\end{eqnarray}
where $M_{\rm ref}$ and $M_{\rm ice}$ are the mass of refractory and icy materials, respectively,
and the former is further decomposed into iron-bearing ($M_{\rm Fe}$) and silicate-bearing ($M_{\rm Si}$) materials.
The mass ratio of refractory and icy materials is often referred to as the ice-to-rock ratio ($f_{\rm ItoR}$), which is given as
\begin{equation}
f_{\rm ItoR} \equiv \frac{M_{\rm ice}}{M_{\rm ref}}.
\end{equation}
The iron mass fraction ($f_{\rm Fe}$) is defined as $M_{\rm Fe}/M_{Z}$, and if the abundance of icy materials are negligible, $f_{\rm Fe}$ is written as
\begin{equation}
f_{\rm Fe} \simeq \frac{ M_{\rm Fe} }{ M_{\rm ref} }.
\end{equation}
For the Earth, $f_{\rm Fe} \simeq 1/3$.

In this work, we assume that gaseous envelopes consist purely of hydrogen and helium, that is,
\begin{eqnarray}
\label{eq:M_comp}
M_{\rm env}    & = & M_{XY}, \\ \nonumber
M_{\rm core}   & = & M_{Z}. 
\end{eqnarray}
This assumption is widely used in previous studies, which will be adopted in the following calculations.

\subsection{Mass-radius relation}

The mass-radius relation is one quantity to infer the composition of planets and hence useful for classifying planets.

The core radius may be expressed as a function of planet mass, which is given as \citep[e.g.,][]{2019PNAS..116.9723Z}
\begin{equation}
\label{eq:Rcore}
\frac{ R_{\rm core} }{ R_{\oplus} } =  f_{\rm R} \left( \frac{M_{\rm p} }{ M_{\oplus} } \right)^{1/3.7},
\end{equation}
where $f_{\rm R}$ is the radius enhancement factor due to inclusion of water ice.
This power-law profile is approximate and effective when planets are less massive than $\sim 10 M_{\oplus}$;
detailed calculations that use empirically derived equations of state \citep[EoS,][]{2007ApJ...669.1279S} show slightly shallower dependence on planet mass for more massive planets.
We, however, have confirmed in Appendix \ref{sec:appned1} that adopting this approximation is valid for the sample used in this work and does not change our finding. 

According to \citet{2019PNAS..116.9723Z}, $f_{\rm R}$ is written as
\begin{equation}
f_{\rm R} =  1+ 0.55 f_{\rm H_2O} - 0.14 f_{\rm H_2O}^2,
\end{equation}
where the mass fraction of water ice ($f_{\rm H_2O}$) in cores is
\begin{equation}
\label{eq:f_H2O}
 f_{\rm H_2O} \equiv  \frac{M_{\rm H_2O}}{M_{\rm core} } 
                        \simeq \frac{M_{\rm ice}}{M_{\rm ref} +  M_{\rm ice}} = \frac{ f_{\rm ItoR} }{1+ f_{\rm ItoR} }.
\end{equation}
The above functional form of $f_{\rm R}$ is the outcome of interior modeling with an updated water ice EoS \citep{2012PhRvL.108i1102K}.
Solar system comets infer that $M_{\rm ice} \simeq M_{\rm H_2O}$ \citep[e.g.,][]{2011ARA&A..49..471M}, 
and it is assumed in equation (\ref{eq:f_H2O}) that the same condition is applicable to exoplanets,
that is, planet-building solid bodies formed predominantly within the CO snow line.

The radius for H/He envelopes can be written as \citep{2014ApJ...792....1L}
\begin{equation}
\label{eq:R_env}
\frac{ R_{\rm env} }{ R_{\oplus} } \simeq 12 \left( \frac{M_{\rm p} }{ M_{\oplus} } \right)^{-0.21}  \left( \frac{M_{XY} }{ M_{\rm p} } \right)^{0.59},
\end{equation}
where planets are assumed to expose stellar flux comparable to the Earth, and their age is about a few Gyr.
In the original equation \citep[i.e., equation (4) in][]{2014ApJ...792....1L},
explicit dependence on insolation flux and system age is derived.
However, the corresponding powers are sufficiently small (i.e., 0.044 and -0.18, respectively),
and we do not consider them in this study.
Also, the notation of variables is adjusted to be consistent with ours (i.e., $f_{\rm env} = M_{\rm XY}/ M_{\rm p} \times 100$ \%).
Finally, we neglect the contribution of the outermost radiative layer to the planet radius as it is typically very small \citep[$\sim 0.1 R_{\oplus}$,][]{2014ApJ...792....1L};
given the uncertainty of radius measurements (up to 8 \%) and that H/He envelopes are present for planets with radius of $\ga 2 R_{\oplus}$ as shown below, 
this omission does not affect our finding considerably.

\subsection{Core composition} \label{sec:core_comp}

The composition of planetary cores is one key information to constrain where the cores formed.
We define three characteristic compositions (Table \ref{table1}), 
which will be used to classify observed exoplanets in the following sections.

\begin{table*}
\begin{minipage}{17cm}
%\begin{table}
\centering
\caption{Three characteristic compositions of planetary cores}
\label{table1}
{%\scriptsize
\begin{tabular}{l|c|c|c|c}
\hline 
Type                 & Composition in mass                                   &  $ f_{\rm ItoR} $     &  $ f_{\rm H_2O} $   &  $ f_{\rm R} $   \\ \hline 
Earth-like rock  & $\sim 1/3$ iron + $\sim 2/3$ silicate            &  $0$                       &   $0$                       &  $1$                 \\ \hline  
Water-rich         & $\sim 3/4$ Earth-like rock + $1/4$ water    &  $ 1/3$                   &   $1/4$                     &  $\sim 1.29$     \\ \hline  
Pure water        & Water only                                                   &  $ \infty $                  &  $ 1 $                    &   $\sim1.41$          \\ \hline        
\end{tabular}
}
%\end{table}
\end{minipage}
\end{table*}

When $ f_{\rm R} =1$ (i.e., $f_{\rm ItoR}= f_{\rm H_2O}=0$), the mass-radius relation reproduces the properties of the Earth.
The composition satisfying such a condition is referred to as Earth-like rock in this work.

Small bodies in the solar system such as asteroids and comets are viewed as the remnant of planet-forming materials.
Their ice-to-rock mass ratio is known to be about 0.3 at most \citep[e.g.,][]{2011ARA&A..49..471M}.
The resulting value of $f_{\rm R}$ becomes 1.129.
The corresponding composition is called water-rich.

While unlikely, considering planets made of pure water (i.e., $f_{\rm H_2O}=1$ and $f_{\rm R}=1.41$) provides one reference.
The composition of such planets is referred to as pure water.

\subsection{Primordial envelope mass}

Observed exoplanets treated in this study cover a wide mass range.
For less massive planets, there is a possibility that they may have not only accreted primordial envelopes, but also lost the envelopes due to disk evolution.
Here, we consider upper and lower bounds on the primordial envelope mass, taking account of these processes.

Gas accretion onto planetary cores has been investigated extensively for a long time \citep[][]{1980PThPh..64..544M,1982P&SS...30..755S,1986Icar...67..391B}.
It is well known that gas accretion begins when hydrostatic equilibrium of envelopes is no longer maintained.
Historically, accretion of planetesimals onto cores and subsequent release of their gravitational energy were recognized as the main heat source 
to counteract gravitational contraction of envelopes.
When cores undergo gas accretion in the vicinity of the host star, however, 
the presence of planetesimals is not guaranteed automatically due to potentially rapid consumption of these bodies.
The resulting envelope mass ($M_{XY, \rm max}$) before the onset of rapid gas accretion is given as \citep{2015ApJ...811...41L} 
\begin{equation}
\label{eq:MXY_max}
M_{XY, \rm max} \simeq 0.3 M_{\oplus} \left( \frac{ M_{\rm core} }{ 5 M_{\oplus} }\right)^{2.7} \left( \frac{\tau_{\rm disk} }{1 \mbox{ Myr} }\right)^{0.4},
\end{equation}
where $\tau_{\rm disk}=10$ Myr is adopted in this work.
The formula is the outcome of standard planetary evolution calculations widely used in the literature, 
except for neglect of planetesimal accretion.
Due to the absence of planetesimals and the adopted disk lifetime, the resulting envelope mass should be viewed as an upper limit.

As shown by previous studies \citep[e.g.,][]{1982P&SS...30..755S,2000ApJ...537.1013I,2006ApJ...648..666R},
accreting envelopes develop radiative layers at their surface regions, which mask inner convective layers.
This leads to the two, well-known results: insensitivity of gas accretion processes to surrounding environments 
and high dependence of the critical core mass on envelopes' opacity, which is dominated by dust opacity.
When deriving equation (\ref{eq:MXY_max}), the solar metallicity is adopted, 
and its dependence on $M_{XY}$ is found to be modest because the corresponding power-law index is 0.4.
We therefore do not consider the dependence in this work.

In the limit of dust-free envelopes or for certain opacities, the radiative layer is approximated to be isothermal 
\citep[e.g.,][]{2006ApJ...648..666R,2012ApJ...747..115O,2022ApJ...935..101H}.
Given that gas accretion is regulated by the properties of envelopes at the boundary between the radiative and convective layers,
the temperature at the isothermal region is often quoted in previous work.
The value of $M_{XY}$ in equation (\ref{eq:MXY_max}) corresponds to the temperature of 2,500 K,
which is an natural outcome of H$_2$ dissociation in the model of \citet{2015ApJ...811...41L}.

On the contrary to giant planets, less massive planets may experience loss of primordial envelopes due to disk evolution.
The process is referred to as spontaneous mass loss \citep{2016ApJ...825...29G} or boil-off \citep{2016ApJ...817..107O}
and caused by reduction of the surrounding gas pressure following disk dispersal \citep{2012ApJ...753...66I,2024MNRAS.529.2716R};
at the interface between planetary envelopes and the surrounding disk gas, pressure equilibrium is established.
Following disk evolution, pressure caused by disk gas decreases, and the interface moves away from planets.
This drives mass loss and is effective only for less massive planets because their gravitational potential is shallower than that of more massive planets.
The remaining envelope mass ($M_{XY, \rm min}$) is written as \citep{2016ApJ...825...29G}
\begin{equation}
\label{eq:MXY_min}
M_{XY, \rm min} \simeq 0.1 M_{\oplus} \left( \frac{ M_{\rm core} }{ 5 M_{\oplus} }\right)^{1.44}  \left( \frac{\tau_{\rm disk} }{1 \mbox{ Myr} }\right)^{0.5}.
\end{equation}
The resulting value of $M_{XY}$ should be viewed as a lower limit 
as it is the minimum value of $M_{XY}$ achieved immediately after gas disks are gone,
and we adopt that $\tau_{\rm disk} = 1$ Myr in this work.

When deriving equation (\ref{eq:MXY_min}), 
dust-free envelopes are considered, and the adopted temperature is $10^3$ K.
Given that its dependence is found to be modest with the power-law index of 0.25, we do not consider possible variation in this work.

Thus, less massive planets formed in gaseous protoplanetary disks may have envelopes with the mass range of
\begin{equation}
\label{eq:Mxy_range}
M_{XY, \rm min} \leq M_{XY} \leq M_{XY, \rm max}.
\end{equation}

\subsection{Classification scheme} \label{sec:class}

The formulation described in the above sections becomes a basis to construct a  classification scheme 
that is applicable to planets emerging immediately after the dispersal of gas disks.

\begin{figure*}
\begin{minipage}{17cm}
%\begin{figure}%[!ht]
\begin{center}
\includegraphics[width=8.4cm]{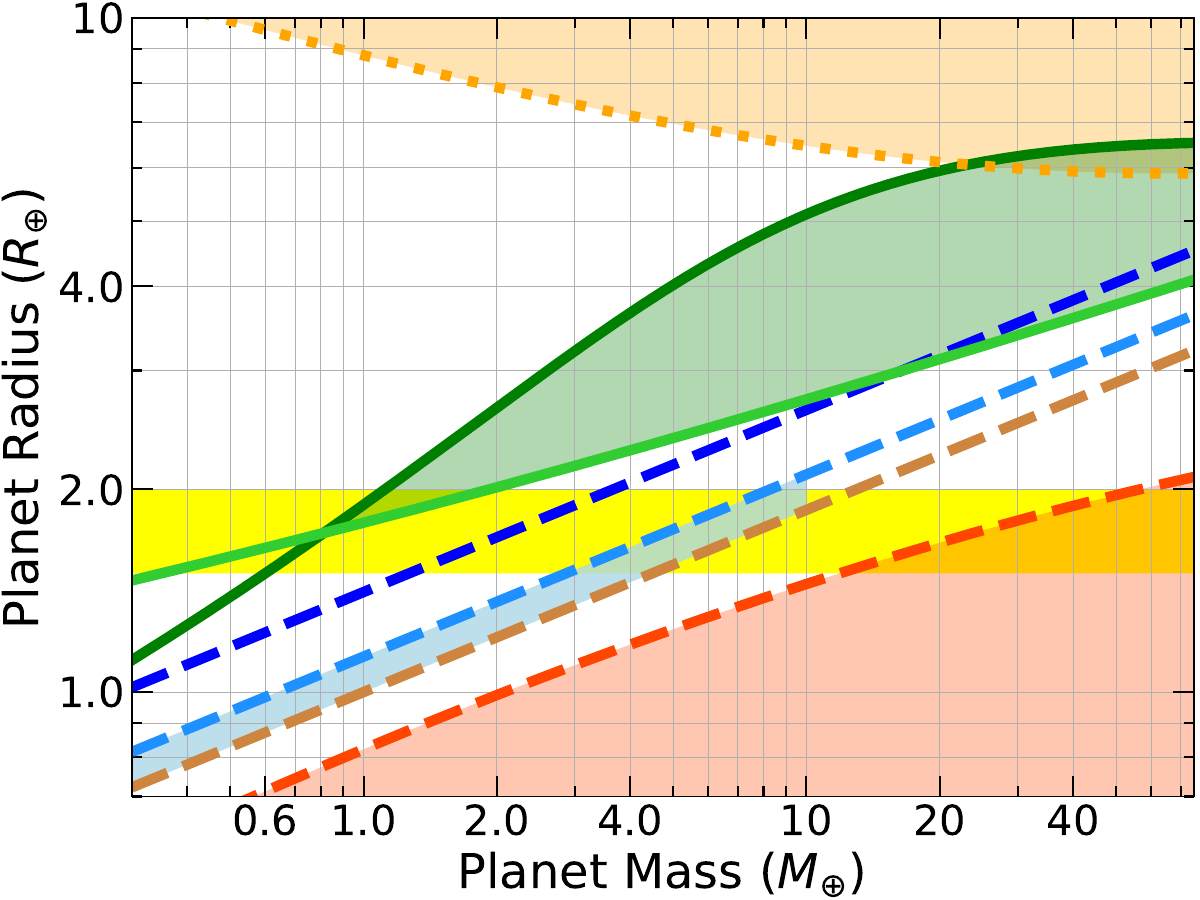}
\includegraphics[width=8.4cm]{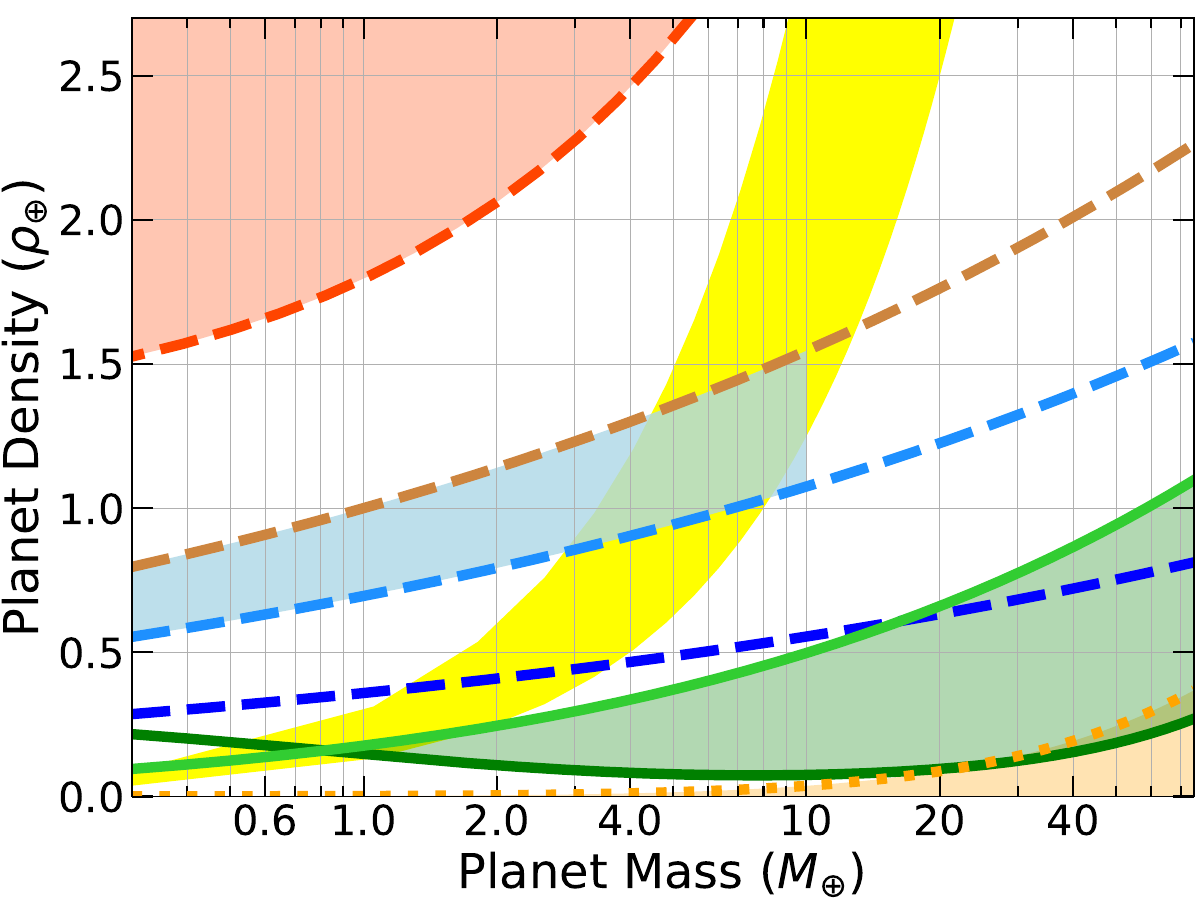}
\caption{Possible parameter spaces filled out by formation processes operating during the gas disk stage 
in the mass-radius and mass-density diagrams on the left and right panels, respectively.
Three characteristic compositions of planetary cores, Earth-like rock, water rich, and pure water, are denoted by the brown, light blue, and blue dashed lines, respectively (Table \ref{table1}).
Given that the ice-to-rock mass ratio of solar system comets is 0.3 at most (i.e., water rich) and that $10 M_{\oplus}$ is one canonical value for the critical core mass,
the area encompassed by the brown and light blue dashed lines with planet mass of $< 10 M_{\oplus}$ is shaded by blue.
For comparison, the mass-radius relation for pure iron planets is denoted by the red dashed line, and the red shaded region is the forbidden parameter space for any planets.
Planetary envelopes due to gas accretion and retention operating in gas disks define possible planet radii, 
which are represented by the green and light green solid lines, respectively 
(equations (\ref{eq:MXY_max}) and (\ref{eq:MXY_min})).
The area covered by these two lines is the parameter space, where planets with gaseous envelopes born out of gas disks reside, and is denoted by the green shaded region.
For comparison, planets that have equal core and envelope masses are denoted by the orange dotted line, beyond which gas giants distribute (the orange shaded region).
Finally, the radius valley that is used as a reference to differentiate between rocky and gaseous planets is shown by the yellow shaded region.}
\label{fig2}
\end{center}
%\end{figure}
\end{minipage}
\end{figure*}

Figure \ref{fig2} shows such a classification scheme in the mass-radius and mass-density diagrams. 
Three characteristic compositions of planetary cores are represented by the dashed lines (Table \ref{table1}).
The parameter space sandwiched by the Earth-like rock and water-rich lines with $M_{\rm p} < 10 M_{\oplus}$ is shaded by blue.
The choice of $10 M_{\oplus}$ is motivated by the critical core mass often quoted in the literature, 
above which gas accretion starts \citep[e.g.,][]{1996Icar..124...62P,2000ApJ...537.1013I}.
Planets without gaseous envelopes are therefore expected to distribute in this shaded region.

Iron is the heaviest element that can contribute to planet mass significantly.
While unlikely, the mass-radius relation for planets composed of pure iron is plotted as a reference in Figure \ref{fig2}.
The red shaded region is thus forbidden parameter space for any planets.

If planets possess gaseous envelopes, their size inflates, and they move to the green shaded region that is defined by equation (\ref{eq:Mxy_range}).
To maximize the area of the region, the composition of cores is chosen as water-rich when computing the planet radius for the case of $M_{XY, \rm max}$,
while Earth-like rock is used for the case of $M_{XY, \rm min}$.
As a result, the green shaded region is the most conservative estimate on the parameter space to be filled out by formation processes operating in gas disks.

For comparison, we also compute the radius of planets that are composed of 50 \% of cores and 50 \% of envelopes in mass, using equation (\ref{eq:R_env}).
We find that the resulting radius profile (the orange dotted line) intersects with the planet radius computed for the case of $M_{XY, \rm max}$ 
(the green solid line, equation (\ref{eq:MXY_max})) at the planet mass and radius of $20 M_{\oplus}$ and $6 R_{\oplus}$, respectively.
Importantly, that value of the planet mass is consistent with a prediction from the canonical picture of core accretion; 
once cores are as massive as $10 M_{\oplus}$, gas accretion starts,
and when the envelope mass becomes comparable to the core mass, rapid gas accretion takes place.
In other words, rapid gas accretion occurs for planets that have the total mass of $20 M_{\oplus}$, 
and for planets more massive than $20 M_{\oplus}$, they should have undergone rapid gas accretion and are classified as gas giants.
On the other hand, planets that are less massive than $20 M_{\oplus}$ did not experience rapid gas accretion and are classified as sub-giants.

In the following, we will apply this classification scheme to observed exoplanets to categorize them.

\subsection{Application to exoplanets} \label{sec:app}

\begin{figure*}
\begin{minipage}{17cm}
%\begin{figure}%[!ht]
\begin{center}
\includegraphics[width=15cm]{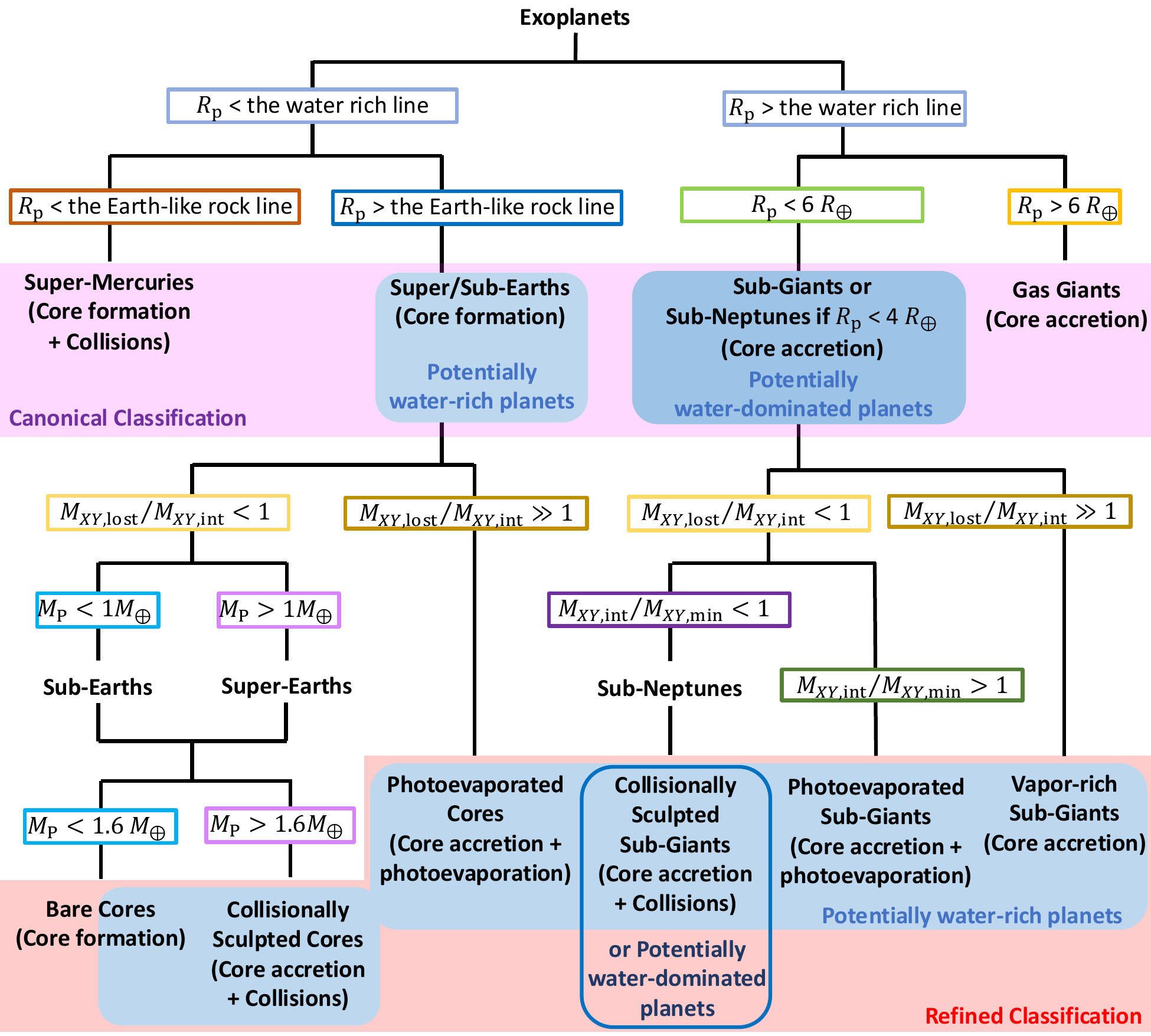}
\caption{Canonical and refined classification schemes of exoplanets.
Key planet properties and relevant physical processes are summarized.
In the canonical classification scheme, four types (super-Mercuries, super/sub-Earths, sub-giants and gas giants) are recovered 
as often used in the literature (Section \ref{sec:app}).
The refined classification scheme expands these types to eight in total due to evolution processes (Section \ref{sec:refine_class}):
super-Mercuries, bare cores, collisionally sculpted cores, photoevaporated cores, 
collisionally sculpted sub-giants, photoevaporated sub-giants, vapor-rich sub-giants, and gas giants.
Among them, massive bare cores, collisionally sculpted cores, photoevaporated cores, 
collisionally sculpted sub-giants, photoevaporated sub-giants, and vapor-rich sub-giants are potentially water-rich.}
\label{fig3}
\end{center}
%\end{figure}
\end{minipage}
\end{figure*}

Observed exoplanets exhibit huge diversity in their properties, 
and classification of them is one key means to constrain their current properties and origins.
We utilize the classification scheme discussed in the above section, 
which is based on various mass-radius relations and formation processes operating in gas disks;
for solid planets without gaseous envelopes, their radius is well defined, 
and application of our formulation is straightforward.
For planets with gaseous envelopes, however, the definition of planet radius depends on many factors.
The radius of most exoplanets is measured by the {\it Kepler} telescope, 
and for such optical observations, where the slant viewing geometry is achieved,
it is appropriate to define a planet's radius at 20 mbar for solar metallicity.
The same definition is adopted when deriving equation (\ref{eq:R_env})
with which the radius of gaseous envelopes is computed \citep{2014ApJ...792....1L}.
We therefore consider that our formulation is directly applicable to observed systems.
As described below, we recover four exoplanet classes widely used in the literature (Figure \ref{fig3}).

First, we divide the full sample into two sub-groups by the water-rich line (Table \ref{table1});
if planets distribute above the line, they are very likely to possess gaseous envelopes,
while if they are below the line, they are composed predominantly of solid materials such as rock and ice.

Second, we further divide these two sub-groups into two sub-classes, namely, the total of four classes;
for planets that have gaseous envelopes, we use $R_{\rm p} = 6 R_{\oplus}$ as the threshold radius,
above which planets are called gas giants, and below which planets are classified as sub-giants.
This nomenclature is motivated by the canonical picture of core accretion as discussed in Section \ref{sec:class}.
Sub-giants can be called sub-Neptunes if their radius is smaller than $4 R_{\oplus}$.
For planets that are likely to be free of envelopes, we use the Earth-like rock line (Table \ref{table1}),
above which they contain some amount of water (up to 25 \% in mass) and are referred to as super/sub-Earths.
Some of these planets are potentially water-rich. 
However, they are distinguished from water-dominated (i.e., nearly pure water) planets.
For planets below the Earth-like rock line, they have higher iron mass fractions than the Earth and are called super-Mercuries.

Once classification is done, we compute the composition of planets.
For super-Mercuries, we estimate the corresponding value of the iron mass fraction, using the look-up table developed by 
\citet{2016ApJ...819..127Z}\footnote{\url{https://lweb.cfa.harvard.edu/~lzeng/planetmodels.html\#mrrelation}};
with given values of planet mass and radius, the required value of  the iron mass fraction is estimated by interpolation.
For super/sub-Earths, the radius enhancement factor ($f_{\rm R}$), that is, the mass fraction of water ice ($f_{\rm H_2O}$) in cores,
is computed directly from equation (\ref{eq:Rcore}) for a given set of planet mass and radius.
For sub-giants, the envelope mass is computed self-consistently from the current value of planet mass and radius, using equations (\ref{eq:Rcore}) and (\ref{eq:R_env}),
if the radius enhancement factor ($f_{\rm R}$, i.e. the water-ice mass fraction of cores) is pre-determined, 
which is not possible due to degeneracy between the envelope radius and the core radius.
We therefore pick up the value of $f_{\rm R}$ randomly from the range between 1.0 to $\sim 1.29$ uniformly (Table \ref{table1});
as described in Section \ref{sec:core_comp}, the choice of the range is motivated by solar system comets.
For gas giants, the envelope mass dominates over the solid mass, and it is hard to reliably compute these masses from the formulation discussed above.
Since many of our samples distribute around $R_{\rm p} = 6 R_{\oplus}$, we assume that $M_{XY} = M_{Z}$.

\begin{figure*}
\begin{minipage}{17cm}
%\begin{figure}%[!ht]
\begin{center}
\includegraphics[width=8.4cm]{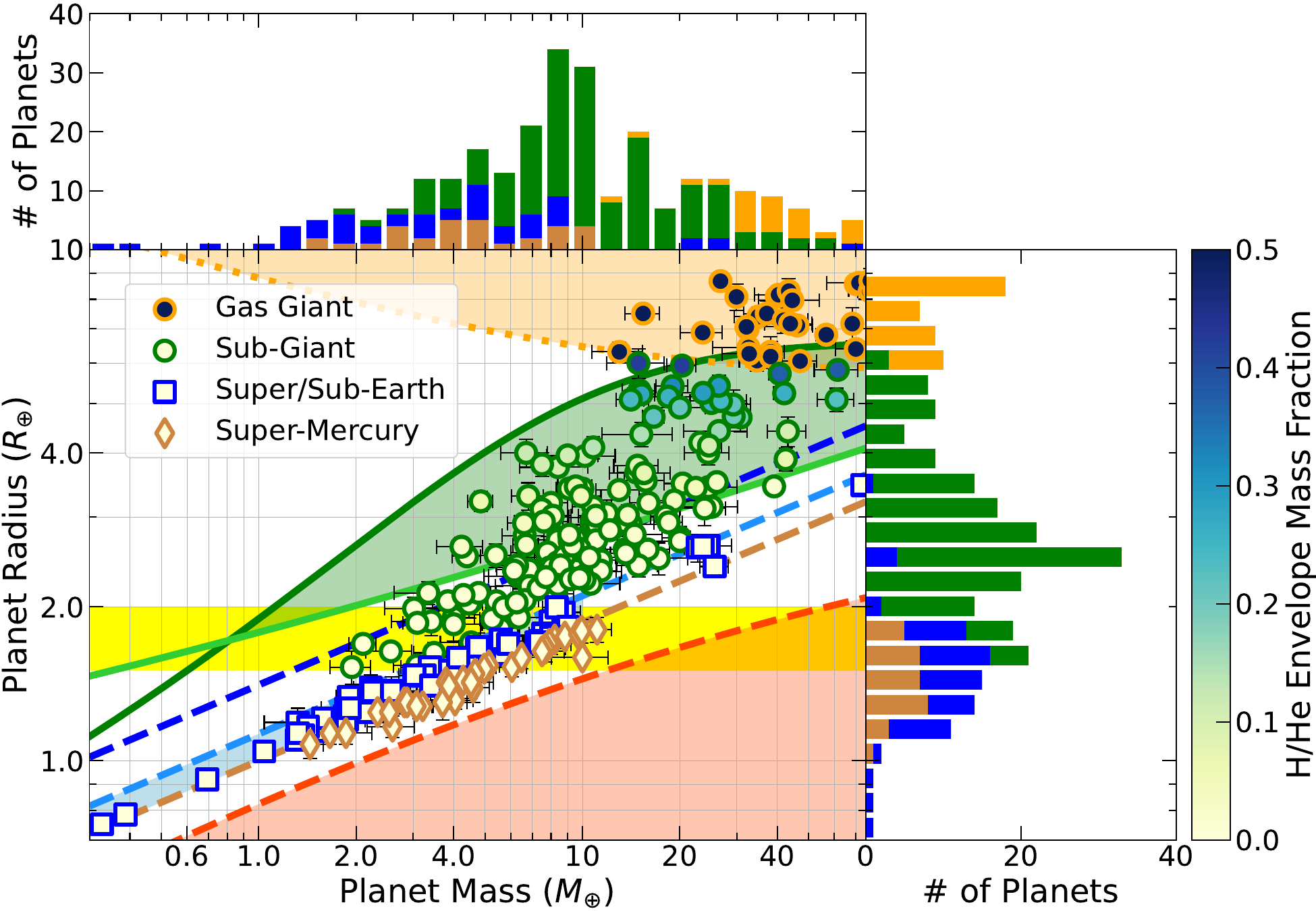}
\includegraphics[width=8.4cm]{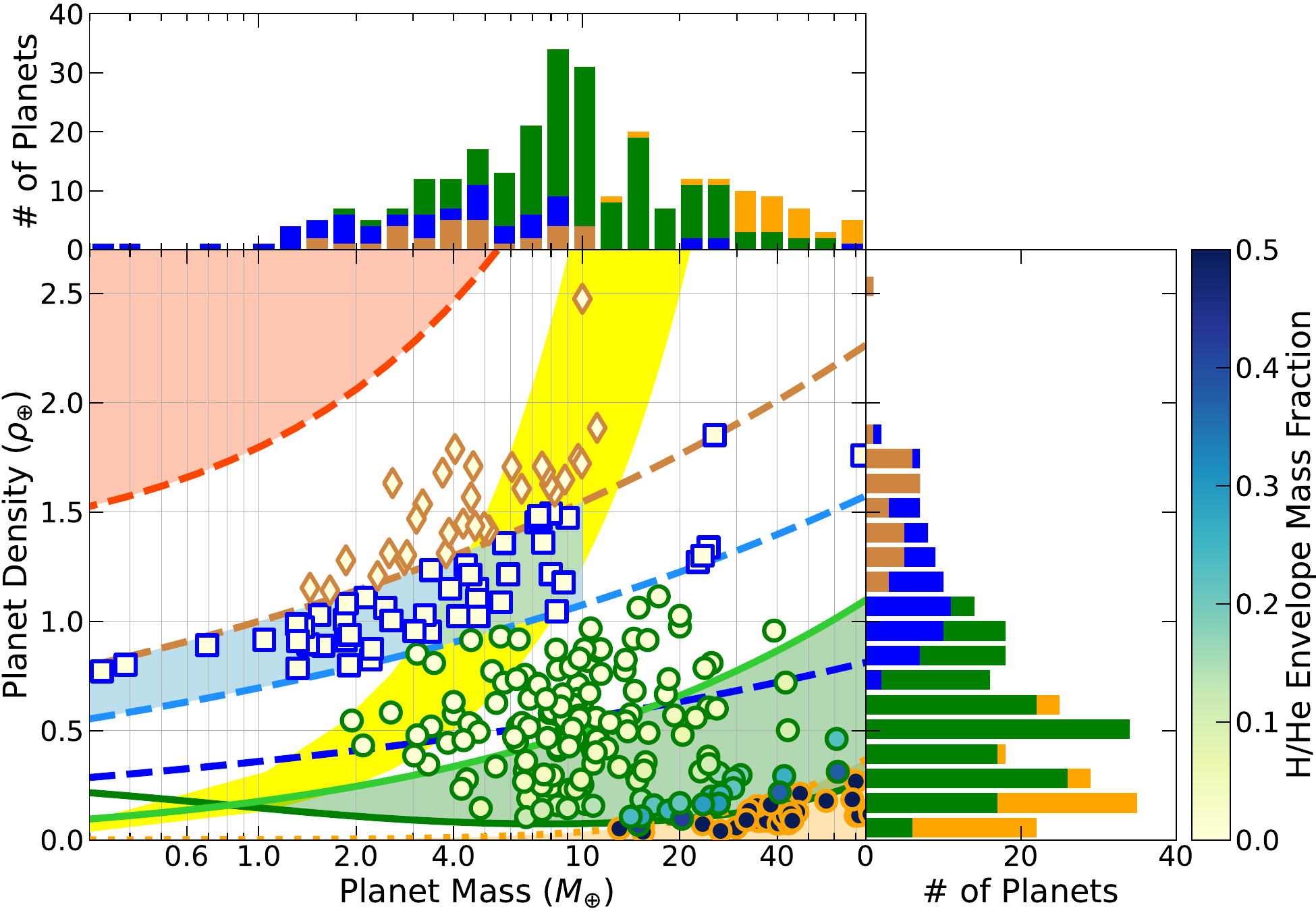}
\includegraphics[width=8.4cm]{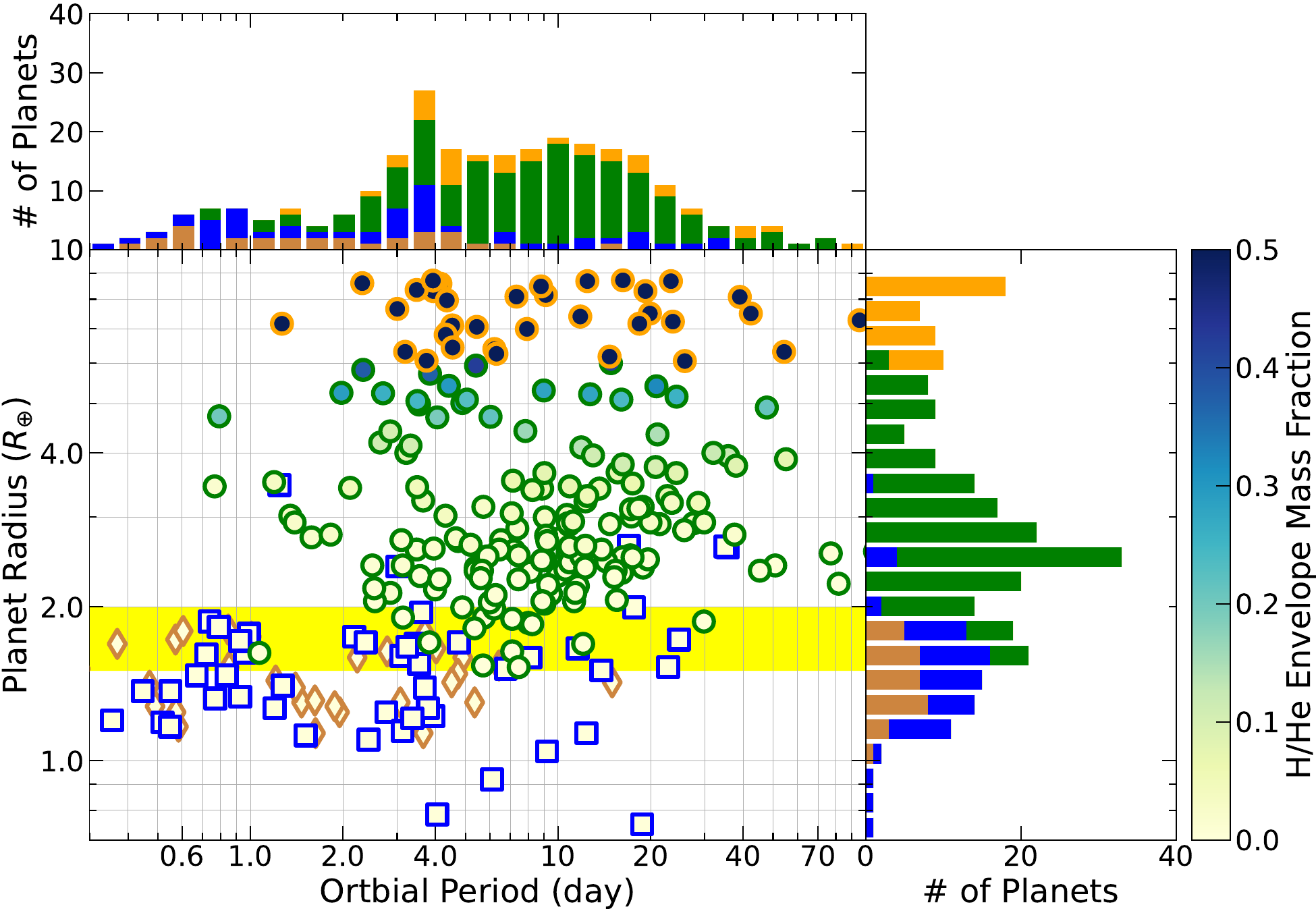}
\includegraphics[width=8.4cm]{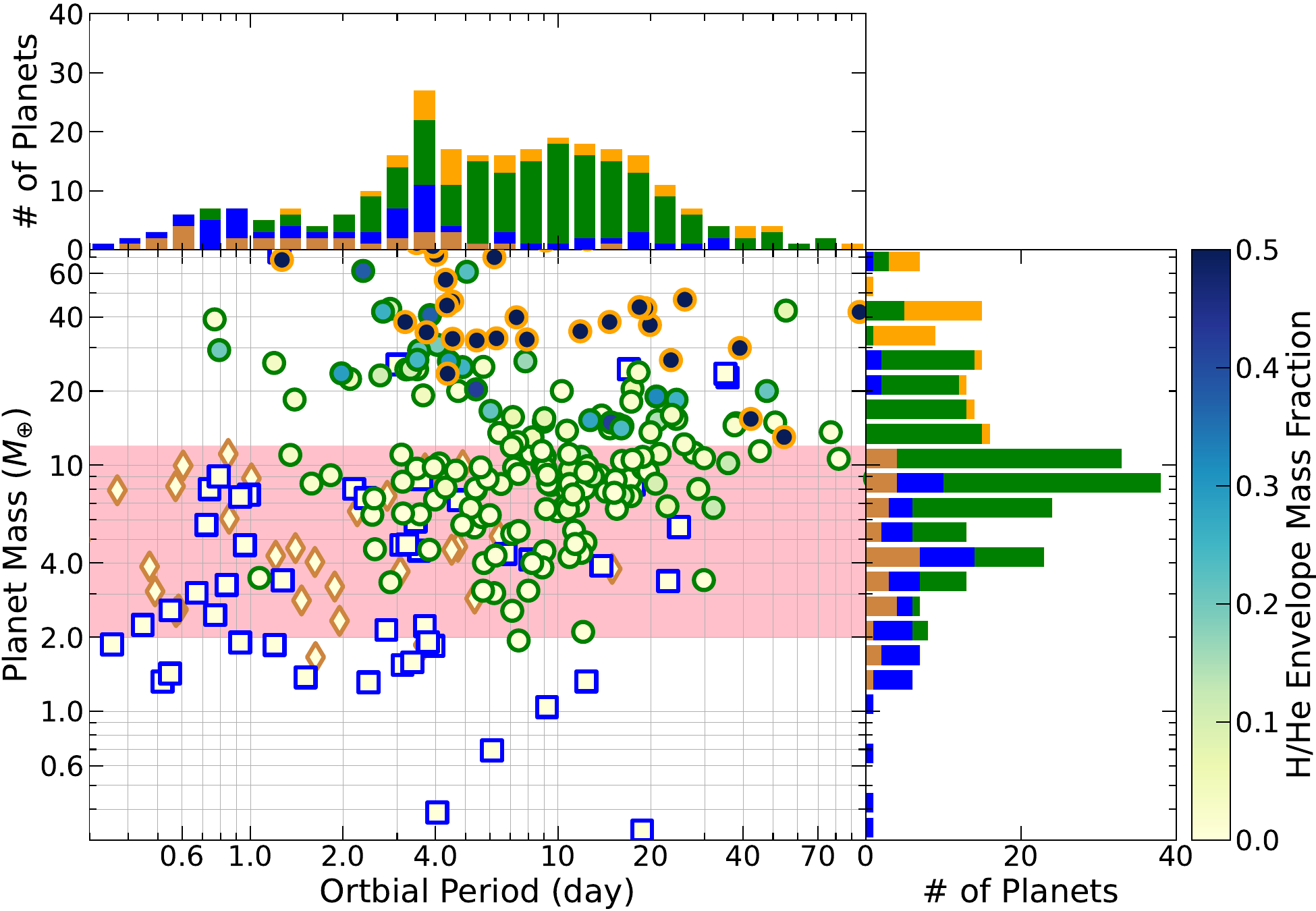}
\caption{Classification of observed exoplanets. 
On the upper left and right panels, the mass-radius and mass-density diagrams are shown (as in Figure \ref{fig2}),
while on the lower left and right panels, the orbital period-radius and orbital period-mass diagrams are exhibited.
On each panel, two histograms are plotted along the x- and y-axes, and the computed value of the envelope mass fraction is shown with the color bar.
On the left and the upper right panels, the radius valley is denoted by the yellow shaded region, 
and on the lower right panel, the red shaded region represents the parameter space where sub-giants overlap with super-Mercuries or super/sub-Earths.
Exoplanets are classified by the current values of mass and radius (Figure \ref{fig3}):
Exoplanets that have radius larger than $6 R_{\oplus}$ are classified as gas giants (the circle with the orange edge).
Exoplanets that are smaller than $6 R_{\oplus}$ but larger than the water rich line (Table \ref{table1}) are called sub-giants (the circle with the green edge).
Exoplanets that are distributed between the water rich and Earth-like rock lines (Table \ref{table1}) are super/sub-Earths (the square with the blue edge).
Exoplanets distributing below the Earth-like rock line are referred to as super-Mercuries (the diamond with the brown edge).
A wide range of the bulk density of super-Mercuries implies the diversity of energetic events to remove silicate mantle,
while that of super-Earths infers an abundant population of water-rich planets.
Remarkably, many sub-giants are distributed between the water rich (Table \ref{table1}) and the envelope retention (equation (\ref{eq:MXY_min})) lines, 
suggesting the importance of subsequent evolution processes that remove primordial gaseous envelopes partially.}
\label{fig4}
\end{center}
%\end{figure}
\end{minipage}
\end{figure*}

Figure \ref{fig4} shows the resulting classification and computed value of the envelope mass fraction (i.e., $M_{XY}/M_{\rm p}$) with the color bar.
It is obvious that the mass-radius diagram is suitable for examining the population of sub-giants and gas giants,
while the mass-density diagram is preferred for exploring the population of super/sub-Earths and super-Mercuries.

Our classification scheme makes it clear that the distribution of observed exoplanets in these diagrams is shaped by both formation and evolution processes.
First, a good number of super-Mercuries are present. 
Removal of silicate mantles is required to achieve such high bulk densities, 
and these planets should have experienced highly energetic events such as catastrophic collisions.
A wide spread in their bulk density infers a large range of impact velocity.
The orbital distribution of super-Mercuries is comparable to that of super/sub-Earths,
implying that the occurrence of the energetic events is probabilistic.

Second, super/sub-Earths, which are free of envelopes, are distributed mainly below the planet mass of $10 M_{\oplus}$,
suggesting that the maximum value of the critical core mass is about $10 M_{\oplus}$.
Also, these planets tend to orbit in the vicinity of the host star, which supports the importance of subsequent envelope loss triggered by radiation from the star.
Planets that are more massive than $1 M_{\oplus}$ have a wide range of bulk density, some of which are water-rich.

Third, many sub-giants and gas giants are distributed in the green and orange shaded regions, respectively.
The current planet radius can thus be explained by the presence of  primordial H/He envelopes.
Remarkably, a good number of sub-giants are also distributed below the green shade region in the mass-radius diagram or above in the mass-density diagram;
such a trend becomes more visible in the latter.
At least three scenarios are considered to understand these planets: 
one is that while they initially possessed more massive gaseous envelopes, so that they distributed in the green shaded region originally,
they lost some of these primordial envelopes by subsequent evolution processes and moved to the current position in these diagrams.
The other scenarios are involved with water, that is,
these planets have higher water abundance in cores (i.e., $f_{\rm H_2O}$), namely water-dominated planets,
or the presence of water in envelopes may account for the observed radius without invoking such planets \citep[e.g.,][]{2023ApJ...953...57K,2025ApJ...988..186A}.
The latter two scenarios cannot be ruled out currently, while there are no water-dominated planets in the solar system.

In the following, we explore the former scenario, that is, the current, very tenuous envelope is the outcome of subsequent evolution processes.

\section{Post-formation processes} \label{sec:post}

Subsequent evolution processes play a critical role in determining the present properties of planets.
One clear example is the discovery of a radius valley \citep{2017AJ....154..109F};
as described above, the valley is carved by removal of primordial envelopes \citep[e.g.,][]{2012ApJ...761...59L,2012MNRAS.425.2931O}.
We here consider photoevaporation driven by host stellar irradiation and collisions between planets that result in mass growth and loss.

\subsection{Radiation-driven envelope loss} \label{sec:post_photo}

We discuss how radiation-driven envelope loss affects the properties of planets and hence their position in the mass-radius and mass-density diagrams.

We here focus on photoevaporation; it is viewed as one leading process to explain the bimodal radius distribution of close-in planets
\citep[e.g.,][]{2012MNRAS.425.2931O,2012ApJ...761...59L}.
Other processes such as core-powered mass loss \citep[e.g.,][]{2018MNRAS.476..759G,2019MNRAS.487...24G} provide important and complementary contributions.
However, photoevaporation eventually comes into play, and close-in planets cannot escape from it.
In fact, previous work confirms that photoevaporation follows spontaneous/boil-off and core-powered mass loss \citep{2024MNRAS.529.2716R}.

The efficiency of photoevaporative mass loss is quantified by calculating how much envelope mass is actually lost by stellar irradiation.
The ratio of computed lost mass ($M_{XY, \rm lost}$) to the initial mass ($M_{XY, \rm int}$) may be written as \citep{2013ApJ...776....2L}
\begin{equation}
\label{eq:massloss_PE}
\frac{M_{XY, \rm lost}}{M_{XY, \rm int}} = 0.5 \left ( \frac{F_{\rm p}}{ F_{\rm th}}\right)^{1.1},
\end{equation}
where $F_{\rm p}$ is the stellar flux planets receive currently, and $F_{\rm th}$ is the threshold flux, which is written as
\begin{equation}
\label{eq:massloss_PE_th}
F_{\rm th} = 0.5 F_{\oplus} \left( \frac{ M_{\rm core} }{ M_{\oplus} } \right)^{2.4} \left( \frac{ \epsilon }{ 0.1} \right)^{-0.7},  
\end{equation}
$F_{\oplus}$ is the solar flux the Earth receives currently, and $\epsilon$ is the photoevaporation efficiency.
Equations (\ref{eq:massloss_PE}) and (\ref{eq:massloss_PE_th}) are derived from the parameter study covering a wide range of model parameters
under the assumption that $M_{XY, \rm int}$ is composed purely of hydrogen and helium.
It should be noted that evolution of both stellar flux and planet radius is taken into account when deriving the formula.
In this study, we adopt that $\epsilon=0.1$.

It is obvious from equation (\ref{eq:massloss_PE}) that if $F_{\rm p} > 2 F_{\rm th}$, 
then the computed envelope mass being lost by photoevaporation exceeds the available mass;
in reality, the value of $M_{XY, \rm lost}$ is capped by that of $M_{XY, \rm int}$.
Thus, the condition that $M_{XY, \rm lost}/M_{XY, \rm int} \gg 1$ represents the strong photoevaporation regime.

On the other hand, if planets currently keep gaseous envelopes even under the action of photoevaporation, 
then the initial envelope mass is retrieved as (equation (\ref{eq:massloss_PE}))
\begin{equation}
\label{eq:MXY_int}
\frac{ M_{XY, \rm int} }{M_{XY}} =  \left[ 1-  0.5 \left ( \frac{F_{\rm p}}{ F_{\rm th}}\right)^{1.1} \right]^{-1},
\end{equation}
where $M_{XY} = M_{XY, \rm int}- M_{XY, \rm lost}$ is the current envelope mass.
In order for photoevaporation to reduce the envelope mass considerably,
$F_{\rm p}$ should be sufficiently strong 
such that the retrieved value of $M_{XY, \rm int}$ from equation (\ref{eq:MXY_int}) becomes much higher than that of $M_{XY}$, that is $M_{XY, \rm int} / M_{XY} \gg 1$.
If photoevaporation is negligible, then $ M_{XY, \rm int}/ M_{XY} \sim 1$.
As described above, possible values of $M_{XY}$ are constrained by gas accretion and retention processes (equation (\ref{eq:Mxy_range})).
Accordingly, the condition that $M_{XY, \rm int} /M_{XY, \rm min} \ll 1$ becomes a stringent lower limit for negligible photoevaporative mass loss.

In summary, two regimes of photoevaporation are considered:
\begin{equation}
\frac{M_{XY, \rm lost}}{M_{XY, \rm int}} \gg 1 \mbox{ for the strong photoevaporation regime,}
\end{equation}
\begin{equation}
\frac{ M_{XY, \rm int} }{M_{XY, \rm min}} \ll 1  \mbox{ for the negligible photoevaporation regime.}
\end{equation}

\begin{figure*}
\begin{minipage}{17cm}
%\begin{figure}%[!ht]
\begin{center}
\includegraphics[width=8.4cm]{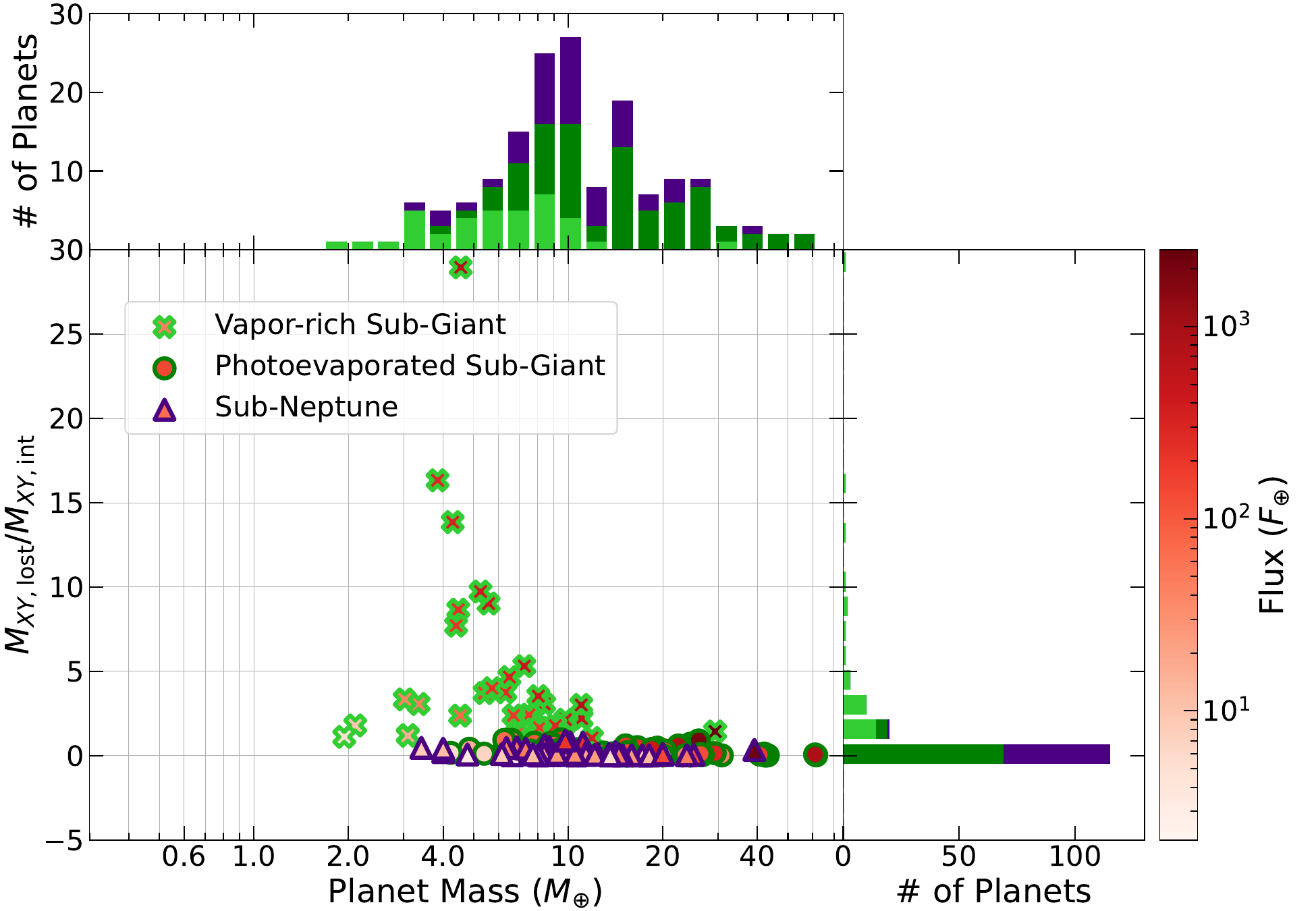}
\includegraphics[width=8.4cm]{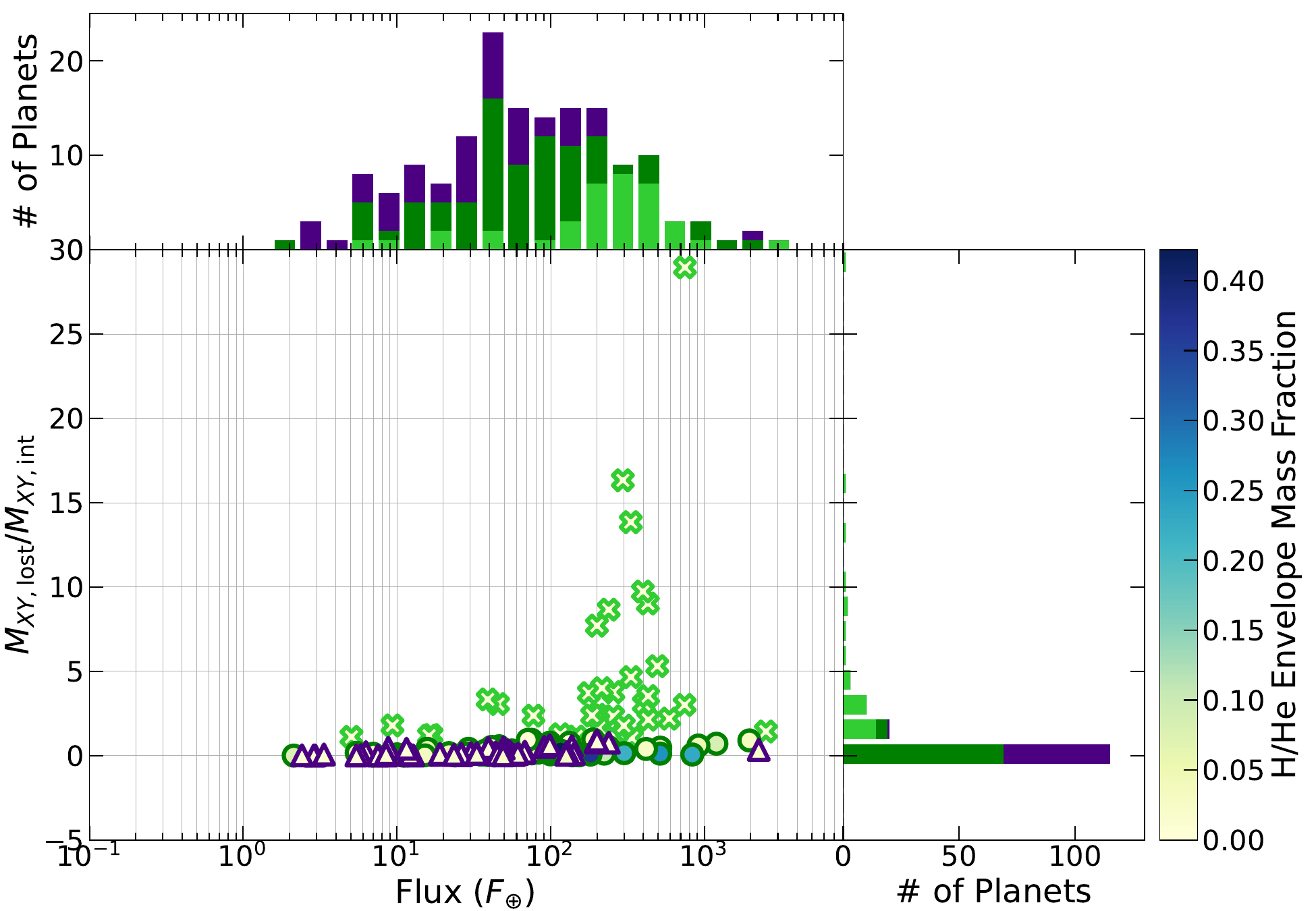}
\includegraphics[width=8.4cm]{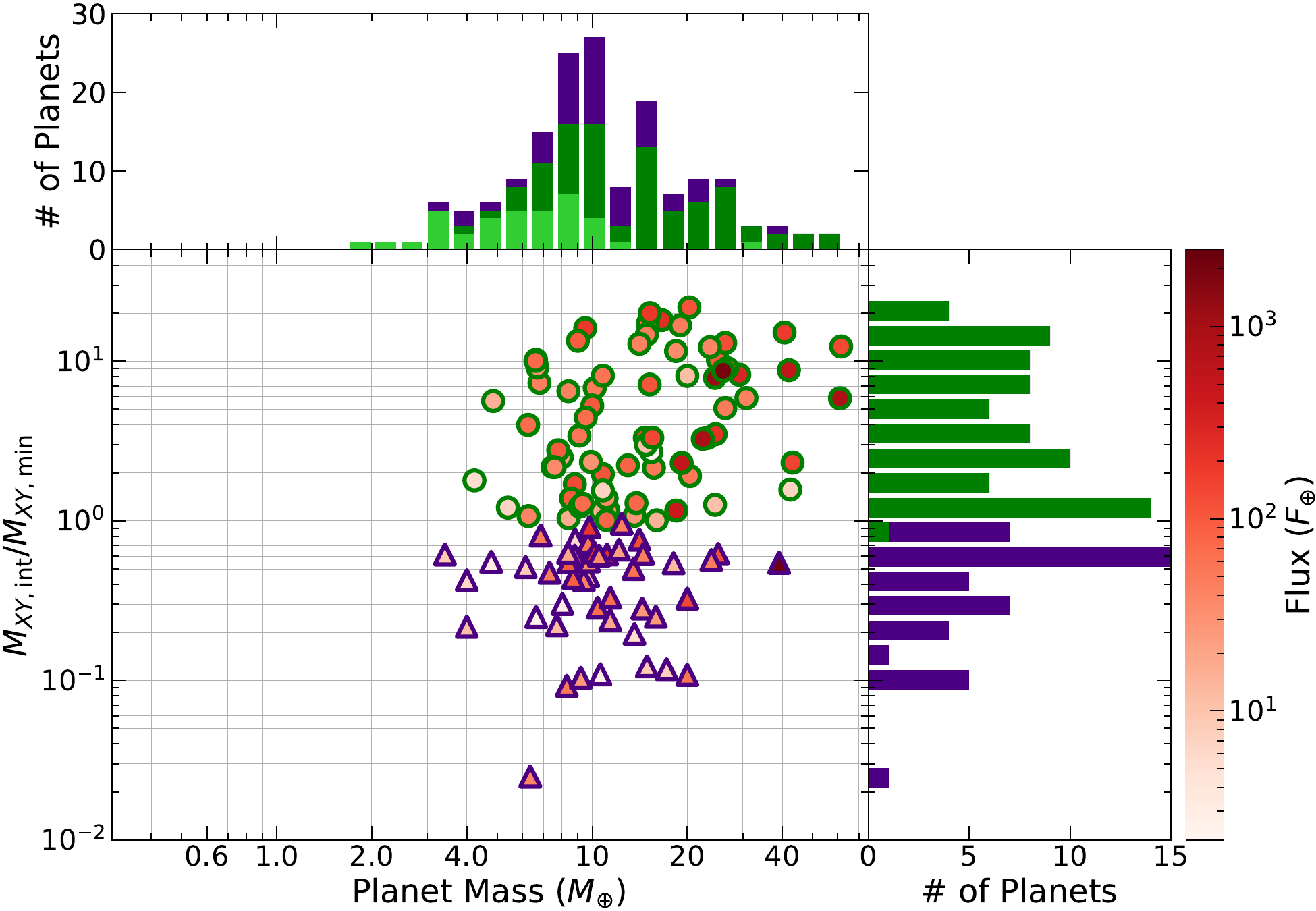}
\includegraphics[width=8.4cm]{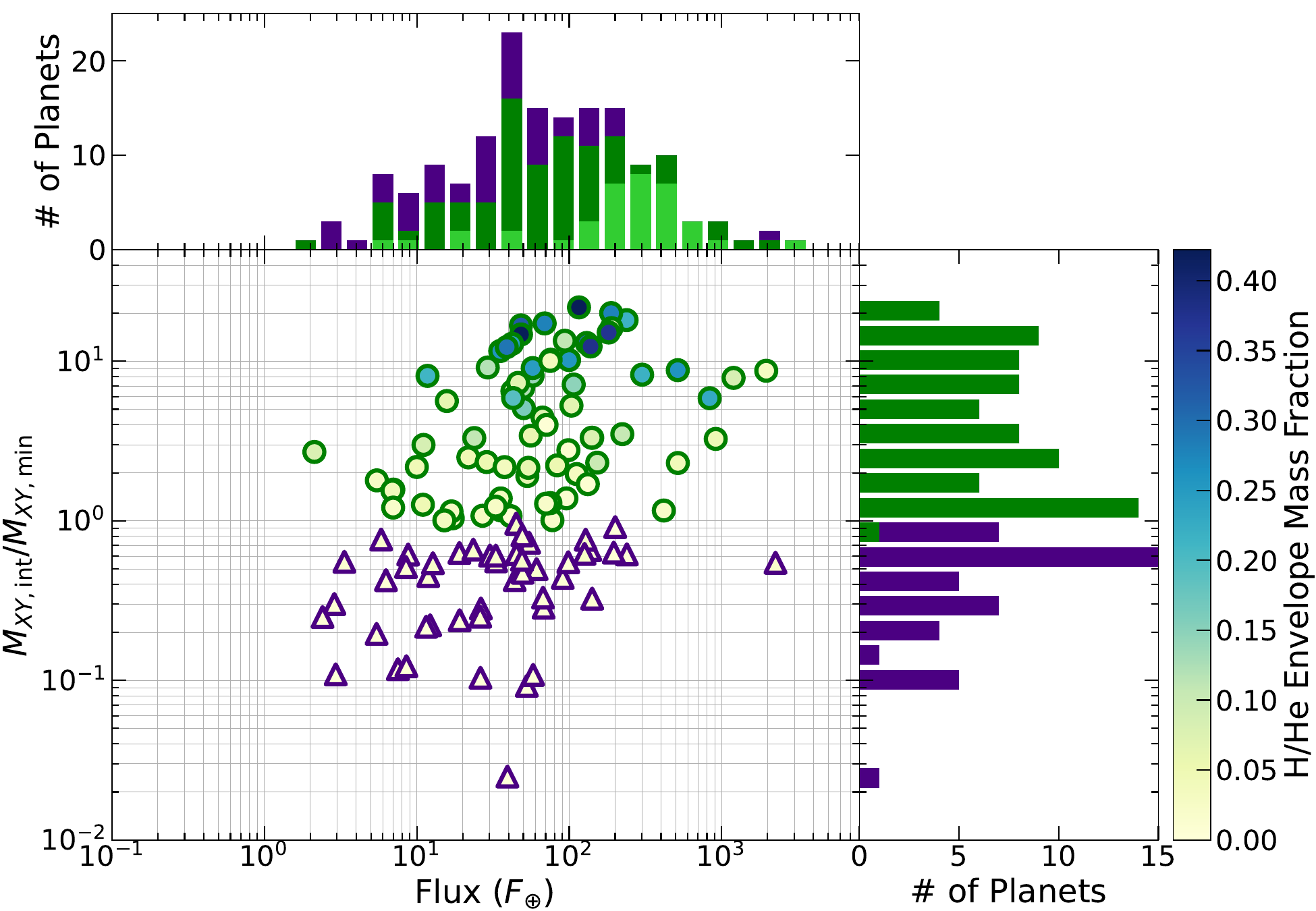}
\caption{The mass ratio of envelopes that are accreted during the gas disk stage and can be lost subsequently by photoevaporation
as a function of planet mass and insolation flux on the left and right panels, respectively.
Exoplanets classified as sub-giants in Figure \ref{fig4} are only considered.
On the upper panels, the computed value of $M_{XY, \rm lost}/M_{XY, \rm int}$ is plotted,
while on the lower panels, the computed value of $M_{XY, \rm int}/M_{XY, \rm min}$ is shown.
The color bar on the left panels represents insolation flux planets receive currently,
and that on the right panels denotes the computed envelope mass fraction.
Sub-giants that have $M_{XY, \rm lost}/M_{XY, \rm int} \gg 1$ are referred to as vapor-rich sub-giants (the cross with the light green edge)
due to the apparent contradiction of the present large planet radius against very high photoevaporation efficiencies.
Sub-giants that satisfy the condition that $M_{XY, \rm lost}/M_{XY, \rm int} <1$ and $M_{XY, \rm int}/M_{XY, \rm min} > 1$ are defined 
as photoevaporated sub-giants (the circle with the green edge)
because they clearly experienced gas accretion and maintain (some of) primordial envelopes even under photoevaporative mass loss.
Sub-giants that have $M_{XY, \rm int}/M_{XY, \rm min} < 1$ are temporarily called sub-Neptunes (the upper triangle with the purple edge).
The origin of these planets needs to be explored further because gas accretion/retention and photoevaporation cannot reproduce their too tenuous envelopes 
if planetary cores were as massive as currently during the gas disk stage (Section \ref{sec:post_impact}).}
\label{fig5}
\end{center}
%\end{figure}
\end{minipage}
\end{figure*}

Figure \ref{fig5} shows how these envelope mass ratios behave for observed exoplanets classified as sub-giants.
We find that sub-giants are divided into three types, which are referred to as 
vapor-rich sub-giants, photoevaporated sub-giants, and sub-Neptunes (Figure \ref{fig3}).
First, vapor-rich sub-giant planets are defined such that they satisfy the condition that $M_{XY, \rm lost}/M_{XY, \rm int} \gg 1$.
The name of vapor-rich sub-giants is motivated by that 
while these planets are expected to currently have gaseous envelopes under the classification scheme developed in Section \ref{sec:current},
very high photoevaporation efficiencies disfavor it.
This (apparent) contradiction may point out invalidity of the assumption that gaseous envelopes are composed purely of hydrogen and helium,
which was used in previous studies;
if other volatiles such as water are present as vapor in the envelopes, photoevaporation efficiencies drop, 
and survival of gaseous envelopes against photoevaporation becomes less problematic to account for the current large planet radius. 

Second, photoevaporated sub-giants are defined by the condition that $M_{XY, \rm lost}/M_{XY, \rm int} < 1$ and  $M_{XY, \rm int}  / M_{XY, \rm min} > 1$;
the latter makes sure that while these planets underwent gas accretion when gas disks were present, 
some factions of gaseous envelopes were lost by photoeveporation.
Among the samples used in this work, planets with the mass of $\gtrsim 4 M_{\oplus}$ are classified as photoevaporated sub-giants,
indicating that the minimum value of the critical core mass is about $4 M_{\oplus}$.

Third, the name of sub-Neptunes is temporarily used for exoplanets fulfilling the condition that $M_{XY, \rm lost}/M_{XY, \rm int} < 1$ and $M_{XY, \rm int}  / M_{XY, \rm min} < 1$.
The presence of such planets requires a more careful examination
because their cores are massive enough to trigger gas accretion and indeed have tenuous envelopes.
However, these envelopes are too tenuous to be explained by gas accretion and retention, photoevaporation, or the combination of these three processes 
if planetary cores would have gained the current mass during the gas disk stage.
We propose that these planets are the outcome of collisions after the gas disk stage and discuss more in Section \ref{sec:post_impact}.

We now turn our attention to super/sub-Earths.
Computation of $ M_{XY, \rm int}/ M_{XY, \rm min}$ is not possible for these planets because they are very unlikely to have envelopes now.
We therefore focus on the ratio of $M_{XY, \rm lost}/M_{XY, \rm int}$.

\begin{figure*}
\begin{minipage}{17cm}
%\begin{figure}%[!ht]
\begin{center}
\includegraphics[width=8.4cm]{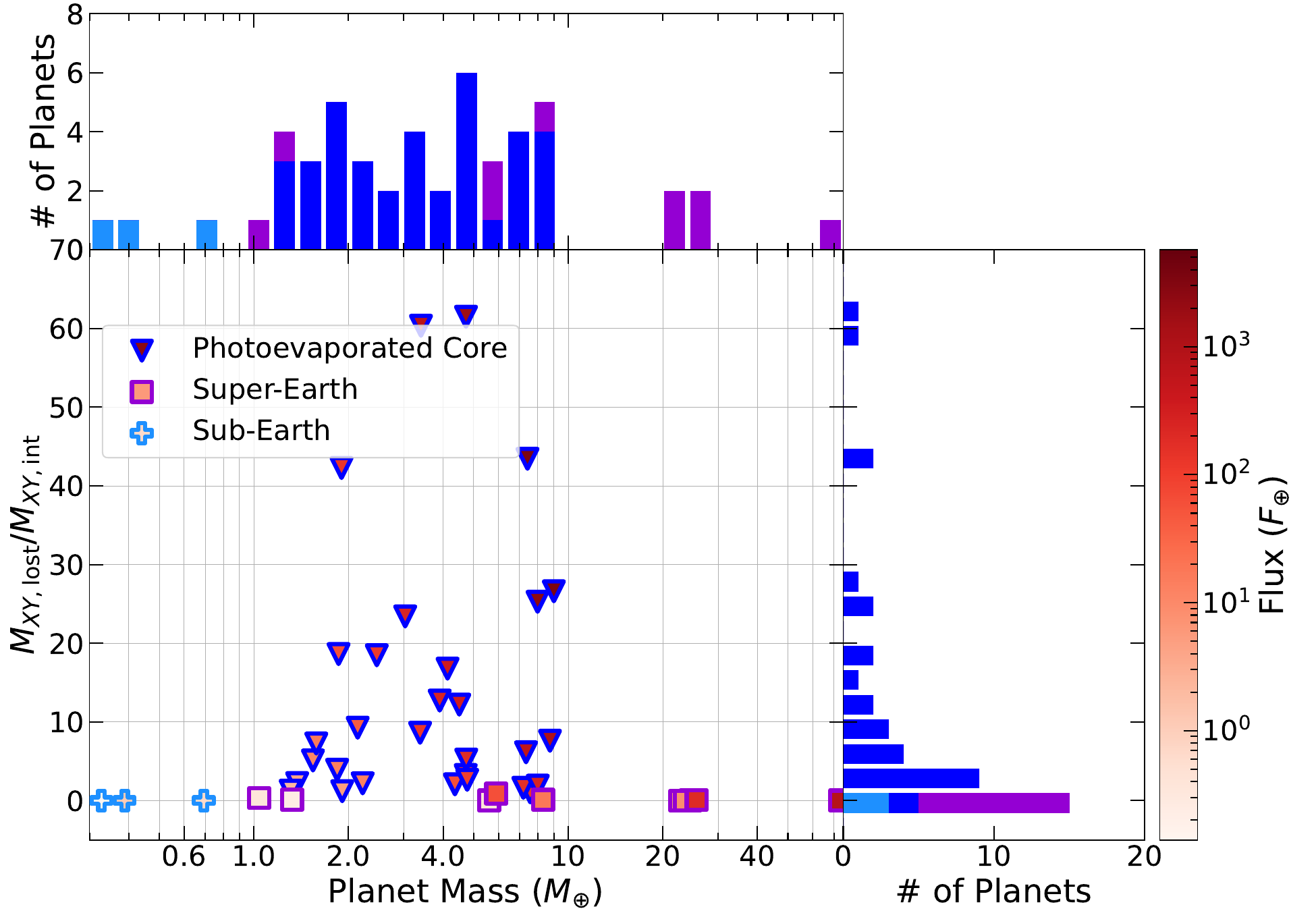}
\includegraphics[width=8.4cm]{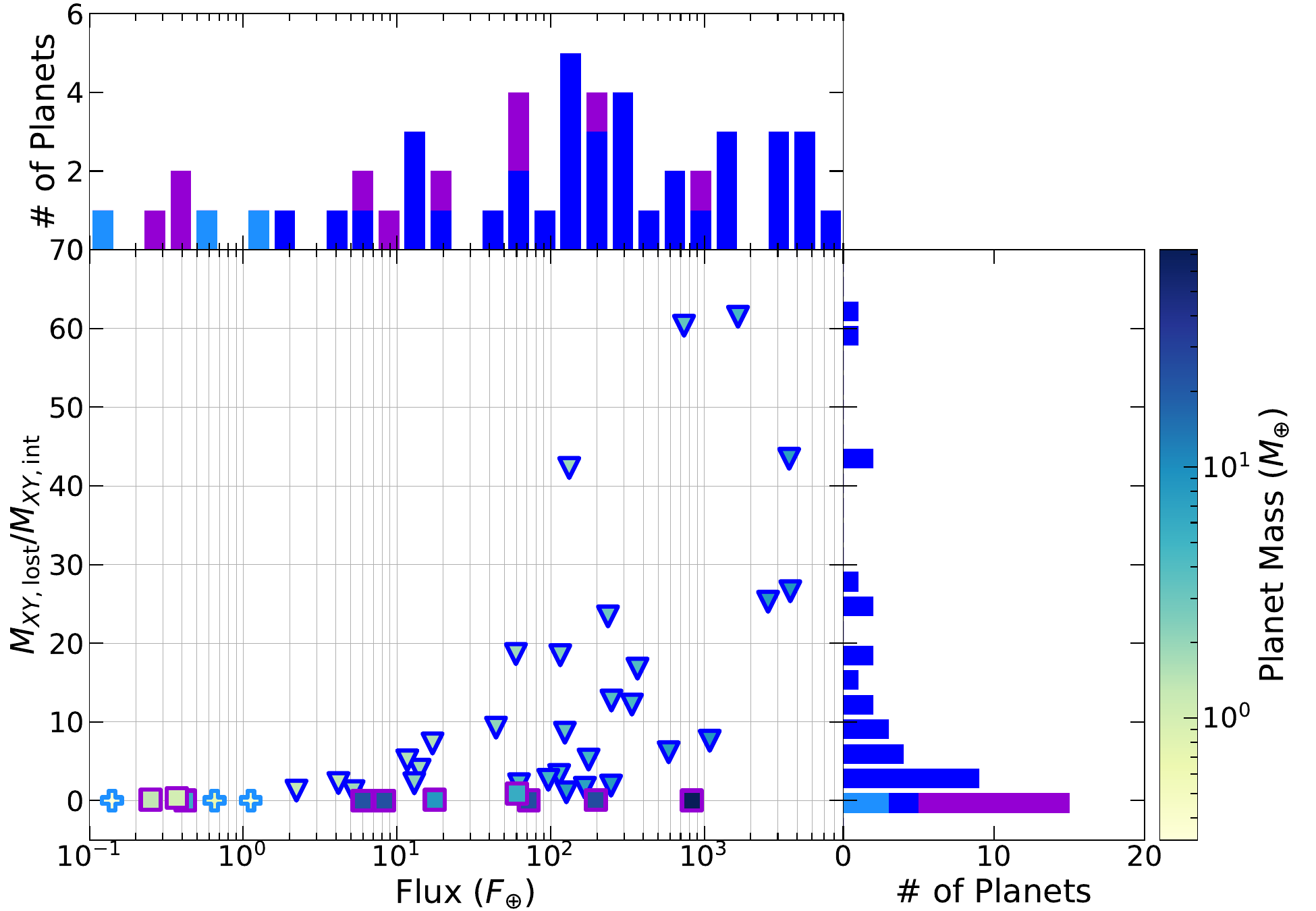}
\includegraphics[width=8.4cm]{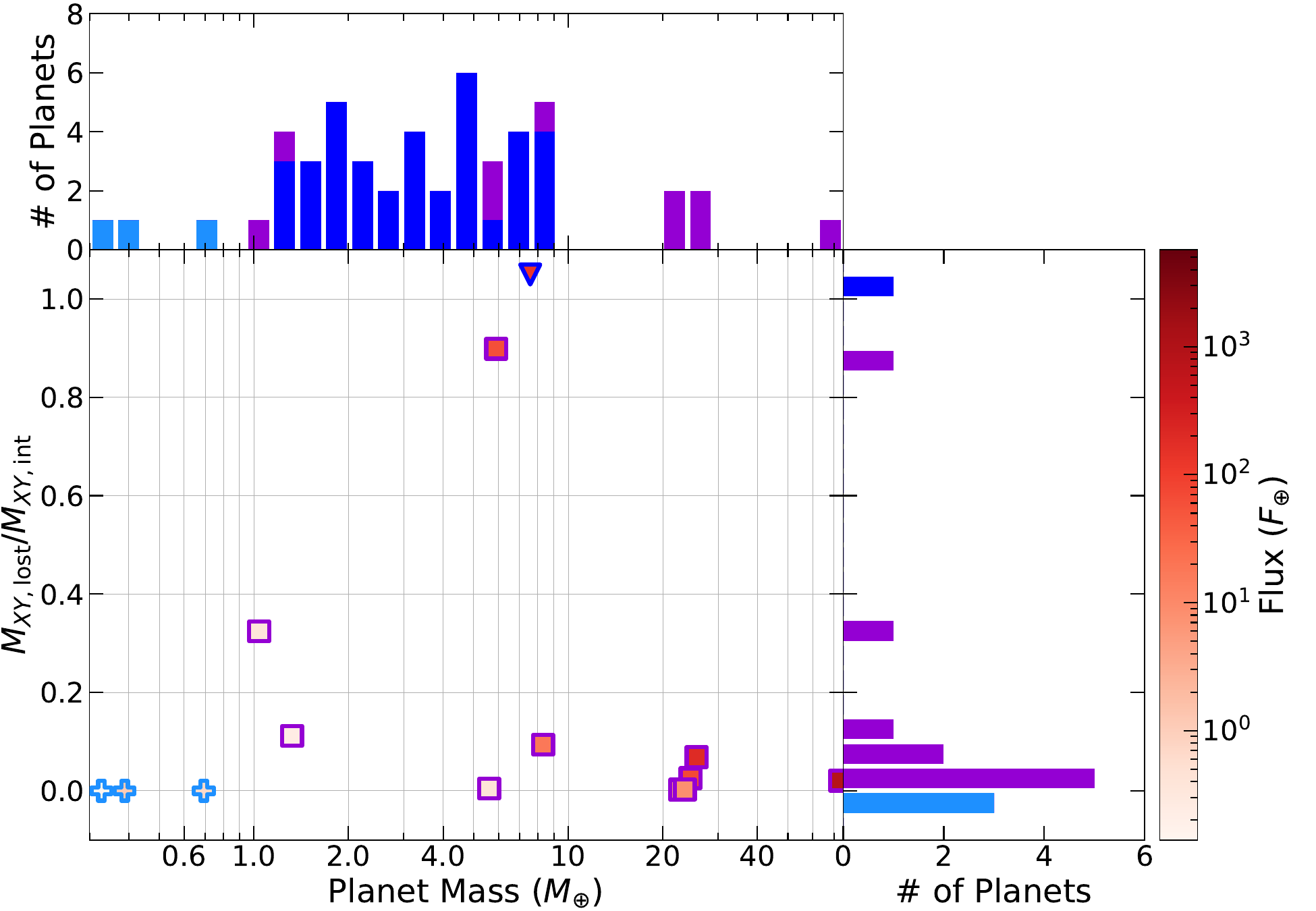}
\includegraphics[width=8.4cm]{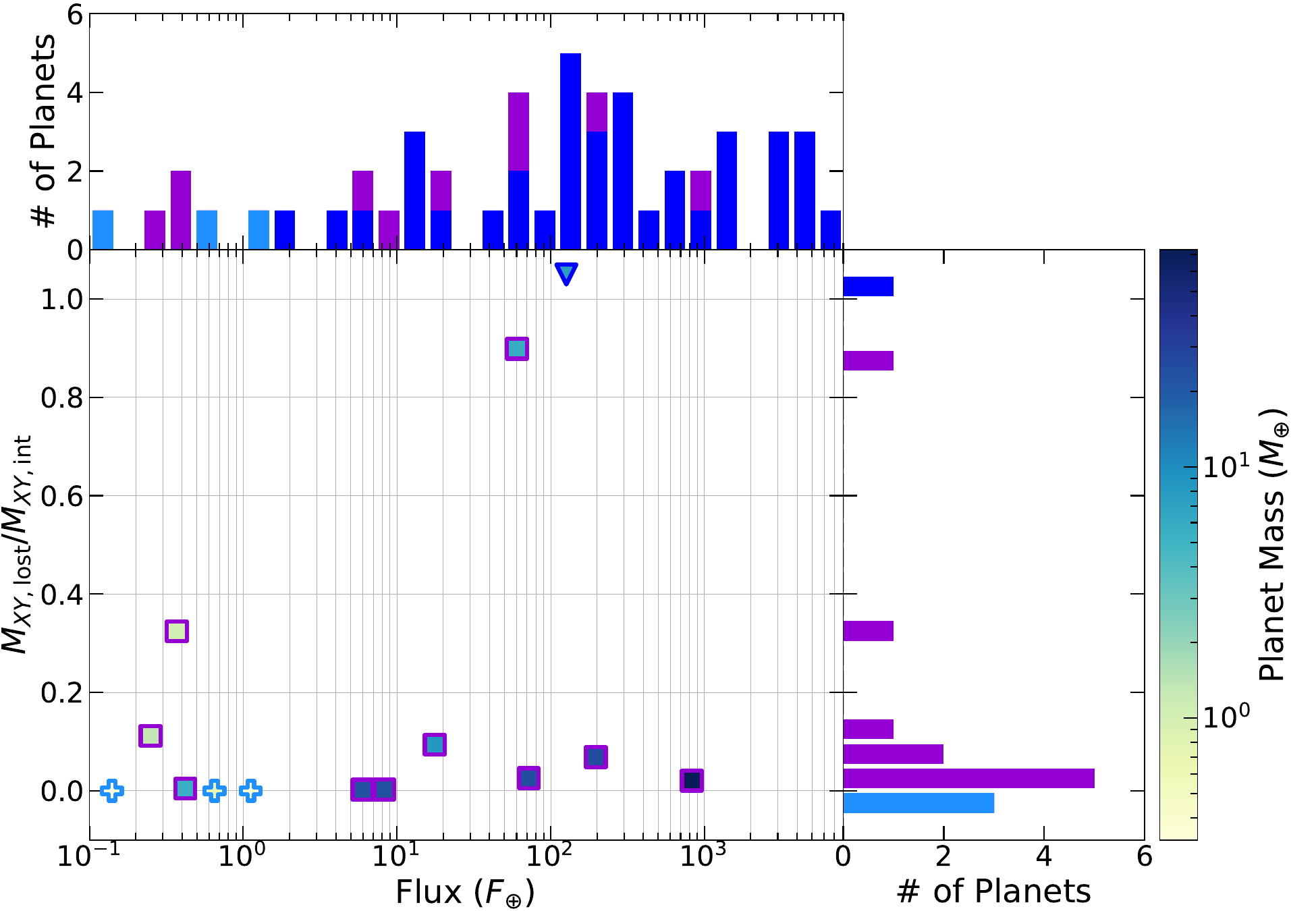}
\caption{The mass ratio of envelopes that are accreted during the gas disk stage and can be lost subsequently by photoevaporation
as a function of planet mass and insolation flux on the left and right panels, respectively (as in Figure \ref{fig5}).
Exoplanets classified as super/sub-Earths in Figure \ref{fig4} are only examined.
The lower panels are the zoom-in version of the upper ones.
Super-Earths that have $M_{XY, \rm lost}/M_{XY, \rm int} \gg 1$ are referred to as photoevaporated cores (the lower triangle with the blue edge)
because their primordial envelopes were likely lost completely by photoevaporation.
Super-Earths that satisfy the condition that $M_{XY, \rm lost}/M_{XY, \rm int} <1$ are temporarily called super-Earths (i.e., $M_{\rm p} \geq 1 M_{\oplus}$, the square with the violet edge),
while the counterpart of sub-Earths is named as sub-Earths again (i.e., $M_{\rm p} < 1 M_{\oplus}$, the plus with the light blue edge).
Inefficiencies of photoevaporation for super-Earths not only constrain the minimum value of the critical core mass, which is $\sim 1.3 M_{\oplus}$, 
but also pose a question about the absence of envelopes around very massive ($\gtrsim 20 M_{\oplus}$) super-Earths.}
\label{fig6}
\end{center}
%\end{figure}
\end{minipage}
\end{figure*}

Figure \ref{fig6} shows the resulting behaviors.
We find that super/sub-Earths are further classified into sub-groups (Figure \ref{fig3}).
First, some of these planets satisfy the condition that $M_{XY, \rm lost}/M_{XY, \rm int} \gg 1$, 
that is, their primordial envelopes were lost by photoevaporation completely.
We call these planets photoevaporated cores.
Second, there are exoplanets that meet the condition that $M_{XY, \rm lost}/M_{XY, \rm int} < 1$.
We tentatively refer these planets as super-Earths and sub-Earths;
the former class is used for planets that are more massive than $1 M_{\oplus}$, and the latter is for planets less massive than that.
While these planets are currently unlikely to have gaseous envelopes,
the photoevaporation efficiency is not high enough to remove the entire envelope.
It is thus possible to deduce from the samples used in this work that planets that are less massive than $\sim 1.3M_{\oplus}$ did not experience gas accretion;
tighter determination of this value is limited by the lack of planets in the mass range of $\sim 1.3M_{\oplus}$ to $\sim 5.5 M_{\oplus}$ in the current dataset.
Given that photoevaporated sub-giants indicate that the minimum value of the critical core mass is about $4 M_{\oplus}$, 
the critical core mass for the onset of gas accretion ranges from $\sim 1.3-4M_{\oplus}$ to $\sim 10 M_{\oplus}$.

Our analysis also manifests the lack of photoevaporative mass loss for very massive ($\gtrsim 20 M_{\oplus}$) super-Earths, while these planets appear to be rare.
Similarly to sub-Neptunes, they are massive enough to accrete a good amount of disk gas, yet the current properties infer the absence of gaseous envelopes.

In the following, we explore how impacts between (proto)planets reconcile this (apparent) contradiction.

\subsection{Impact-driven mass growth and envelope loss} \label{sec:post_impact}

As described above, the current properties of sub-Neptunes and massive super-Earths are not consistent with predictions from gas accretion/retention, 
even taking account of photoevaporation.
One possibility to resolve this inconsistency is that planets were less massive when gas disks were present; 
after gas disks dissipated, planets grew in mass and achieved the current properties.

When there is no disk gas, the eccentricity of (proto)planets and solid planet-forming materials increases readily, resulting in collisions among these bodies.
Collisions produce various outcomes.
Given that the composition of super-Earths is between Earth-like rock andwater-rich (Figure \ref{fig4}),
it is natural to assume that the corresponding collisions lead to mergers without noticeable solid mass loss.
Instead, these collisions drive potentially significant mass loss of gaseous envelopes.
In this section, we focus on super-Earths and sub-Neptunes defined in Section \ref{sec:post_photo} and explore these mergers and the resulting envelope mass loss.
We will consider catastrophic collisions to form super-Mercuries in the following section.

\begin{figure*}
\begin{minipage}{17cm}
%\begin{figure}%[!ht]
\begin{center}
\includegraphics[width=8.4cm]{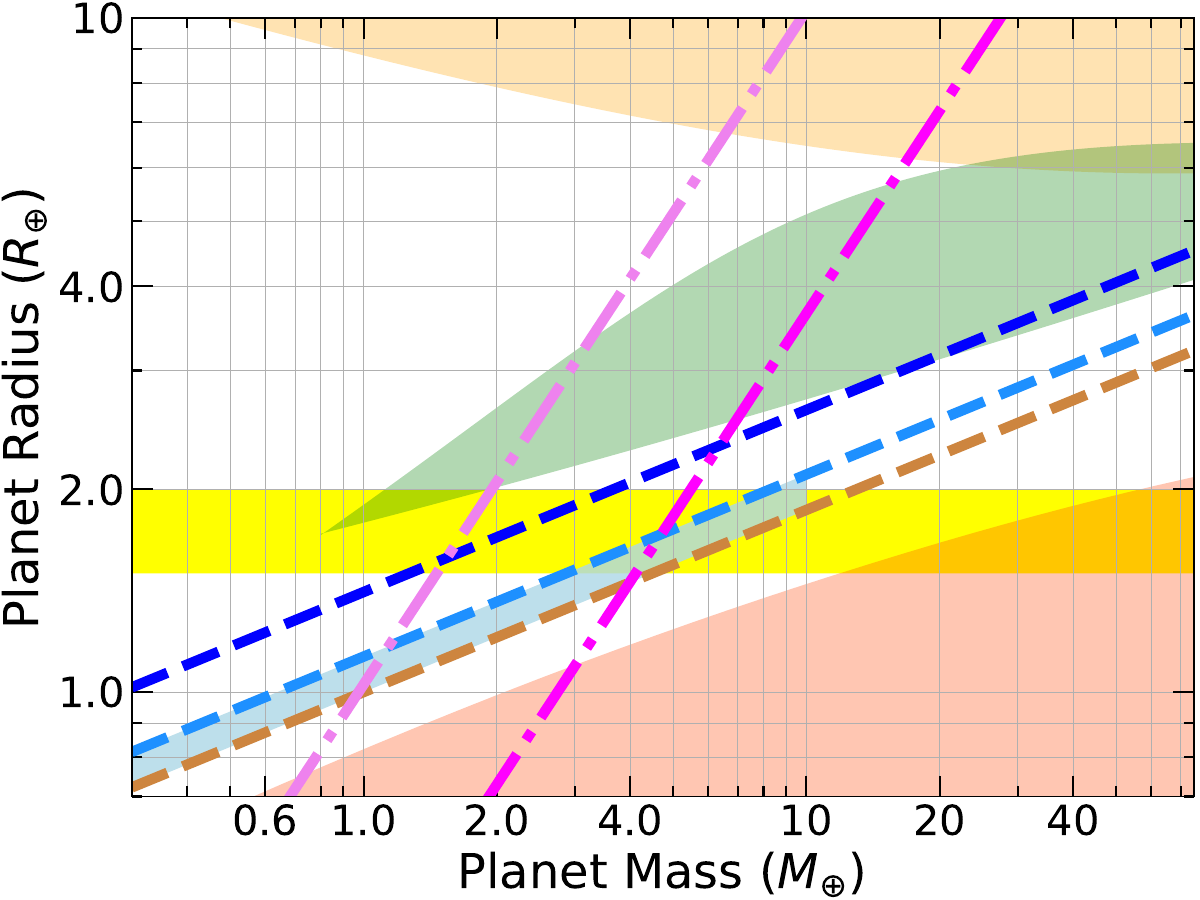}
\includegraphics[width=8.4cm]{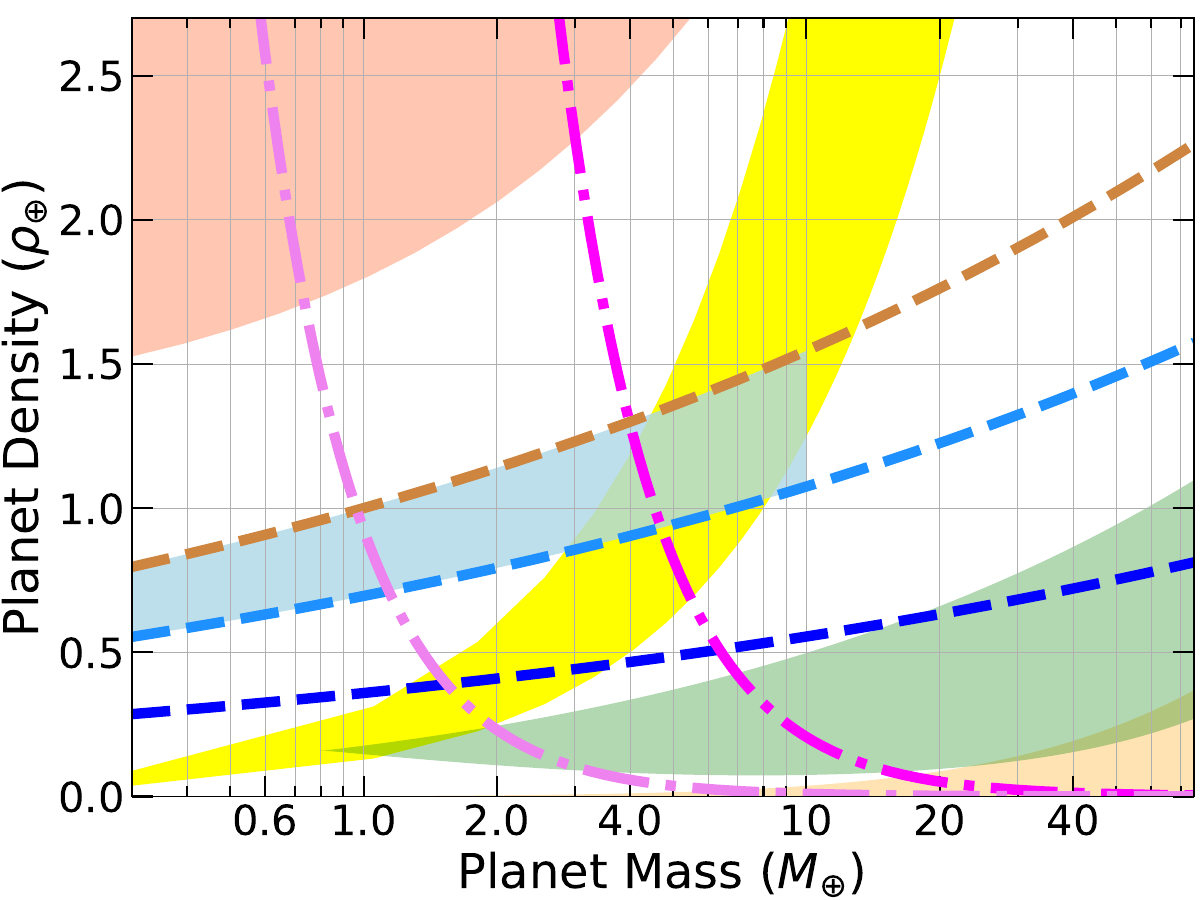}
\caption{The behavior of collisional envelope mass loss in the mass-radius and mass-density diagrams on the left and right panels, respectively (as in Figure \ref{fig2}).
The magenta dash-dotted line corresponds to the case that $M_{XY, \rm lost}/M_{XY, \rm int}=1$, i.e., the complete envelope loss,
while the pink dash-dotted line is for the case that $M_{XY, \rm lost}/M_{XY, \rm int}=0.5$, i.e., the 50 \% loss of envelope mass (equation (\ref{eq:R_p_cr})).
The entire envelope can be removed by single head-on collisions between identical planets with mass of $\sim4 M_{\oplus}$ (and radius of $\sim 1.5 R_{\oplus}$) or higher,
namely below the magenta dash-dotted line in the mass-radius diagram and above in mass-density diagram.}
\label{fig7}
\end{center}
%\end{figure}
\end{minipage}
\end{figure*}

The envelope mass loss by collisions is investigated by SPH (Smoothed-particle hydrodynamics) simulations 
\citep[e.g.,][]{2020ApJ...897..161K,2020MNRAS.496.1166D,2022MNRAS.513.1680D}.
These simulations derive empirical relationships between the lost mass and collision parameters,
one of which is given as \citep{2020ApJ...897..161K}
\begin{equation}
\label{eq:Mxy_impact}
\frac{M_{XY, \rm lost}}{M_{XY, \rm int}} \simeq 7.7 \times 10^{-6} \left( \frac{ Q }{ 10^4 \mbox{ erg g}^{-1} } \right)^{0.67},
\end{equation}
where
\begin{equation}
Q = \frac{1}{2} (1-b)^2 (1+2b) \frac{ M_{\rm i} M_{\rm t} }{ (M_{\rm i} + M_{\rm t})^2 } v_{\rm imp}^2,
\end{equation}
$b$ is the impact parameter, $M_{\rm i}$ and $M_{\rm t}$ are the mass of impactors and targets, respectively, and $v_{\rm imp}$ is the impact velocity.
The above formula is applicable for less energetic impacts, that is, $v_{\rm imp} \simeq v_{\rm esc} $, 
where $v_{\rm esc} $ is the escape velocity of colliding bodies.

More detailed formulae are available in the literature \citep[e.g.,][]{2023ApJ...954..196K}.
However, such formulae contain more fitting parameters and require more detailed properties (e.g., internal energy) of planets, 
which are not necessarily obtained readily for observed exoplanets with high confidence.
We therefore adopt the above equation in this work.

If head-on collisions (i.e., $b=0$) between two identical bodies (i.e., $M_{\rm i} = M_{\rm t}=M_{\rm p}$) are considered, equation (\ref{eq:Mxy_impact}) is re-written as
\begin{equation}
\frac{M_{XY, \rm lost}}{M_{XY, \rm int}} \simeq 5.1 \times 10^{-1} \left( \frac{ M_{\rm p} }{M_{\oplus}} \right)^{0.67} \left( \frac{ R_{\rm p} }{R_{\oplus}} \right)^{-0.67},
\end{equation}
where $v_{\rm imp}= v_{\rm esc} (\equiv \sqrt{2GM_{\rm p}/R_{\rm p}})$ is adopted.
Then, the critical planet radius ($R_{\rm p, cr}$) below which primordial envelopes will be reduced by $M_{XY, \rm lost}$ due to single head-on collisions between identical bodies is
\begin{equation}
\label{eq:R_p_cr}
R_{\rm p, cr} \simeq 1.8 R_{\oplus} \left( \frac{ M_{XY, \rm lost}/M_{XY, \rm int} }{1} \right)^{-1/0.67} \left( \frac{ M_{\rm p} }{5 M_{\oplus}} \right),
\end{equation}
where $R_{\rm p, cr}$ is normalized by the case that $M_{XY, \rm lost}/M_{XY, \rm int} =1$.
This equation confirms that collisions between planets in the super-Earth regime or above can remove their gaseous envelopes significantly.

Figure \ref{fig7} shows the behavior of $R_{\rm p, cr}$ in the mass-radius and mass-density diagrams.
Two values of $M_{XY, \rm lost}/M_{XY, \rm int}$ are considered: 
$M_{XY, \rm lost}/M_{XY, \rm int}=1$, that is, the entire envelope is lost (the magenta dash-dotted line), 
and $M_{XY, \rm lost}/M_{XY, \rm int}=0.5$, that is, half of envelopes are lost (the pink dash-dotted line).
We find that planets with mass of $\sim 4 M_{\oplus}$ (the corresponding radius of $\sim 1.5 R_{\oplus}$) or higher can lose their entire envelopes by collisions.
Interestingly, this mass range overlaps with that of the critical core mass discussed in Section \ref{sec:post_photo}.
Thus, for planets distributed below the magenta dash-dotted line in the mass-radius diagram or above in the mass-density diagram, 
collisions can affect the gaseous envelope mass significantly.
On the other hand, for planets above the magenta dash-dotted line in the mass-radius diagram or below in the mass-density diagram,
the effect of collisions on the gaseous envelope mass may be modest or even negligible.

Equation (\ref{eq:Mxy_impact}) also enables computing the properties of progenitors if head-on collisions between two identical bodies are assumed.
We apply such calculations to sub-Neptunes and super-Earths defined in Section \ref{sec:post_photo} and 
examine how collisions resolve the existence of very tenuous envelopes around sub-Neptunes and their absence around super-Earths.

\begin{figure*}
\begin{minipage}{17cm}
%\begin{figure}%[!ht]
\begin{center}
\includegraphics[width=8.4cm]{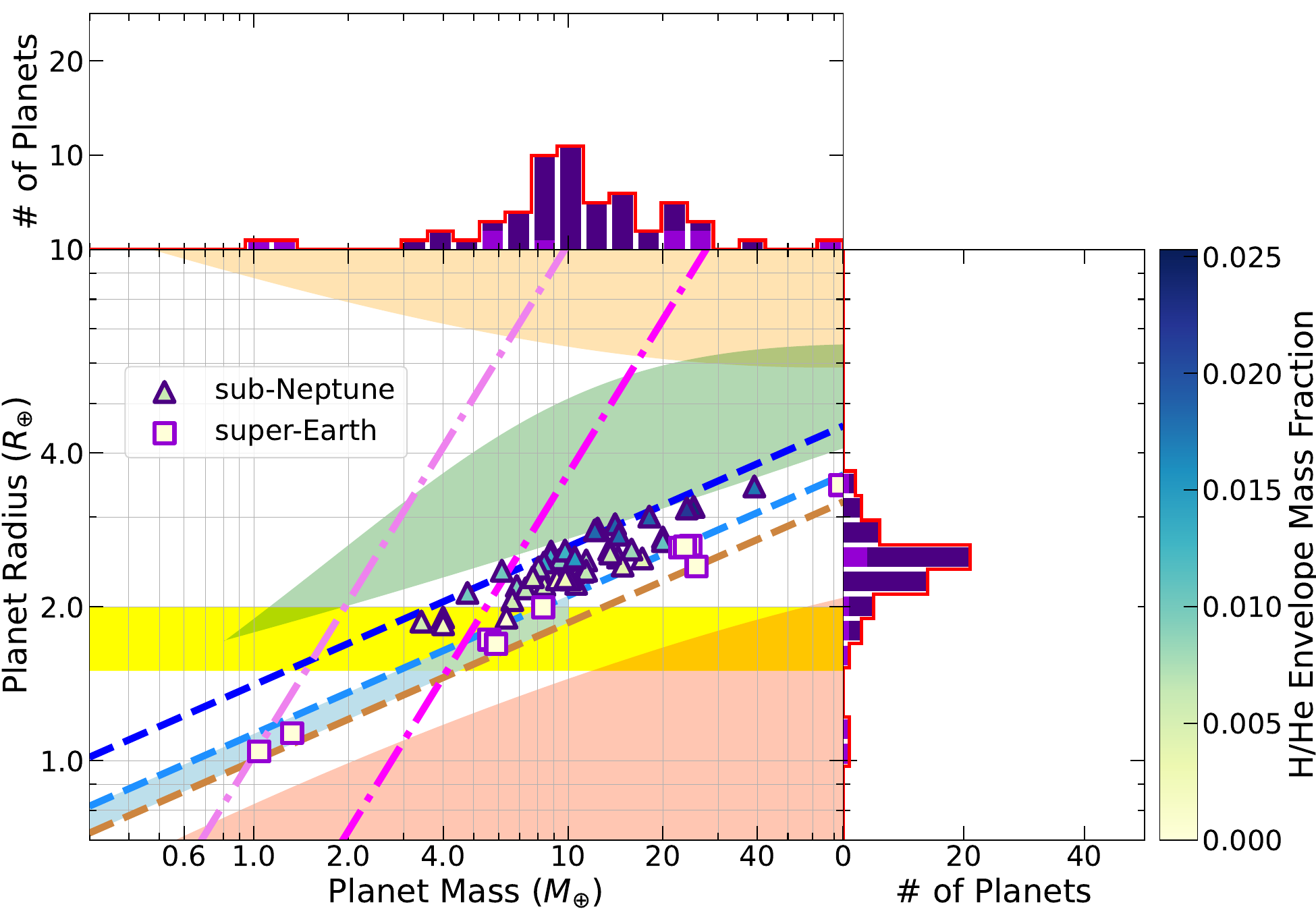}
\includegraphics[width=8.4cm]{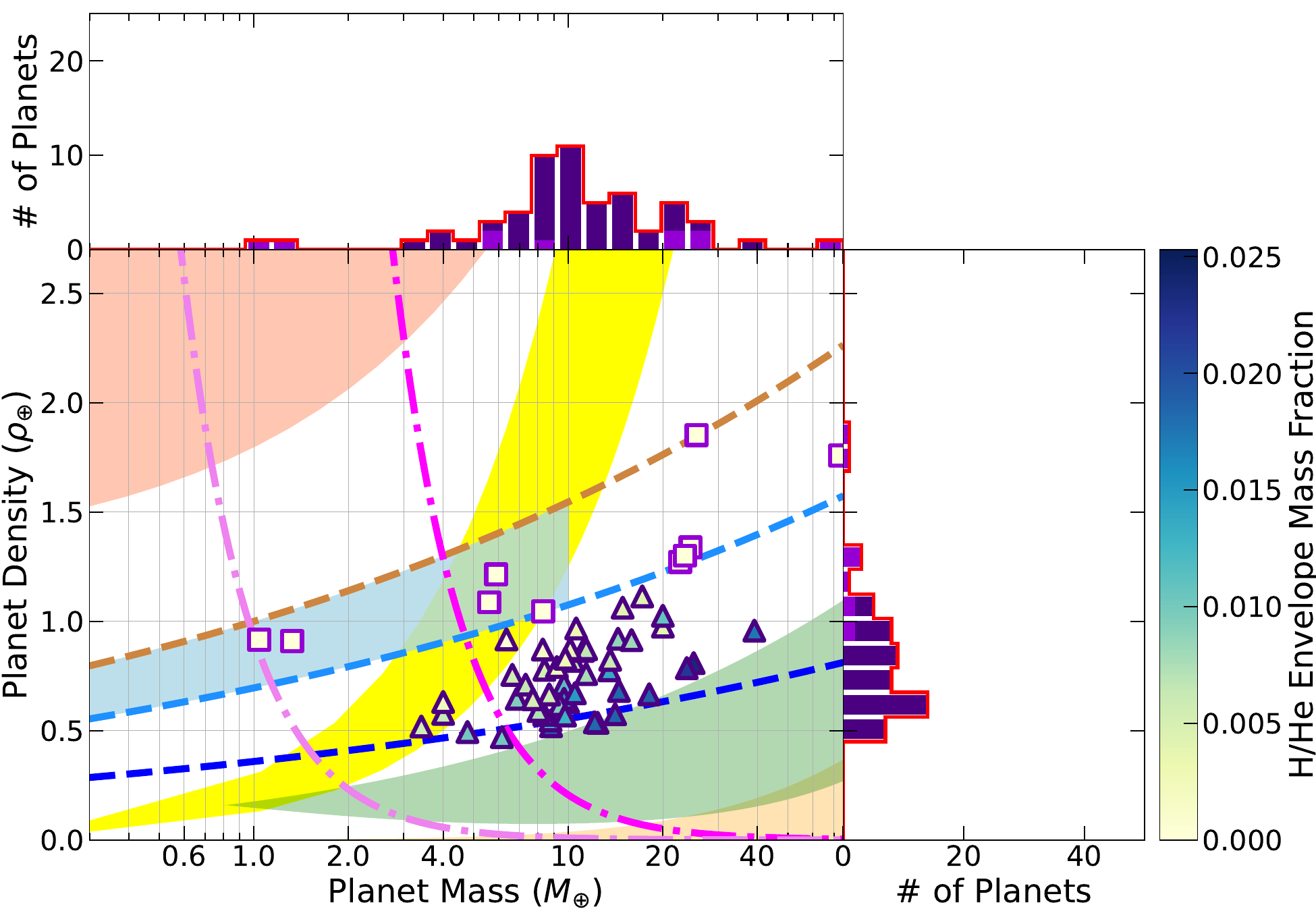}
\includegraphics[width=8.4cm]{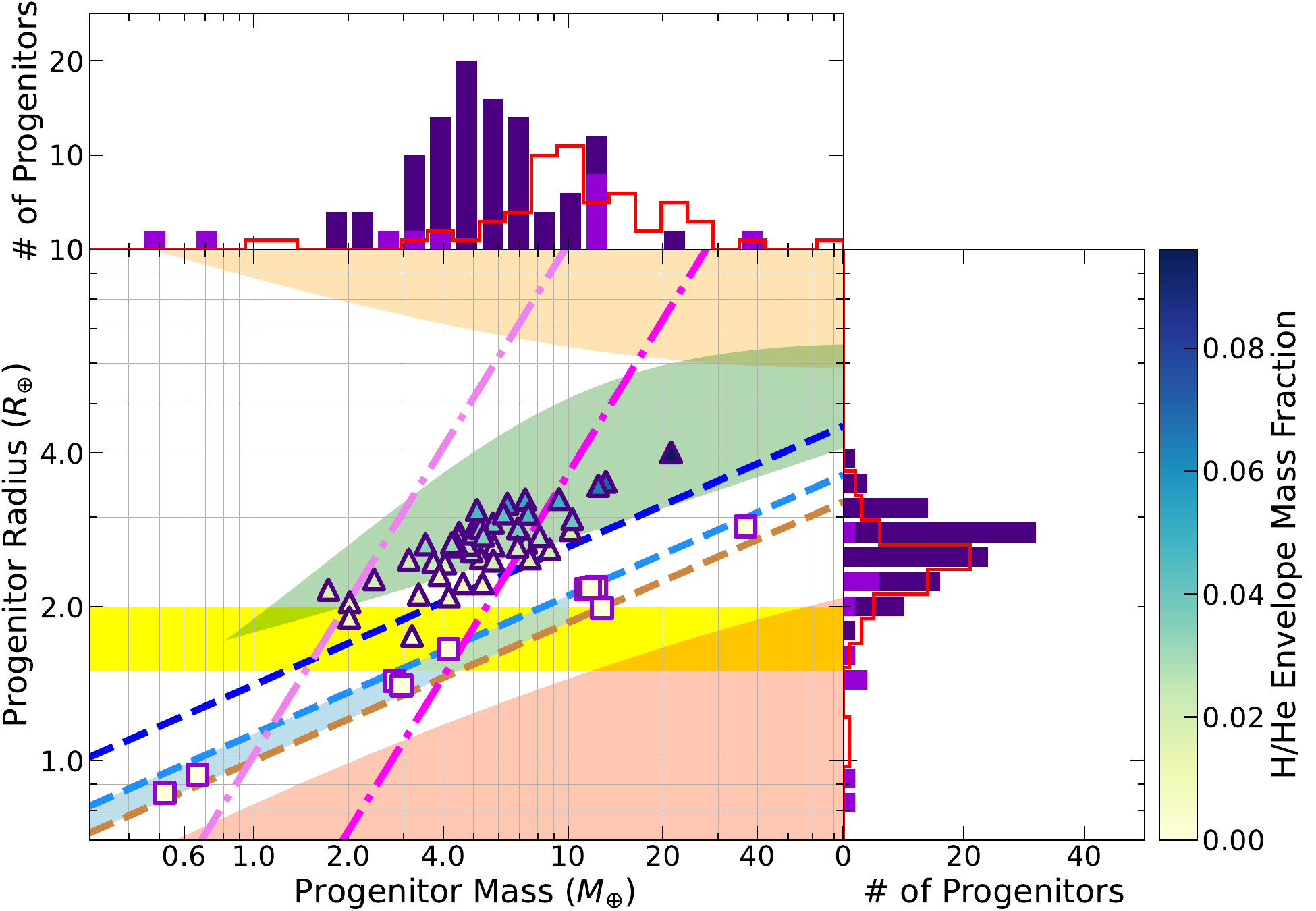}
\includegraphics[width=8.4cm]{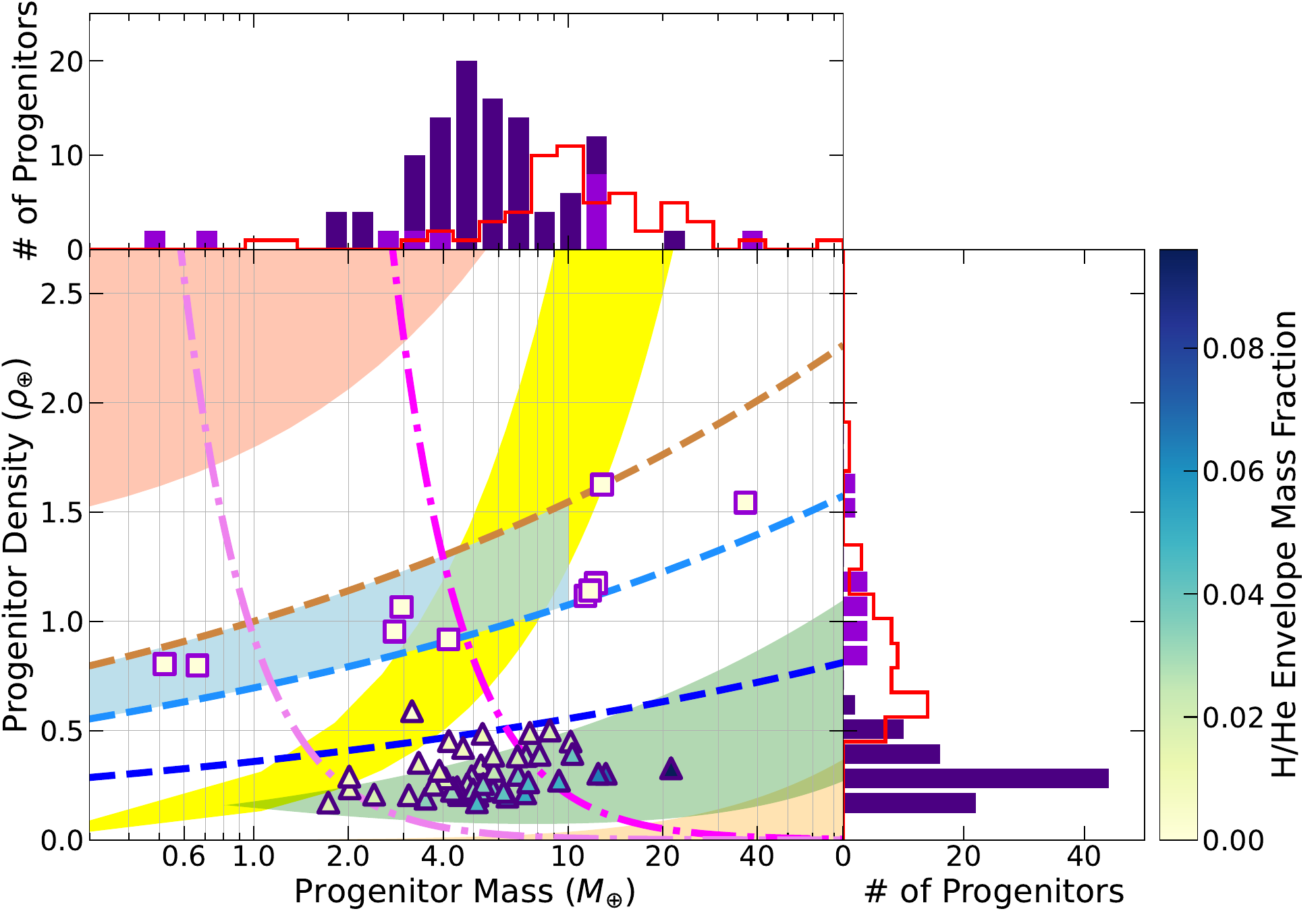}
\includegraphics[width=8.4cm]{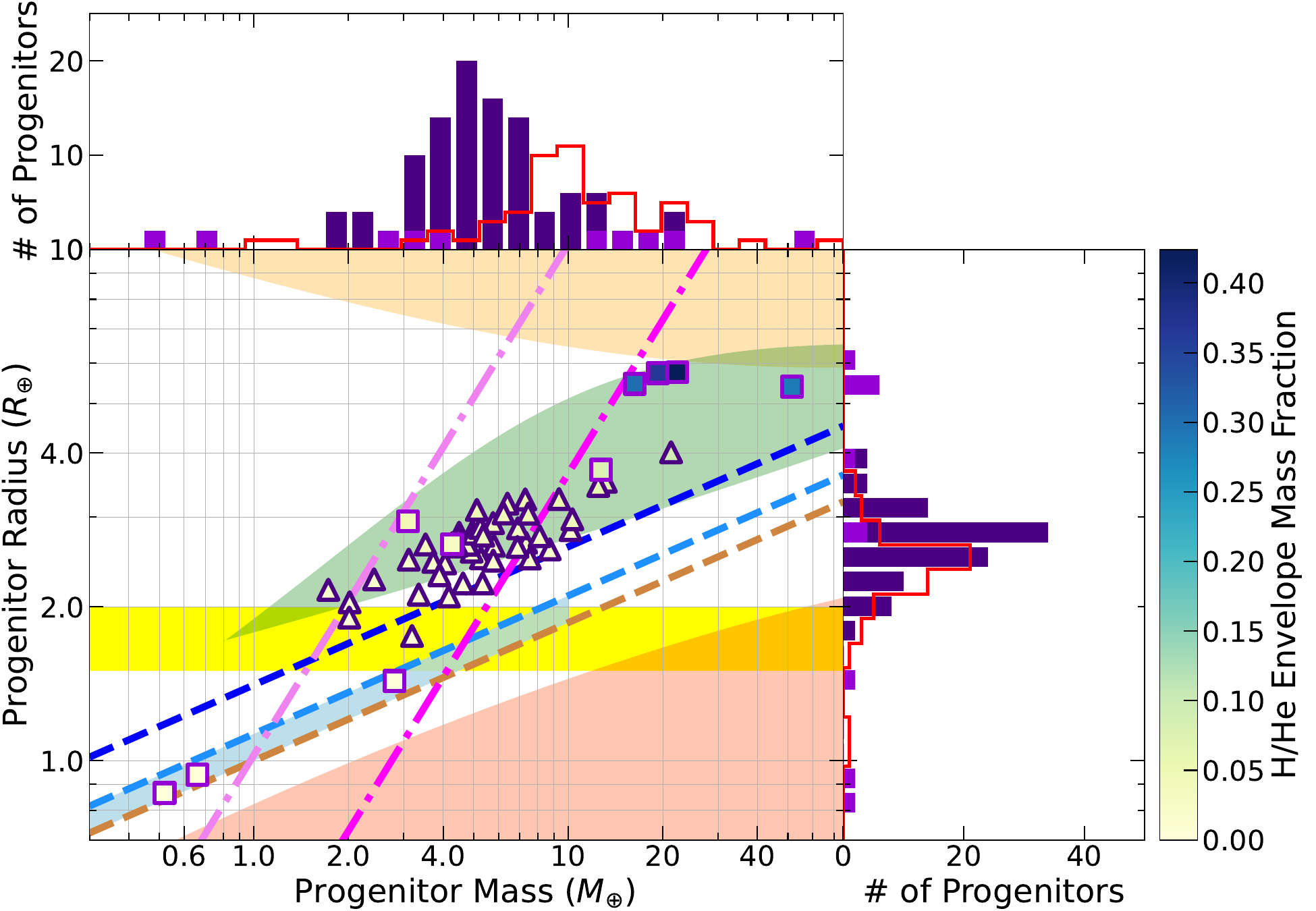}
\includegraphics[width=8.4cm]{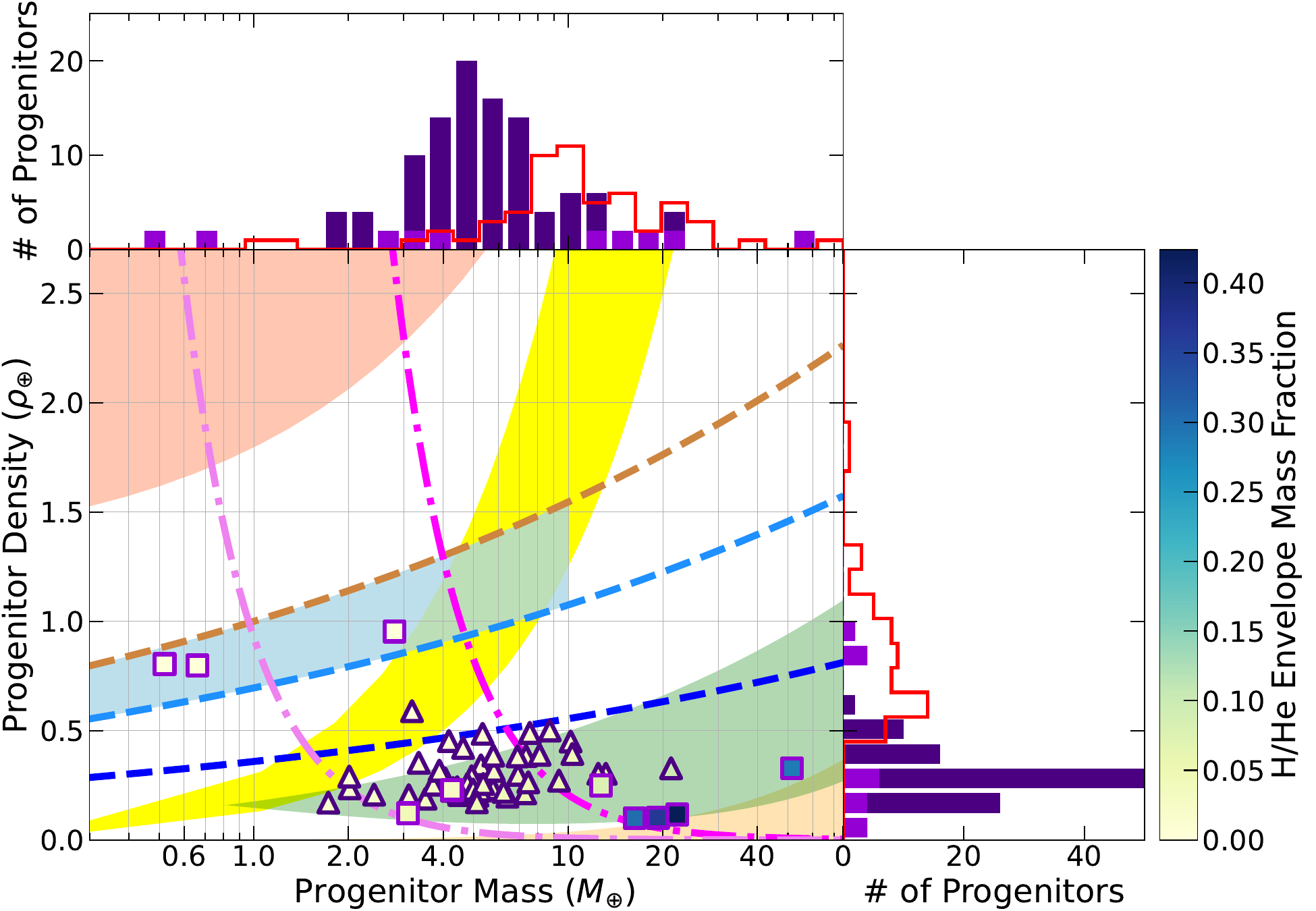}
\caption{The effect of collisional envelope mass loss in the mass-radius and mass-density diagrams on the left and right panels, respectively (as in Figures \ref{fig4} and \ref{fig7}).
In these plots, sub-Neptunes and super-Earths defined in Figures \ref{fig5} and \ref{fig6} and their progenitors are only considered.
For comparison, the current properties of sub-Neptunes and super-Earths are shown on the top panels.
On the middle panels, the properties of their progenitors are plotted, assuming head-on collisions between two identical bodies.
For the progenitors of sub-Neptunes, their primordial envelope mass is computed, using equation (\ref{eq:Mxy_impact}).
On the contrary to sub-Neptunes and super-Earths, most of their progenitors distribute in the green and blue shaded regions,
suggesting that standard core accretion is applicable to these progenitors, 
and the presence/absence of too tenuous envelopes is understood by collisional envelope mass loss.
For comparison, the red line in two histograms denotes the distribution of sub-Neptunes and super-Earths.
On the bottom panels, the primordial properties of progenitors are shown; 
the initial envelope mass of progenitors of super-Earths is computed, assuming that the minimum value of the critical core mass is about $3 M_{\oplus}$ (Figure \ref{fig9}),
that is, gas accretion was possible for planetary cores more massive than $3 M_{\oplus}$.}
\label{fig8}
\end{center}
%\end{figure}
\end{minipage}
\end{figure*}

Figure \ref{fig8} shows the resulting properties of progenitors.
For comparison, the current properties of sub-Neptunes and super-Earths are plotted on the top panels;
as described above, their positions in the mass-radius and mass-density diagrams deviate 
both from predictions of gas accretion and removal and from constraints imposed by the composition of solar system bodies.
On the middle panels, the computed properties of progenitors are shown.
In such calculations, the core mass of progenitors is taken to be a half of the core mass of the current planets,
and the population of progenitors doubles.
For sub-Neptunes, the initial envelope mass of progenitors is computed self-consistently, using equations (\ref{eq:R_env}) and (\ref{eq:Mxy_impact}),
under the assumption that two identical progenitors lost the same amount of gaseous envelopes by a collision.
Remarkably, most of the progenitors of sub-Neptunes, which might be classified as water-dominated planets, are now distributed in the green shaded region;
equivalently, standard core accretion accounts for the properties of these progenitors.
Some progenitors are distributed below the magenta dash-dotted line in the mass-radius diagram or above the line in the mass-density diagram.
Given that the line defines the threshold for the entire envelope mass loss by collisions,
the presence of these progenitors implies limitation of applicability of the assumption that gaseous envelopes are composed purely of hydrogen and helium,
which was adopted in previous studies.
Such progenitors and the corresponding sub-Neptunes are preferred to be classified as vapor-rich sub-giants, as discussed in Section \ref{sec:post_photo}.
Thus, collisions likely occurred for vapor-rich sub-giants as well.

For super-Earths, retrieval of the initial envelope mass from equation (\ref{eq:Mxy_impact}) is not possible due to the lack of the current envelope.
Instead, the absence of the current envelope provides a hint of the minimum value of the critical core mass.
Figure \ref{fig8} (the middle panels) shows that some progenitors distribute above the magenta dash-dotted line in the mass-radius diagram or below the line in the mass-density diagram.
This suggests that collisional removal of the entire primordial envelope did not operate for these progenitors
and implies that they did not possess gaseous envelope originally.
In other words, the minimum value of the critical core mass is about $\sim 4 M_{\oplus}$.

One might then argue that the above consideration is based on an implicit assumption that collisions occurred earlier than photoevaporation;
if the opposite would be the case, then the critical core mass might be much smaller.
We examine this possibility by computing the ratio of $M_{XY, \rm lost}/M_{XY, \rm int}$ as done in Section \ref{sec:post_photo}.

\begin{figure*}
\begin{minipage}{17cm}
%\begin{figure}%[!ht]
\begin{center}
\includegraphics[width=8.4cm]{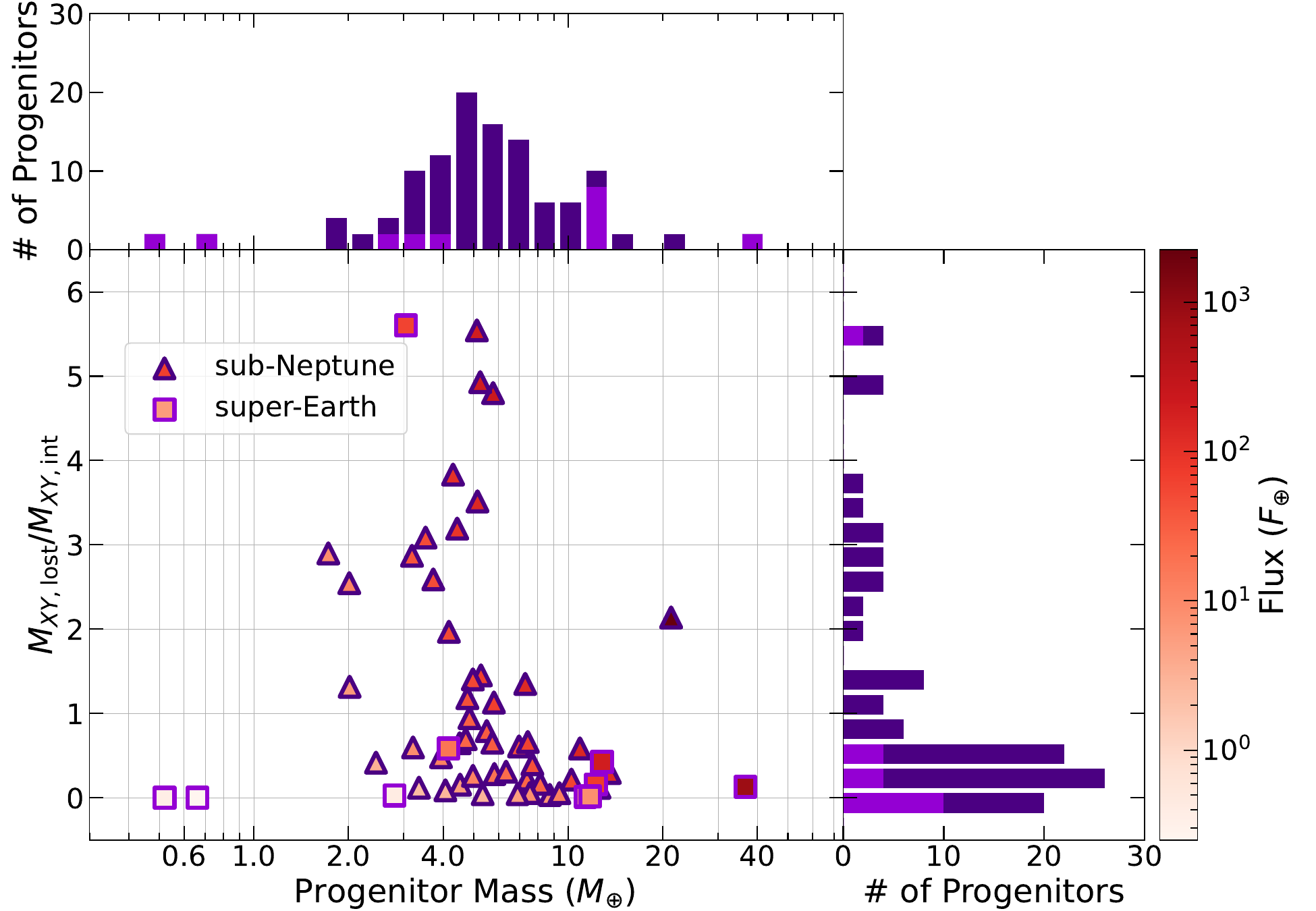}
\includegraphics[width=8.4cm]{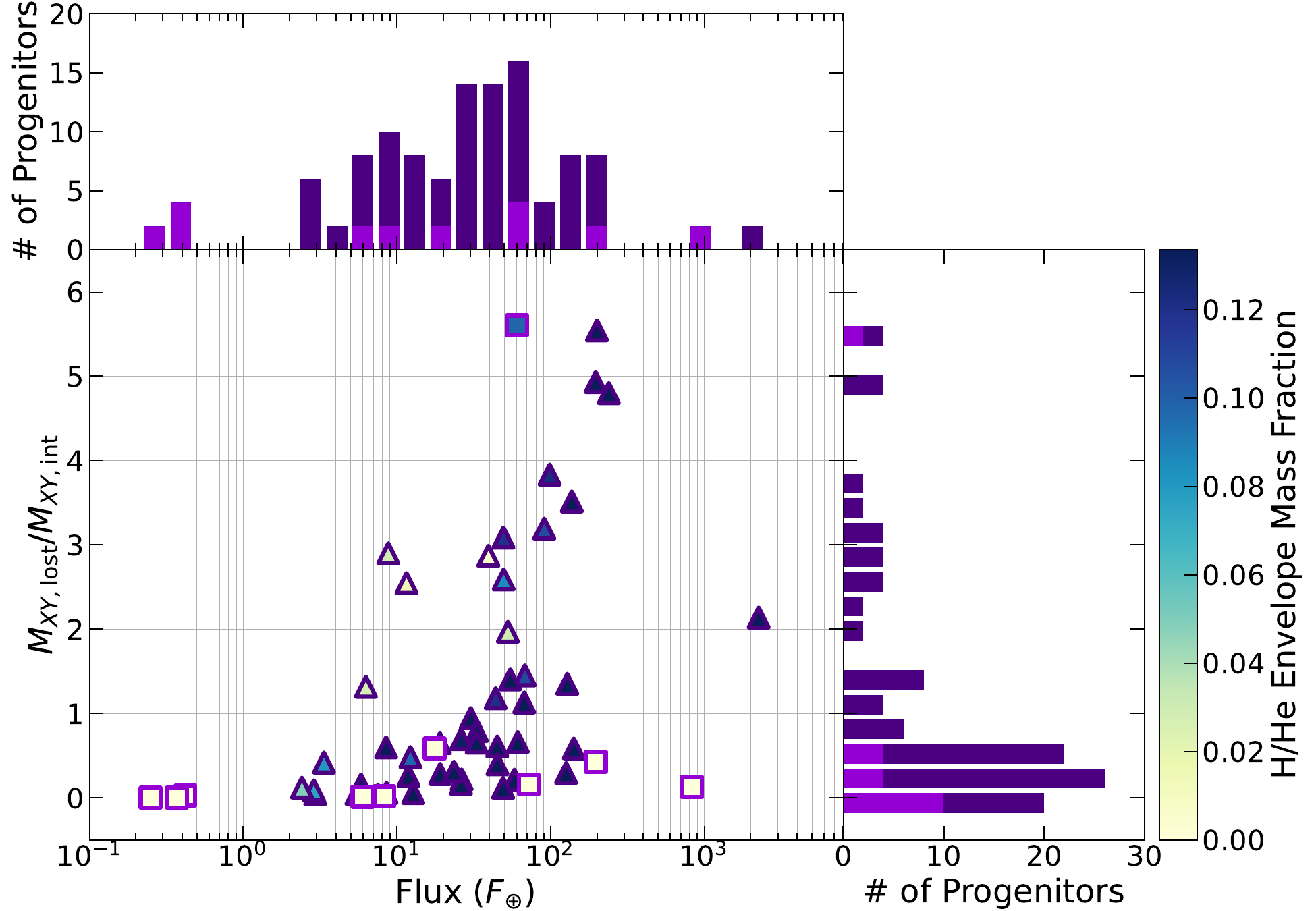}
\caption{The mass ratio of envelopes that are accreted during the gas disk stage and can be lost subsequently by photoevaporation
as a function of planet mass and insolation flux on the left and right panels, respectively (as in Figure \ref{fig6}).
The progenitors of sub-Neptunes and super-Earths defined in Figures \ref{fig5} and \ref{fig6} are only considered.
For the progenitors of super-Earths, high photoevaporation efficiencies suggest a possibility that photoevaporation is followed by collisional envelope loss,
and the critical core mass goes down to $\sim 3 M_{\oplus}$.
Many progenitors of sub-Neptunes predict high photoevaporation efficiencies as well, 
and hence the opposite possibility (i.e., collisions followed by photoevaporation) is very likely.
The critical core mass reduces even down to $\sim 1.6 M_{\oplus}$ for them.}
\label{fig9}
\end{center}
%\end{figure}
\end{minipage}
\end{figure*}

Figure \ref{fig9} shows the results of $M_{XY, \rm lost}/M_{XY, \rm int}$ for the progenitors of super-Earths.
For completeness, the value of $M_{XY, \rm lost}/M_{XY, \rm int}$ is computed for the progenitors of sub-Neptunes as well.
We find that the value of $M_{XY, \rm lost}/M_{XY, \rm int}$ exceeds unity when the progenitors of super-Earths are more massive than $\sim 3 M_{\oplus}$.
It is therefore possible that these progenitors initially possessed primordial envelopes and subsequently lost them by photoevaporation before collisions occurred.
On the other hand, the photoevaporation efficiency of less massive progenitors is nearly zero for the samples used in this work.
These suggest that the minimum value of the critical core mass is about $\sim 3 M_{\oplus}$ or higher for the progenitors of super-Earths.

Figure \ref{fig9} also depicts that many progenitors of sub-Neptunes exhibit the value of $M_{XY, \rm lost}/M_{XY, \rm int}$ that is higher than unity,
and hence two possibilities are considered:
the one is that collisions should have occurred earlier than photoevaporation for them; otherwise, the presence of the current envelope cannot be explained.
The other possibility is that they are vapor-rich sub-giants, and the estimate of photoevaporative mass loss is not accurate as discussed above.
For both possibilities, collisions are viable to reproduce the current properties of sub-Neptunes.
At the same time, the opposite situation (i.e., photoevaporation followed by collisions) cannot be ruled out for some progenitors due to lower values of $M_{XY, \rm lost}/M_{XY, \rm int}$.
The current sub-Neptune samples thus infer that the minimum value of the critical core mass would be $\sim 1.6 M_{\oplus}$ 
above which the presence of primordial envelopes is expected.

In summary, various values of the critical core mass are expected, especially for lower bounds; 
when evolution processes such as photoevaporation and collisional envelope loss are taken into account,
the minimum value of the critical core mass ranges from $\sim 1.6 M_{\oplus}$ to $\sim 3 M_{\oplus}$.

Figure \ref{fig8} (the bottom panels) synthesizes the above considerations and shows the {\it primordial} properties of progenitors;
in order to take account of gas accretion onto the progenitors of super-Earths, the initial envelope mass is computed 
by randomly picking up a value from the mass range (equation(\ref{eq:Mxy_range})) uniformly,
if they are more massive than $\sim 3 M_{\oplus}$.
The progenitor properties of sub-Neptunes do not change under the assumption either that collisions occurred before photoevaporation or that they are vapor-rich sub-giants.
It is clear that most progenitors distribute in the blue and green shaded regions that are specified by planet formation processes,
while a small number of outliers are present.

\subsection{Impact-driven mantle stripping}

The above section focuses on less energetic collisions (i.e., $v_{\rm imp} \simeq v_{\rm esc}$), and hence there is no solid mass loss.
As described in Section \ref{sec:app}, some exoplanets exhibit higher bulk densities than Earth-like rock and are referred to as super-Mercuries.
We here consider the properties of catastrophic collisions (i.e., $v_{\rm imp} > v_{\rm esc}$) that achieve solid mass loss and reproduce the current properties of super-Mercuries.

The outcome of catastrophic collisions is explored by SPH simulations \citep[e.g.,][]{2010ApJ...712L..73M,2018E&PSL.484..276C,2024MNRAS.529.2577D}.
As with the case of collisional envelope mass loss, 
these simulations are used to derive empirical relationships between collision parameters and the properties of planets formed by collisions.
For catastrophic collisions, impact velocity ($v_{\rm imp}$) and the iron mass fraction of the largest remnant ($f_{\rm Fe}$) are two key parameters 
and have the following relationship \citep{2010ApJ...712L..73M}:
\begin{eqnarray}
\label{eq:v_imp}
v_{\rm imp} & \simeq  &18.7 \mbox{ km s}^{-1} ( f_{\rm Fe} - 0.33)^{0.5} \\ \nonumber
                    & \times  & \left( \frac{ (1+ \mu)^{2.4} }{\mu} \right)^{1/1.2} \left( \frac{ M_{\rm t} }{ M_{\oplus} } \right)^{1/3},
\end{eqnarray}
where $\mu = M_{\rm i}/M_{\rm t}$ is the mass ratio of impactors to targets.
The above equation is derived under the assumption that the iron mass fraction of progenitors is  Earth-like, i.e., 0.33
and hence is valid for planets that have $f_{\rm Fe} > 0.33$ after collisions.
The resulting mass of the largest remnant ($M_{\rm lr}$) is given as \citep{2010ApJ...712L..73M}
\begin{equation}
M_{\rm lr} = [1 -1.2 (f_{\rm Fe} - 0.33)^{1/1.65}] (1 + \mu) M_{\rm t}. 
\end{equation}

It is convenient to normalize impact velocity by the escape velocity of colliding bodies.
If collisions between two identical bodies are considered, then equation (\ref{eq:v_imp}) is re-written as
\begin{equation}
\label{eq:ratio_vimp_v_esc}
\frac{ v_{\rm imp} }{ v_{\rm esc} } = 4.1 \left( \frac{ f_{\rm Fe} - 0.33 } { 0.37 }\right)^{0.5} \left( \frac{ M_{\rm p} }{ M_{\oplus} } \right)^{-1/6} \left( \frac{ R_{\rm p} }{ R_{\oplus} } \right)^{1/2},
\end{equation}
where the ratio of $v_{\rm imp} / v_{\rm esc} $ is scaled by the case that $f_{\rm Fe} = 0.7$,
and $M_{\rm p} (=M_{\rm t})$ and $R_{\rm p}$ are the mass and radius of progenitors.
For certain values of $f_{\rm Fe}$ and $v_{\rm imp} / v_{\rm esc}$, 
the corresponding radius of progenitors ($R_{\rm p, cat}$) that satisfies equation (\ref{eq:ratio_vimp_v_esc}) is expressed as
\begin{eqnarray}
\label{eq:Rp_cat}
R_{\rm p, cat} & = &  9.7 \times 10^{-1} R_{\oplus}  \left( \frac{ f_{\rm Fe} - 0.33 }{ 0.37 } \right)^{-1} \\ \nonumber
                       & \times &  \left( \frac{ v_{\rm imp} / v_{\rm esc} }{ 4 } \right)^2 \left( \frac{ M_{\rm p} }{ M_{\oplus} } \right)^{1/3},
\end{eqnarray}
where $R_{\rm p, cat}$ is scaled by the case that $f_{\rm Fe} = 0.7$ and $v_{\rm imp} / v_{\rm esc} =4 $.

\begin{figure*}
\begin{minipage}{17cm}
%\begin{figure}%[!ht]
\begin{center}
\includegraphics[width=8.4cm]{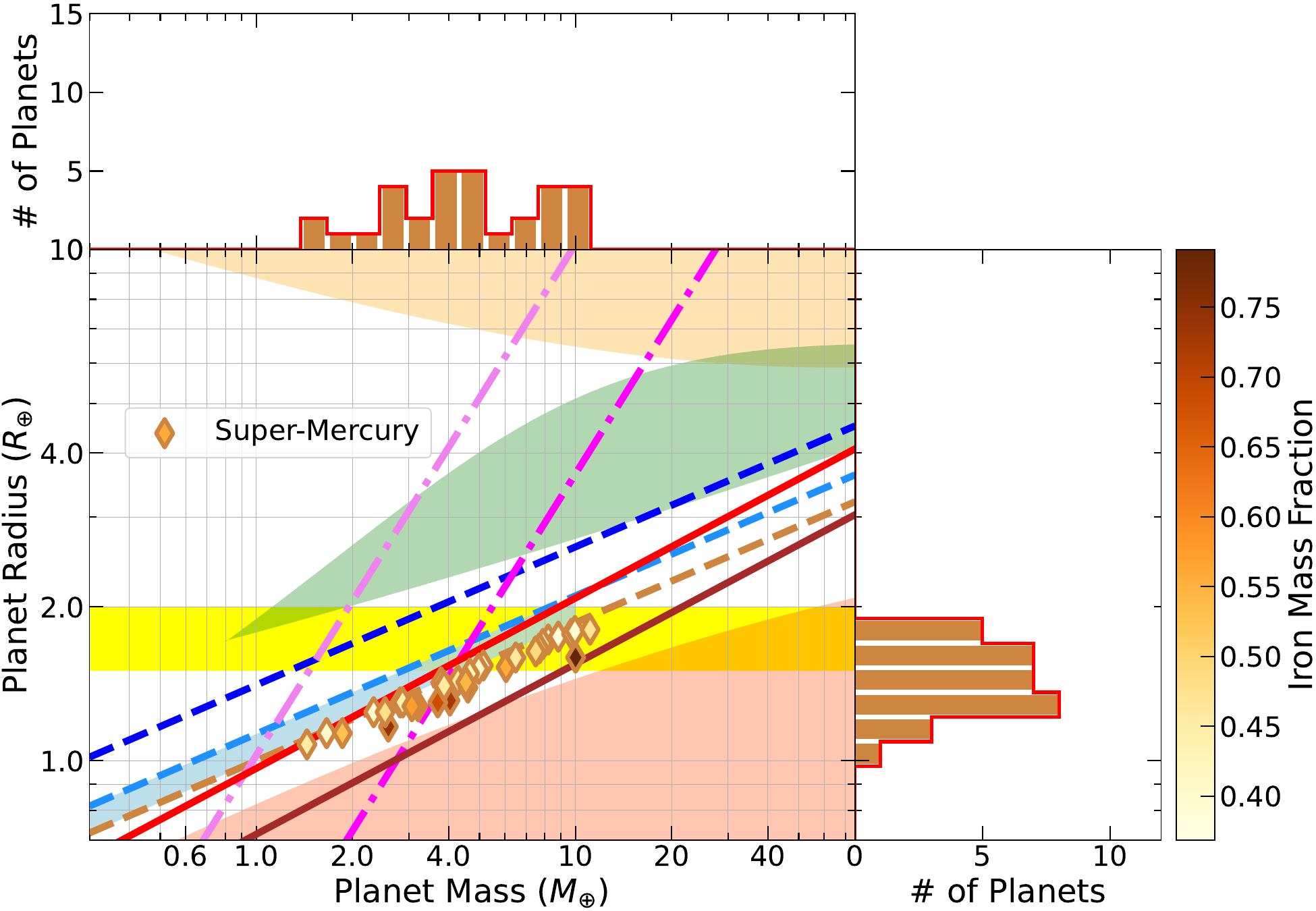}
\includegraphics[width=8.4cm]{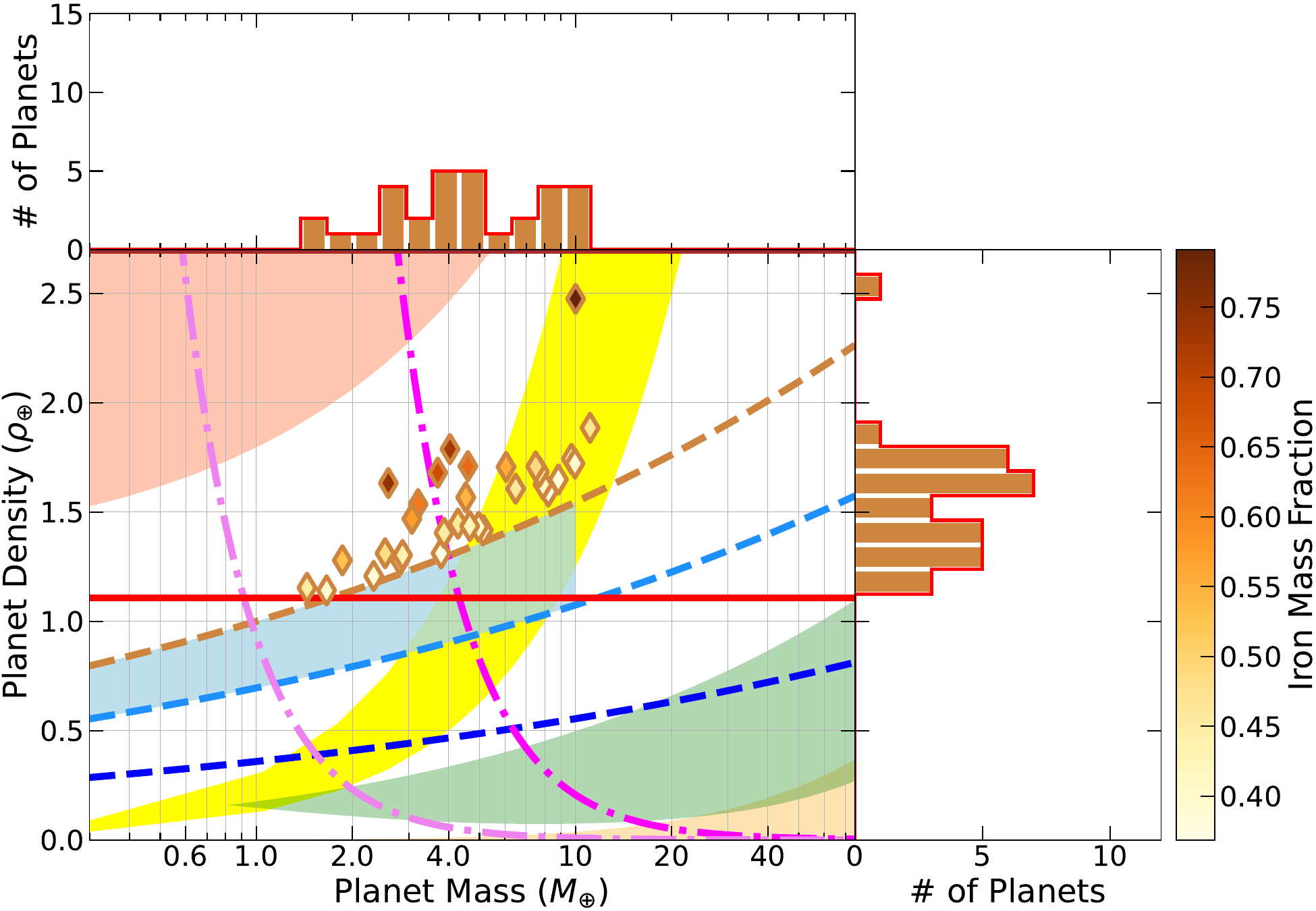}
\includegraphics[width=8.4cm]{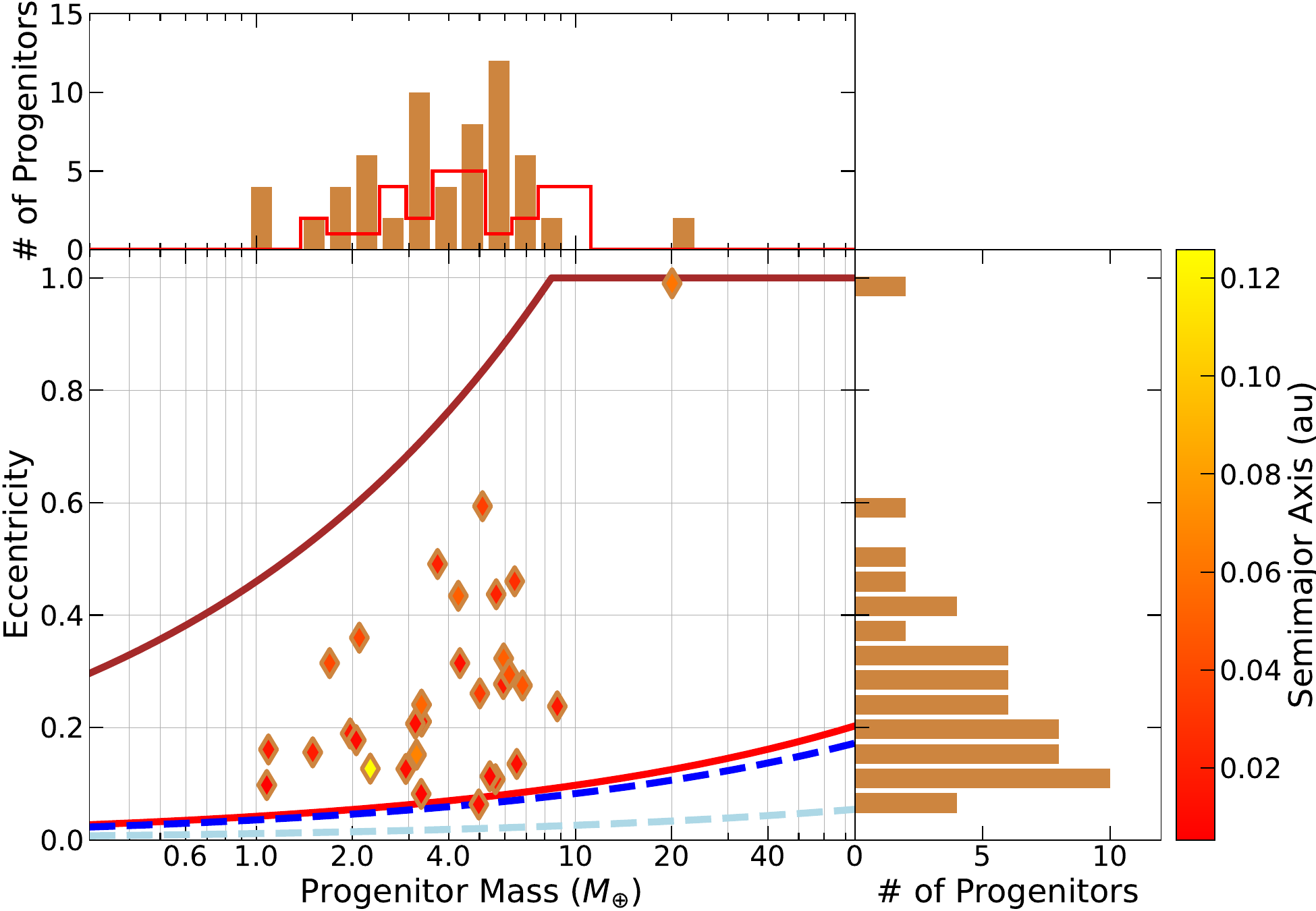}
\includegraphics[width=8.4cm]{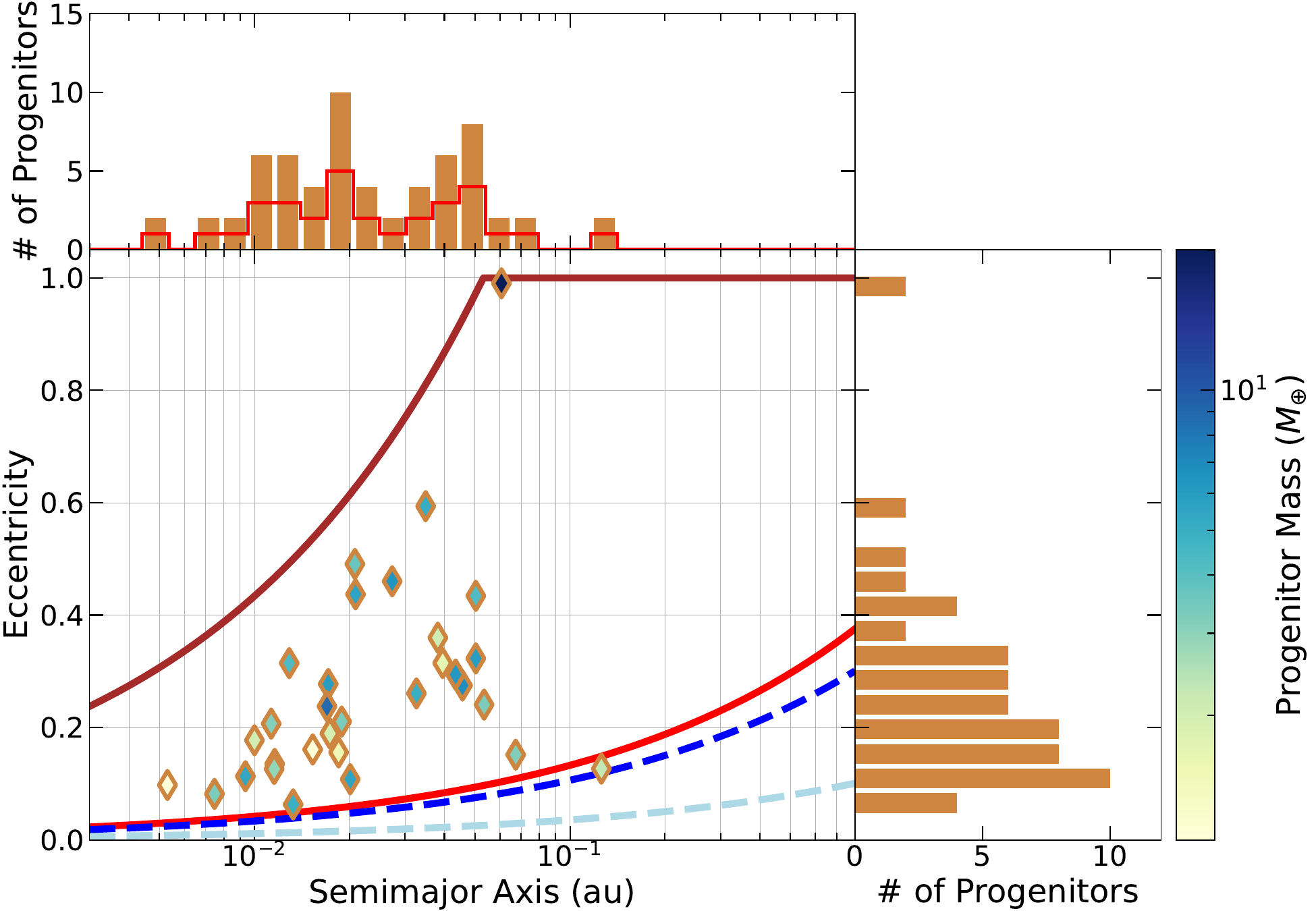}
\caption{The properties of catastrophic collisions that account for the present properties of super-Mercuries.
On the upper panels, the behavior of planet radius ($R_{\rm p, cat}$, equation (\ref{eq:Rp_cat})) that satisfies equation (\ref{eq:ratio_vimp_v_esc}) is shown 
in the mass-radius and mass-density diagrams for two sets of $f_{\rm Fe}$ and $v_{\rm imp} / v_{\rm esc} $.
The red solid line denotes the case that ($f_{\rm Fe}, v_{\rm imp} / v_{\rm esc} $) = (0.7, 4.0), 
while the brown solid line is for the case that ($f_{\rm Fe}, v_{\rm imp} / v_{\rm esc} $) = (0.4, 1.5).
For comparison, the estimated value of the iron mass fraction of super-Mercuries is plotted with the color bar and 
confirms that all the current samples are distributed between these two lines.
On the lower panels, the eccentricity (equation (\ref{eq:ecc})) required for catastrophic collisions to achieve a certain value of impact velocity is plotted 
as a function of planet mass and semimajor axis on the left and right panels, respectively.
The color bar denotes the other quantify for each panel.
On the lower left panel, the red solid line corresponds to the case that ($v_{\rm imp} / v_{\rm esc}, r_{\rm p}$) = (1.5, 0.01 au), 
and the brown solid line is for the case that ($v_{\rm imp} / v_{\rm esc}, r_{\rm p}$) =(4.0, 0.1 au).
On the lower right panel, the red solid line represents the case ($v_{\rm imp} / v_{\rm esc}, M_{\rm p}$) = (1.5, $1 M_{\oplus}$),
and the brown solid line is for the case that  ($v_{\rm imp} / v_{\rm esc}, M_{\rm p}$) = (4.0, $20 M_{\oplus}$).
For comparison, the eccentricity profile predicted from giant impacts is plotted; 
on the left panel, the light blue and blue dashed lines denote the cases that $r_{\rm p} = 0.01$ au and $r_{\rm p} = 0.1$ au, respectively,
while on the right panel, the light blue and blue dashed lines are for the cases that $M_{\rm p} = 1 M_{\oplus}$ and $M_{\rm p} = 20 M_{\oplus}$, respectively.
The resulting eccentricity profiles and our samples suggest that the eccentricity of the progenitors of super-Mercuries needs to be pumped up after formation by giant impacts.}
\label{fig10}
\end{center}
%\end{figure}
\end{minipage}
\end{figure*}

Figure \ref{fig10} (the top panels) shows how $R_{\rm p, cat}$ behaves in the mass-radius and mass-density diagrams.
Two sets of $f_{\rm Fe}$ and $v_{\rm imp} / v_{\rm esc} $ are considered: ($f_{\rm Fe}, v_{\rm imp} / v_{\rm esc} $) = (0.4, 1.5) and (0.7, 4.0).
It is obvious from equation (\ref{eq:Rp_cat}) that $R_{\rm p, cat}  \propto M_{\rm p}^{1/3}$ for a given set of ($f_{\rm Fe}, v_{\rm imp} / v_{\rm esc} $).
As a result, the resulting profile becomes the horizontal line in the mass-density diagram.
For comparison, the iron mass fraction of super-Mercuries is computed from the look-up table \citep{2016ApJ...819..127Z} as described in Section \ref{sec:app}.

Catastrophic collisions require $v_{\rm imp} / v_{\rm esc} > 1$ for planetary bodies, and hence the impact velocity is decomposed into:
\begin{equation}
\frac{ v_{\rm imp} }{ v_{\rm esc} } = \left[ 1 + \left( \frac{ v_{\rm rel} }{ v_{\rm esc} } \right)^2 \right]^{1/2}
                                                     \simeq  \left[ 1 + \left( \frac{ e v_{\rm Kep} }{ v_{\rm esc} } \right)^2 \right]^{1/2},
\end{equation}
where $v_{\rm rel} (\simeq e v_{\rm Kep})$ is the relative velocity between colliding bodies,
$e$ is the corresponding eccentricity, and $v_{\rm Kep}$ is the Keplerian velocity.
Under the assumption that the position of colliding bodies was roughly comparable to the position of the produced planets,
the eccentricity is written as
\begin{eqnarray}
\label{eq:ecc}
e & =         & \left[ \left( \frac{ v_{\rm imp} }{ v_{\rm esc} } \right)^2 -1 \right]^{1/2} \frac{  v_{\rm esc}  }{ v_{\rm Kep} } \\ \nonumber
   & \simeq & 4.7 \times 10^{-1} \left( \frac{ v_{\rm imp} / v_{\rm esc} }{ 4 } \right) \\ \nonumber
   & \times  &  \left( \frac{ M_{\rm p} }{ M_{\oplus} } \right)^{1/2}  \left( \frac{ R_{\rm p} }{ R_{\oplus} } \right)^{-1/2}  \left( \frac{ r_{\rm p} }{0.1 \mbox{ au} } \right)^{1/2}, 
\end{eqnarray}
where the host stellar mass of $1 M_{\odot}$ is adopted in the above equation.
This is the eccentricity value needed to achieve catastrophic collisions with a certain value of impact velocity for a given set of planet mass, radius, and position.

Eccentricity is lower than unity for bound planets.
This requirement constrains planet position, where catastrophic collisions may occur.
Using equation (\ref{eq:ecc}), the maximum semimajor axis ($r_{\rm p, max}$) within which catastrophic collisions occur and drive mantle stripping is given as 
\begin{eqnarray}
\label{eq:rp_max}
r_{\rm p, max} & \simeq & 8.4  \times 10^{-2} \mbox{ au} \left( \frac{ v_{\rm imp} / v_{\rm esc} }{ 4 } \right)^{-2} \\ \nonumber
                        & \times  & \left( \frac{ M_{\rm p} }{ 10 M_{\oplus} } \right)^{-1}  \left( \frac{ R_{\rm p} }{ 1.9 R_{\oplus} } \right), 
\end{eqnarray}
where the host stellar mass of $1 M_{\odot}$ is used, and the case that $v_{\rm imp} / v_{\rm esc} = 4$ is considered.
The corresponding orbital period is about 9 days.
Catastrophic collisions are thus expected to occur 
if progenitors orbit around the G-type host star with semimajor axis of $\lesssim 0.1$ au or orbital period of $\lesssim 10$ days.

Figure \ref{fig10} (the bottom panels) visualizes the profile of equation (\ref{eq:ecc}) as a function of progenitor mass and semimajor axis on the left and right panels, respectively.
For the former, two sets of $v_{\rm imp} / v_{\rm esc} $ and $r_{\rm p}$ are considered: ($v_{\rm imp} / v_{\rm esc}, r_{\rm p}$) = (1.5, 0.01 au) and (4.0, 0.1 au).
For the latter, two sets of $v_{\rm imp} / v_{\rm esc} $ and $M_{\rm p}$ are adopted: ($v_{\rm imp} / v_{\rm esc}, M_{\rm p}$) = (1.5, $1 M_{\oplus}$) and (4.0, $20 M_{\oplus}$).
The resulting eccentricity profiles are compared with the eccentricity of the progenitor of super-Mercuries that is computed
under the assumption of single collisions between two identical bodies, using equations (\ref{eq:v_imp}) and (\ref{eq:ecc}).
We find that these profiles encompass the computed eccentricities, 
supporting not only our choices of the parameter values but also the possibility that catastrophic collisions are one viable process to account for the current properties of super-Mercuries. 
The distribution of the computed eccentricity for the progenitors of super-Mercuries is also consistent broadly with the prediction of $r_{\rm p, max}$ 
(equation (\ref{eq:rp_max}), the right bottom panel of Figure \ref{fig10}).

The eccentricity of progenitors may reflect their formation mechanism.
Recent $N-$body simulations show that if planets are formed by giant impacts,
then the final value of the mean eccentricity ($e_{\rm GI}$) of the system can be expressed as \citep{2025ApJ...991L..49K}
\begin{eqnarray}
e_{\rm GI} & \simeq & 0.3 \frac{ v_{\rm esc} }{ v_{\rm Kep} }  \\ \nonumber
   & \simeq &  3.6 \times 10^{-2} \left( \frac{ M_{\rm p} }{ M_{\oplus} } \right)^{1/2}  \left( \frac{ R_{\rm p} }{ R_{\oplus} } \right)^{-1/2}  \left( \frac{ r_{\rm p} }{0.1 \mbox{ au} } \right)^{1/2},
\end{eqnarray}
where the two-equal mass system orbiting around the host star with $1M_{\odot}$ is considered.

Figure \ref{fig10} (the bottom panels) plots such an eccentricity profile as well;
on the left panel, two values of $r_{\rm p}$ are used: $r_{\rm p} = 0.01$ au and $r_{\rm p} = 0.1$ au,
while on the right panel, two values of $M_{\rm p}$ are considered: $M_{\rm p} = 1 M_{\oplus}$ and $M_{\rm p} = 20 M_{\oplus}$.
It is clear that the computed eccentricities are generally higher than these profiles.
This indicates that catastrophic collisions do not occur naturally after the formation of progenitors by giant impacts.
Instead, additional processes to pump up the eccentricity of progenitors are required.

While it is beyond the scope of this work to identify its origins,
previous studies already list up possible mechanisms such as long-term dynamical metastability \citep[e.g.,][]{2015ApJ...806L..26V,2024MNRAS.527...79G} and 
secular gravitational interactions from currently detected or undetected planets in the system \citep[e.g.,][]{2006ApJ...649..992A,2021MNRAS.508..597P} 
or distant, inclined stellar companions that may or may not be observed \citep[e.g.,][]{1997AJ....113.1915I,2019MNRAS.482.4146D}.
The importance of long-term dynamical instability on the formation of super-Mercuries is consistent with the implication discussed in Section \ref{sec:app};
given that the orbital distributions of super/sub-Earths and super-Mercuries are comparable, the occurrence of energetic events like catastrophic collisions is probabilistic (Figure \ref{fig4}).

\begin{figure*}
\begin{minipage}{17cm}
%\begin{figure}%[!ht]
\begin{center}
\includegraphics[width=8.4cm]{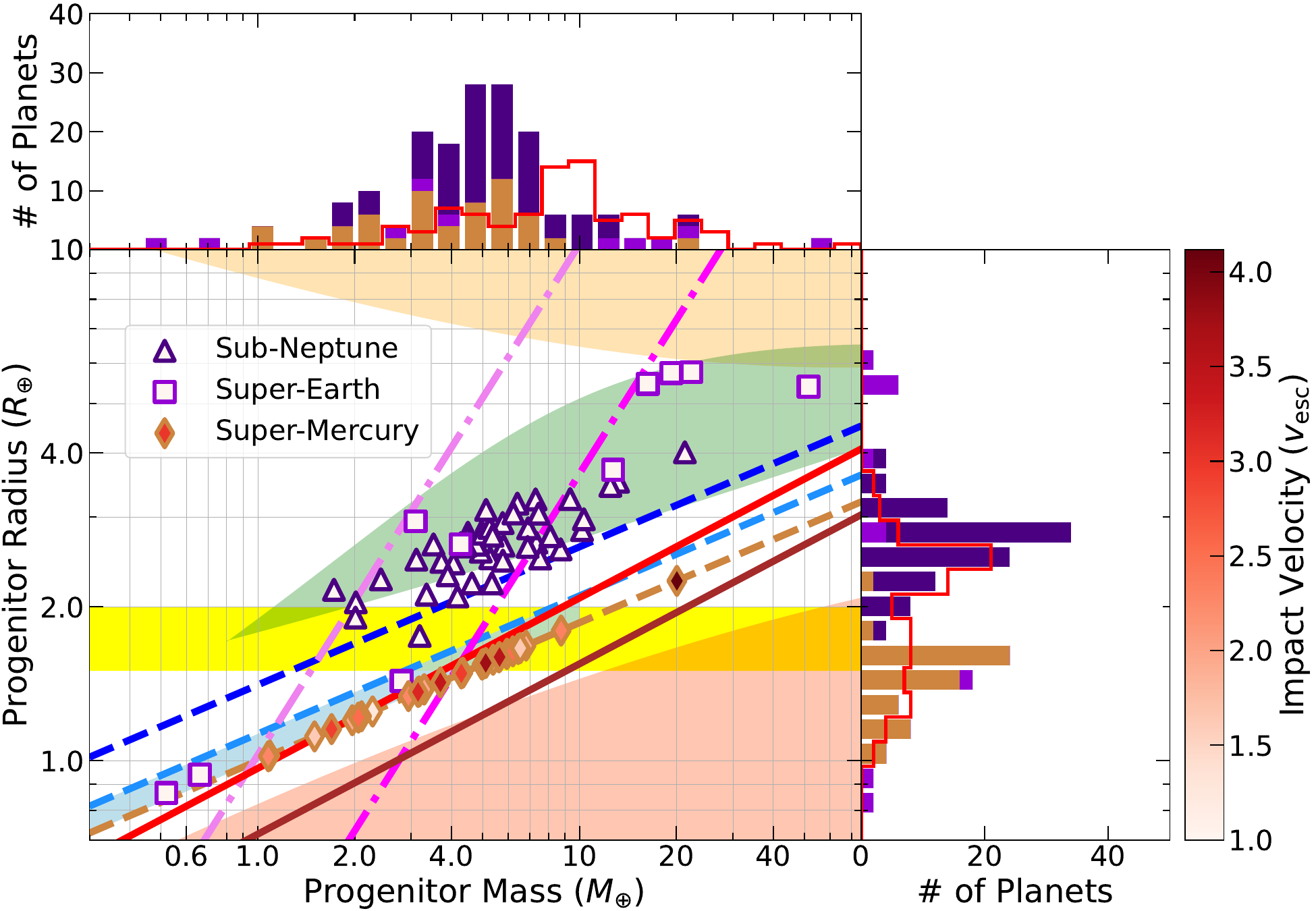}
\includegraphics[width=8.4cm]{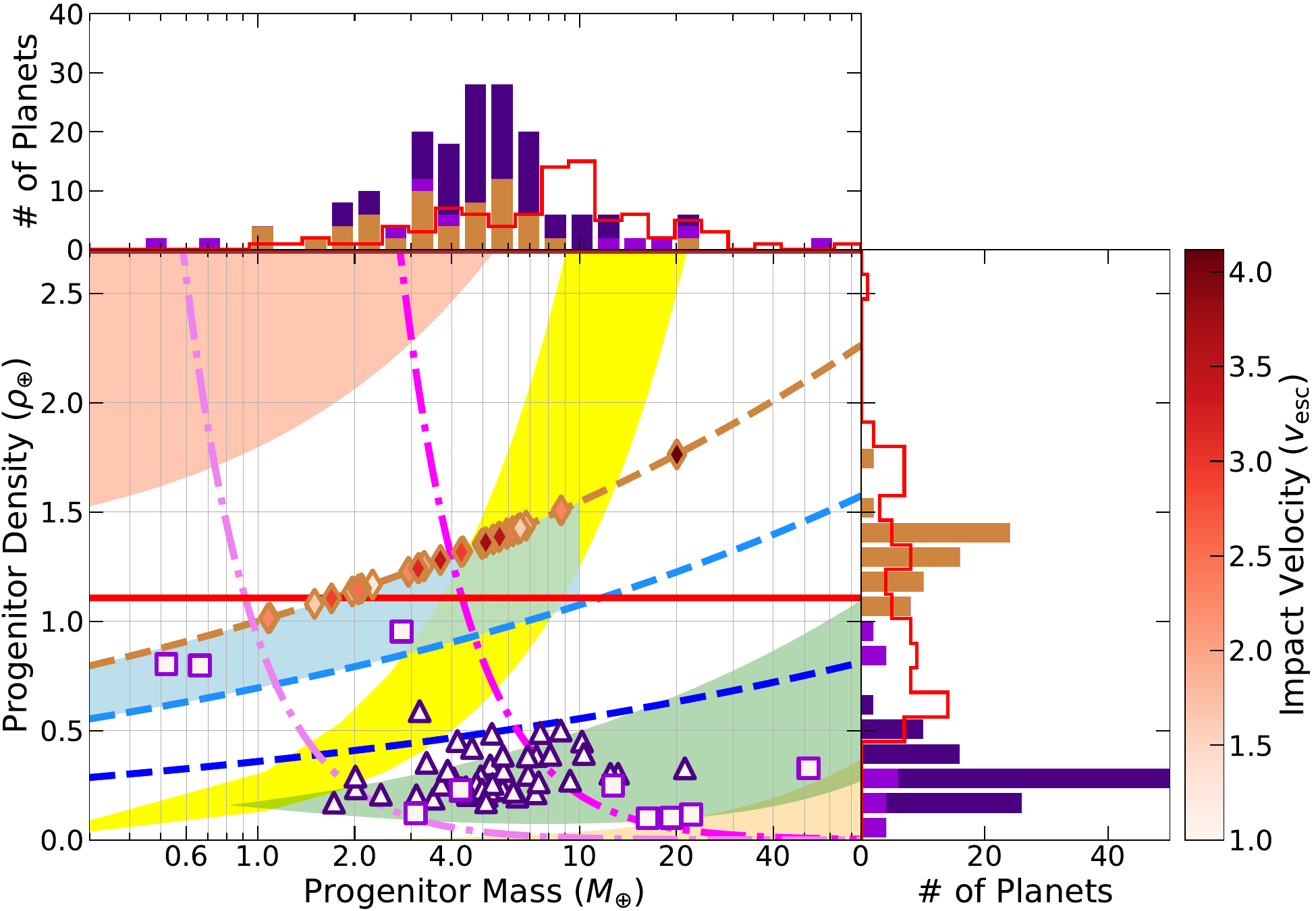}
\includegraphics[width=8.4cm]{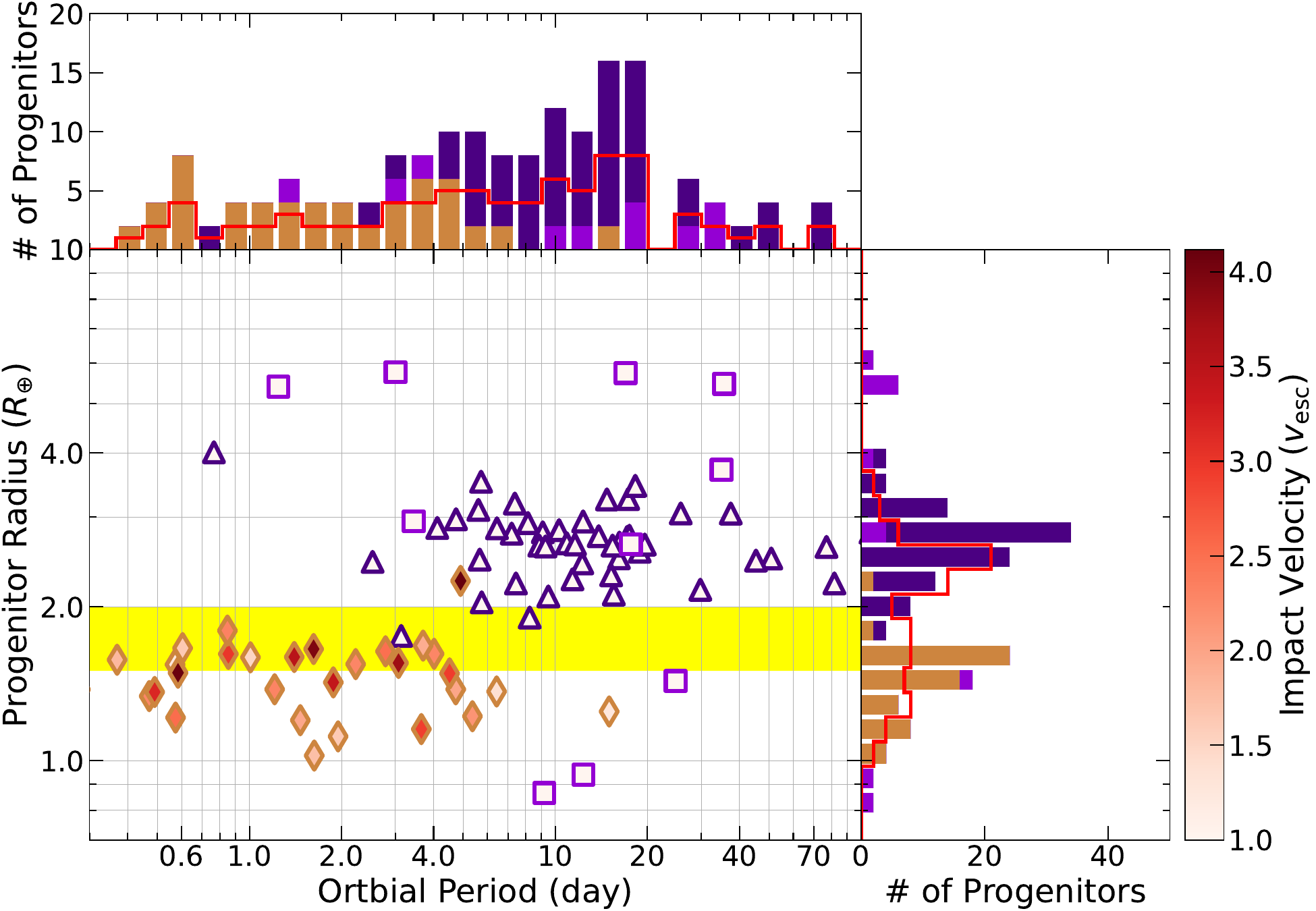}
\includegraphics[width=8.4cm]{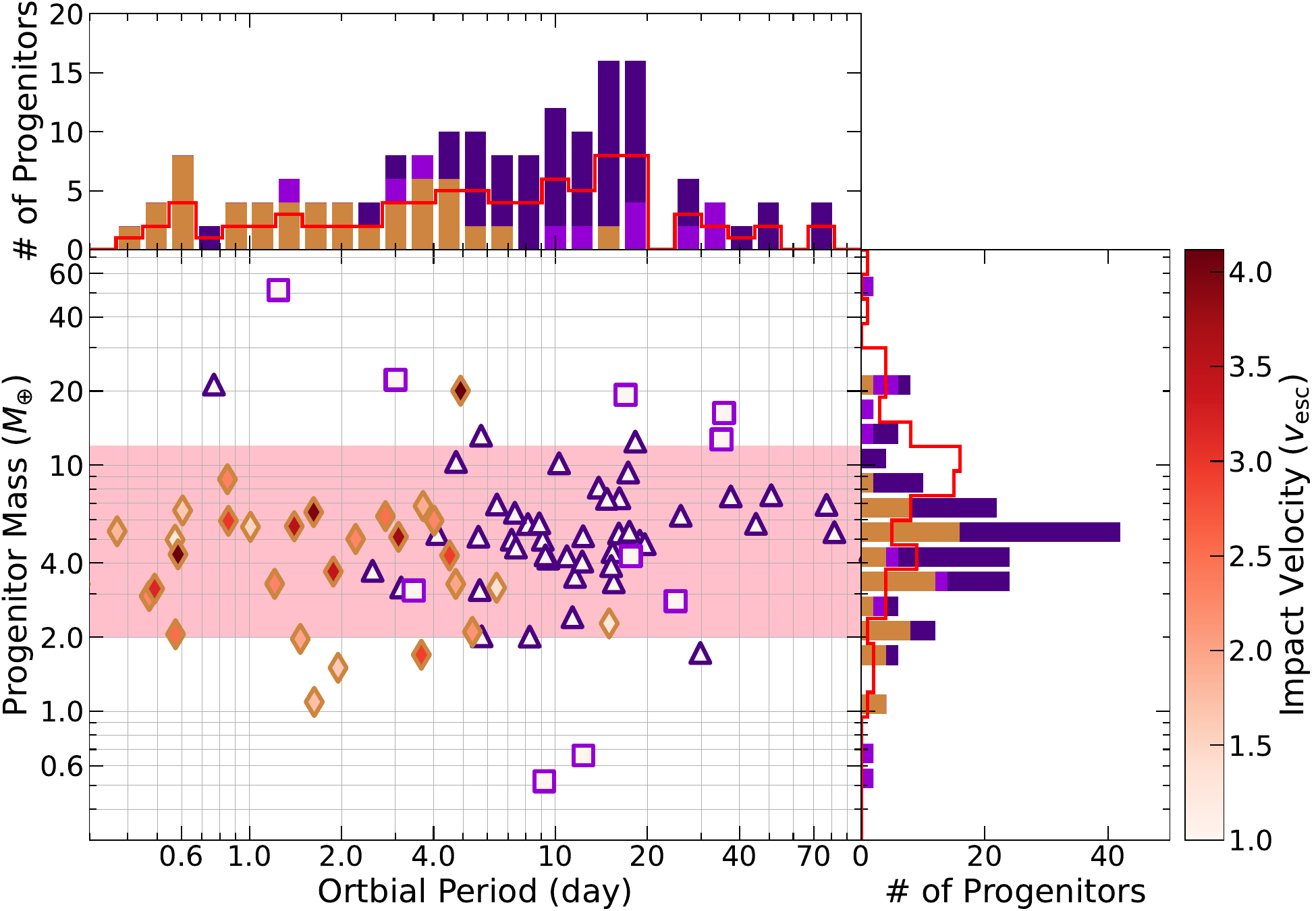}
\caption{The primordial properties of progenitors of exoplanets that likely experienced collisions 
(i.e., sub-Neptunes, super-Earths, and super-Mercuries, as in Figures \ref{fig8} and \ref{fig10}).
On each panel, the color bar represents impact velocity, 
and the red line in the histograms shows the distribution of their descendants (i.e., sub-Neptunes, super-Earths, and super-Mercuries).
The mass-radius and mass-density diagrams (the upper panels) confirm that standard core accretion accounts for the properties of these progenitors,
and subsequent collisions and the resulting mass growth and loss provide a natural explanation for the additional diversity noticeable for observed exoplanets.
The orbital period-radius and orbital period-mass diagrams (the lower panels) show that 
catastrophic collisions occurred predominantly for very close-in (i.e., $\lesssim 10$ days) planets.}
\label{fig11}
\end{center}
%\end{figure}
\end{minipage}
\end{figure*}

Figure \ref{fig11} summarizes the progenitor properties of planets that likely experienced collisions, namely, sub-Neptunes, super-Earths, and super-Mercuries.
As expected, they distribute in the green and blue shaded regions in the mass-radius and mass-density diagrams with a small number of outliers.
Thus, standard core accretion is applicable to most of these progenitors, and subsequent collisions determine the current properties of these planets eventually.

\section{Theoretical predictions} \label{sec:test}

In the above sections, we explore how formation and evolution processes shape the current properties of observed exoplanets.
In this section, we discuss what predictions are derived from such classification to further examine the classification.

\subsection{Link with giant planet formation}

We have so far focused on relatively small-sized exoplanets (i.e., $R_{\rm p} \lesssim 6 R_{\oplus}$).
In this section, we consider large-sized planets (i.e., sub-giants and gas giants) and explore how giant planet formation is constrained by our characterization of observed exoplanets.

First, we already discuss the possible mass range of the critical core mass ($M_{\rm c, crit}$), which is
\begin{equation}
\label{eq:M_c_crit_range}
1.6  - 3 M_{\oplus} \lesssim M_{\rm c, crit} \lesssim 10 M_{\oplus}.
\end{equation}
This is derived from the properties of super-Earths and sub-Neptunes (Figure \ref{fig9}).

The critical core mass is the key quantity to determine when gas accretion begins and known to be a function of solid accretion rate onto cores ($\dot{M}_{\rm core}$) 
and the dust opacity ($ \kappa_{\rm grain}$) of gaseous envelopes.
A mathematical expression of $M_{\rm c, crit}$ is given as \citep{2000ApJ...537.1013I}
\begin{equation}
\label{eq:M_c_crit}
M_{\rm c, crit} \simeq 10 M_{\oplus}  \left( \frac{ \kappa_{\rm grain} }{ \kappa_{\rm grain}^{\rm ISM} } \right)^a \left( \frac{ \dot{M}_{\rm core} }{1 0^{-6} M_{\oplus} \mbox{yr}^{-1} } \right)^b,
\end{equation}
where we set that $a \simeq b \simeq 0.25$, following \citet{2004ApJ...604..388I}, 
and $\kappa_{\rm grain}^{\rm ISM} = 1$ g cm$^{-2}$ is the grain opacity of the interstellar medium (ISM).
The mass range of the critical core mass therefore enables constraining these physical quantities.

%\begin{figure*}
%\begin{minipage}{17cm}
\begin{figure}%[!ht]
\begin{center}
\includegraphics[width=8.4cm]{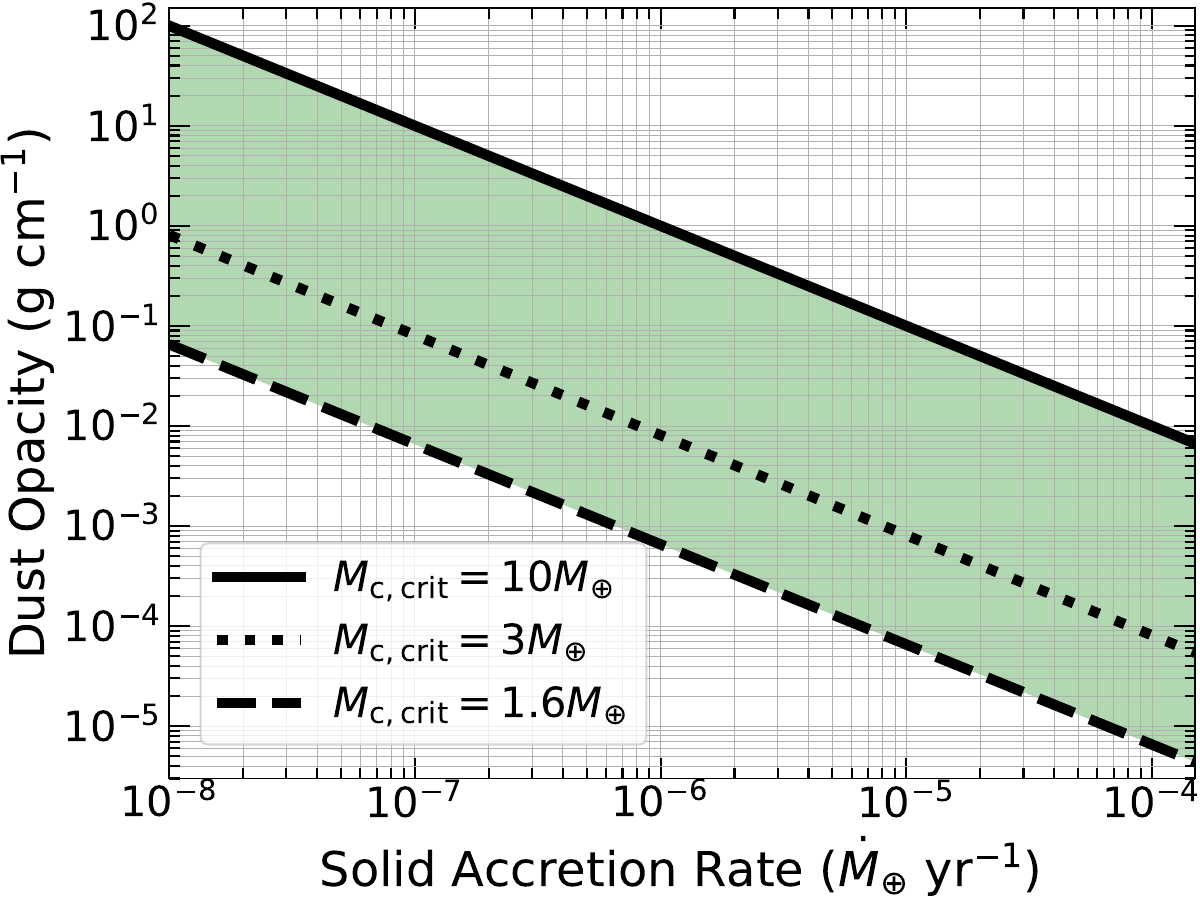}
\caption{The possible combination of dust opacities and solid accretion rates that reproduces the mass range of the critical core mass (equation (\ref{eq:M_c_crit_range})).
The profiles are plotted, using equation (\ref{eq:M_c_crit}), 
and three values of the critical core mass are considered ($M_{\rm c, crit}= 1.6 M_{\oplus}, 3 M_{\oplus}$, and $10 M_{\oplus}$).
The resulting wide parameter space (the green shaded region) makes it difficult to tightly constrain these quantities.}
\label{fig12}
\end{center}
\end{figure}
%\end{minipage}
%\end{figure*}

Figure \ref{fig12} shows the possible range of $ \kappa_{\rm grain}$ as a function of $\dot{M}_{\rm core}$ for given values of the critical core mass.
These profiles are obtained by combining equations (\ref{eq:M_c_crit_range}) and (\ref{eq:M_c_crit}).
Clearly, a wide range of $ \kappa_{\rm grain}$ and $\dot{M}_{\rm core}$ are acceptable, and no tight constraints are obtained.

It should be noted that equation (\ref{eq:M_c_crit_range}) is obtained from observed super-Earths and sub-Neptunes,
and hence it likely represents one conservative range;
the critical core mass can exceed $10 M_{\oplus}$ if dust opacity or solid accretion rates are very high (equation (\ref{eq:M_c_crit})),
increasing the possibility for accreting planets to become gas giants.
For such a case, the shaded region in Figure \ref{fig12} expands further.
Also, the effect of the mean molecular weight of envelopes is not included,
which affects the critical core mass as well \citep[e.g.,][]{2011MNRAS.416.1419H}.

Second, we specify what is the main driver to differentiate between sub-giants and gas giants.
As described above (e.g., the lower panel of Figure \ref{fig5}), photoevaporated sub-giants currently keep a good amount of primordial envelopes, 
confirming that they surely underwent gas accretion.
If their gaseous envelope was always less massive than the core during the gas accretion phase, then rapid gas accretion was not activated, and they avoided turning into gas giants.
This becomes possible if concurrent accretion of gas and solids occurs; (some of) accreted solids participate into cores, 
so that not only the envelope mass, but also the core (or solid) mass increases simultaneously, resulting in $M_{\rm core} > M_{\rm env}$ all the time.
We examine this possibility by focusing on photoevaporated sub-giants and their initial envelope mass and the mass of solids accreted after core formation.

\begin{figure*}
\begin{minipage}{17cm}
%\begin{figure}%[!ht]
\begin{center}
\includegraphics[width=8.4cm]{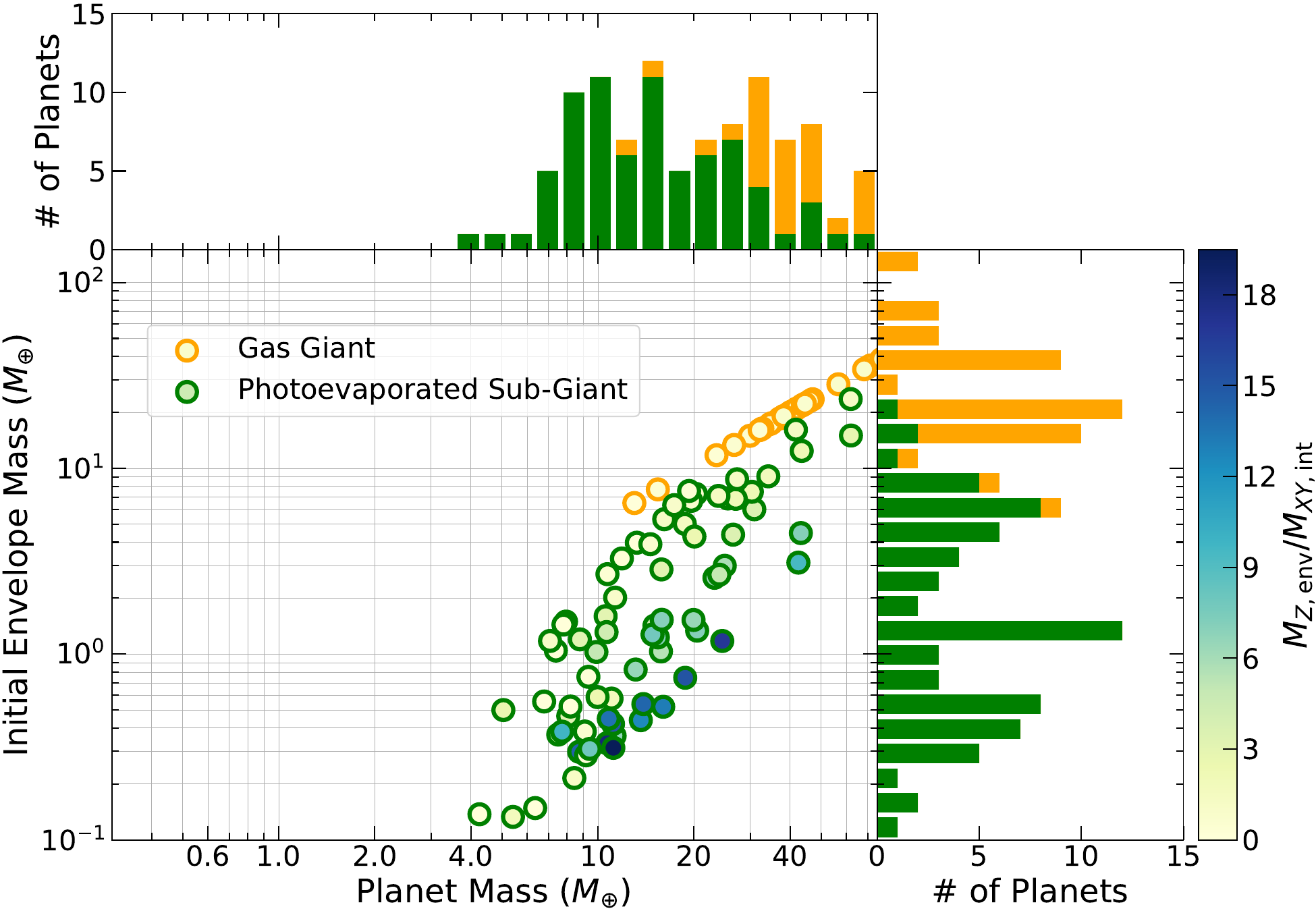}
\includegraphics[width=8.4cm]{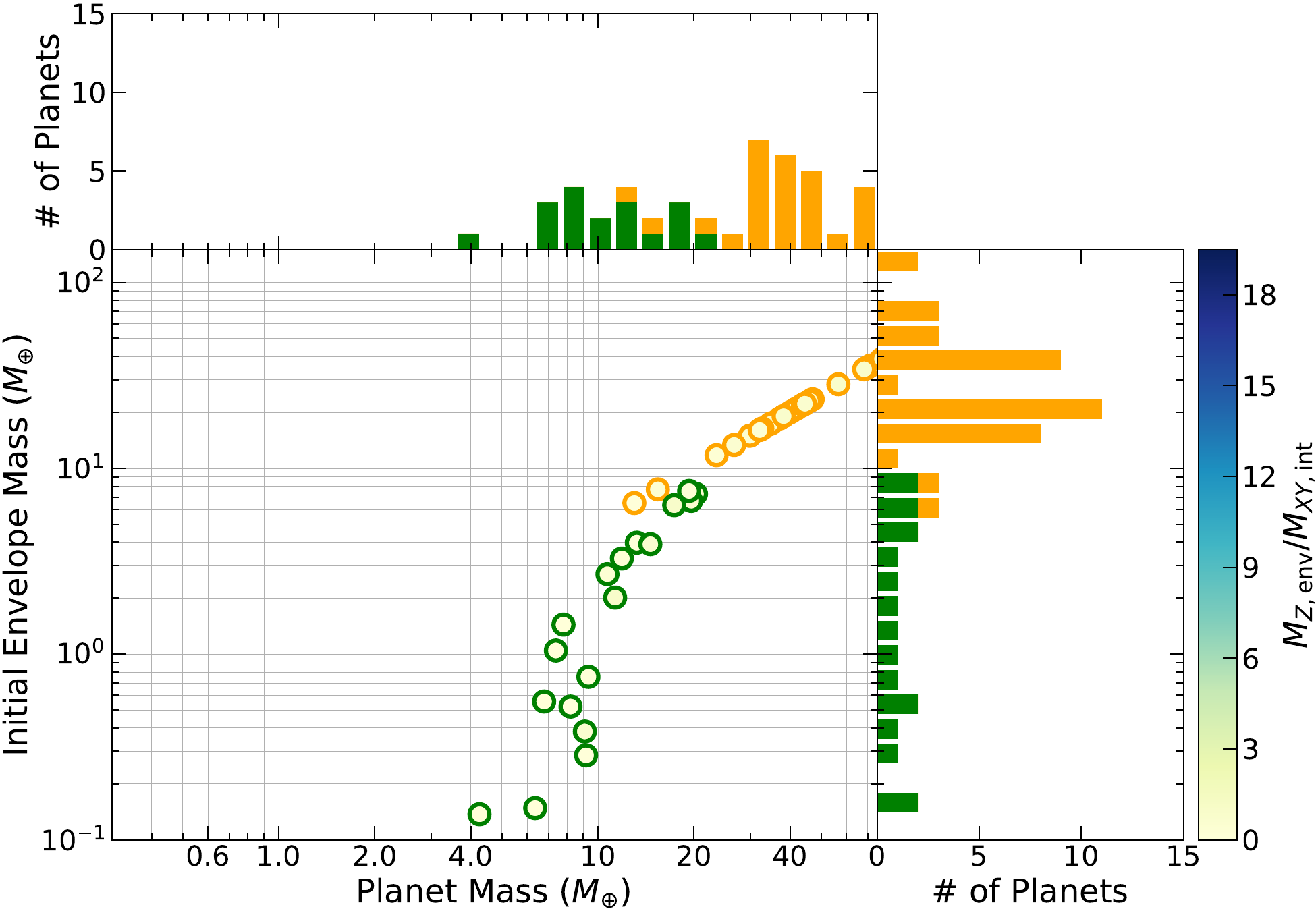}
\caption{The computed value of the initial envelope mass as a function of planet mass for gas giants and photoevaporated sub-giants.
On the left panel, all the exoplanets classified as gas giants and photoevaporated sub-giants in Figures \ref{fig4} and \ref{fig5} respectively are plotted,
while on the right panel, photoevaporated sub-giants that meet the condition that $M_{Z, \rm env}/M_{XY, \rm int} > 1$ are excluded.
On the color bar, the computed value of $M_{Z, \rm env}/M_{XY, \rm int}$ is denoted.
The current samples used in this work clearly exhibit that photoevaporated sub-giants that are more massive than $10M_{\oplus}$ tend to have high values of  $M_{Z, \rm env}/M_{XY, \rm int}$,
supporting the possibility that they did not experience rapid gas accretion due to concurrent accretion of gas and solid.}
\label{fig13}
\end{center}
%\end{figure}
\end{minipage}
\end{figure*}

Figure \ref{fig13} shows the initial (i.e., before photoevaporation) envelope mass ($M_{XY, \rm int}$) as a function of planet mass for photoevaporated sub-giants and gas giants.
On the left panel, all the samples classified into these two types are plotted.
As described in Section \ref{sec:app}, the current envelope mass ($M_{XY}$) for photoevaporated sub-giants is computed from the present planet mass and radius self-consistently,
and hence the total solid mass ($M_Z$) is specified by equation (\ref{eq:Mp_comp}).
The value of $M_{XY, \rm int}$ for photoevaporated sub-giants is retrieved, taking account of photoevaporation as done in Section \ref{sec:post_photo}.
For gas giants, we again assume that $M_{XY} = M_{Z}$ since they distribute around the 50 \% core-50 \% envelope line (Figure \ref{fig4}).
Due to this approximation, the computed value of $M_{XY, \rm int}$ is not accurate.
We therefore assume that $M_{XY, \rm int} = M_{XY} $ for gas giants.

In order to examine the effect of concurrent gas and solid accretion after core formation, 
we decompose $M_Z$ into (c.f., equation (\ref{eq:M_comp}))
\begin{equation}
M_Z = M_{\rm core} + M_{Z, \rm env},
\end{equation}
where $M_{Z, \rm env}$ is the mass of solids accreted during gas accretion.
We apply the above equation to photoevaporated sub-giants and gas giants by randomly picking up the core mass from the mass range (equation (\ref{eq:M_c_crit_range})) uniformly;
we set $3 M_{\oplus}$ for the lower bound and confirmed that the results do not change very much even if $1.6 M_{\oplus}$ is used.
The color bar represents the ratio of $M_{Z, \rm env}/M_{XY, \rm int}$.
We find that photoevaporated sub-giants, especially those more massive than 10 $M_{\oplus}$, have higher values of $M_{Z, \rm env}/M_{XY, \rm int}$, 
supporting the possibility that concurrent accretion of gas and solid operated for these sub-giants.

Figure \ref{fig13} (the right panel) exhibits this trend clearly. 
In this plot, photoevaporated sub-giants that satisfy the condition that $M_{Z, \rm env}/M_{XY, \rm int} > 1$ are excluded.
A continuous transition from these sub-giants to gas giants is identified,
and noticeably, the transition occurs rapidly around the planet mass of $10M_{\oplus}$, the upper bound of the critical core mass.

Thus, current observed exoplanets affirm the importance of concurrent accretion of gas and solid after core formation to differentiate between sub-giants and gas giants,
especially if sub-giants are more massive than $10M_{\oplus}$.

\subsection{Refined classification} \label{sec:refine_class}

We propose a refined classification scheme.

Figure \ref{fig3} summarizes the classification scheme with the corresponding conditions and relevant physical processes.
As described in Sections \ref{sec:current} and \ref{sec:post}, it is based on the current planet mass and radius and insolation flux.
Following the mass range of the critical core mass (equation (\ref{eq:M_c_crit_range})),
planets that are less massive than $1.6 M_{\oplus}$ are now re-defined as bare cores.
On the other hand, super-Earths that are more massive than $1.6 M_{\oplus}$ are called collisionally sculpted cores,
and sub-Neptunes collisionally sculpted sub-giants.
Observed exoplanets are categorized into eight types in total: 
super-Mercuries, bare cores, collisionally sculpted cores, photoevaporated cores, 
collisionally sculpted sub-giants, photoevaporated sub-giants, vapor-rich sub-giants, and gas giants.
Among them, massive bare cores, collisionally sculpted cores, photoevaporated cores, 
collisionally sculpted sub-giants, photoevaporated sub-giants, and vapor-rich sub-giants are potentially water-rich.
We do not discuss the effect of orbital evolution yet, which is briefly touched on the following section.

\subsection{Formation and evolution pathways} \label{sec:pathways}

We are now in the position to discuss formation and evolution pathways of observed exoplanets.
We divide these pathways into four steps: core formation, gas accretion, collisional mass growth and loss, and photoevaporation.
As discussed in Section \ref{sec:post_impact}, different sequences between collisional mass growth and loss and photoevaporation are possible for different planets.
We here assume that collisional mass growth and loss are followed by photoevaporation since it is required for many collisionally sculpted sub-giants
 (i.e., sub-Neptunes in Figure \ref{fig9}).
To readily trace the sequence of planet formation and evolution processes, only the mass-radius and mass-density diagrams are shown here;
detailed properties of planets at each stage are summarized in Appendix \ref{sec:appned2}.

\begin{figure*}
\begin{minipage}{17cm}
%\begin{figure}%[!ht]
\begin{center}
\includegraphics[height=5.1cm]{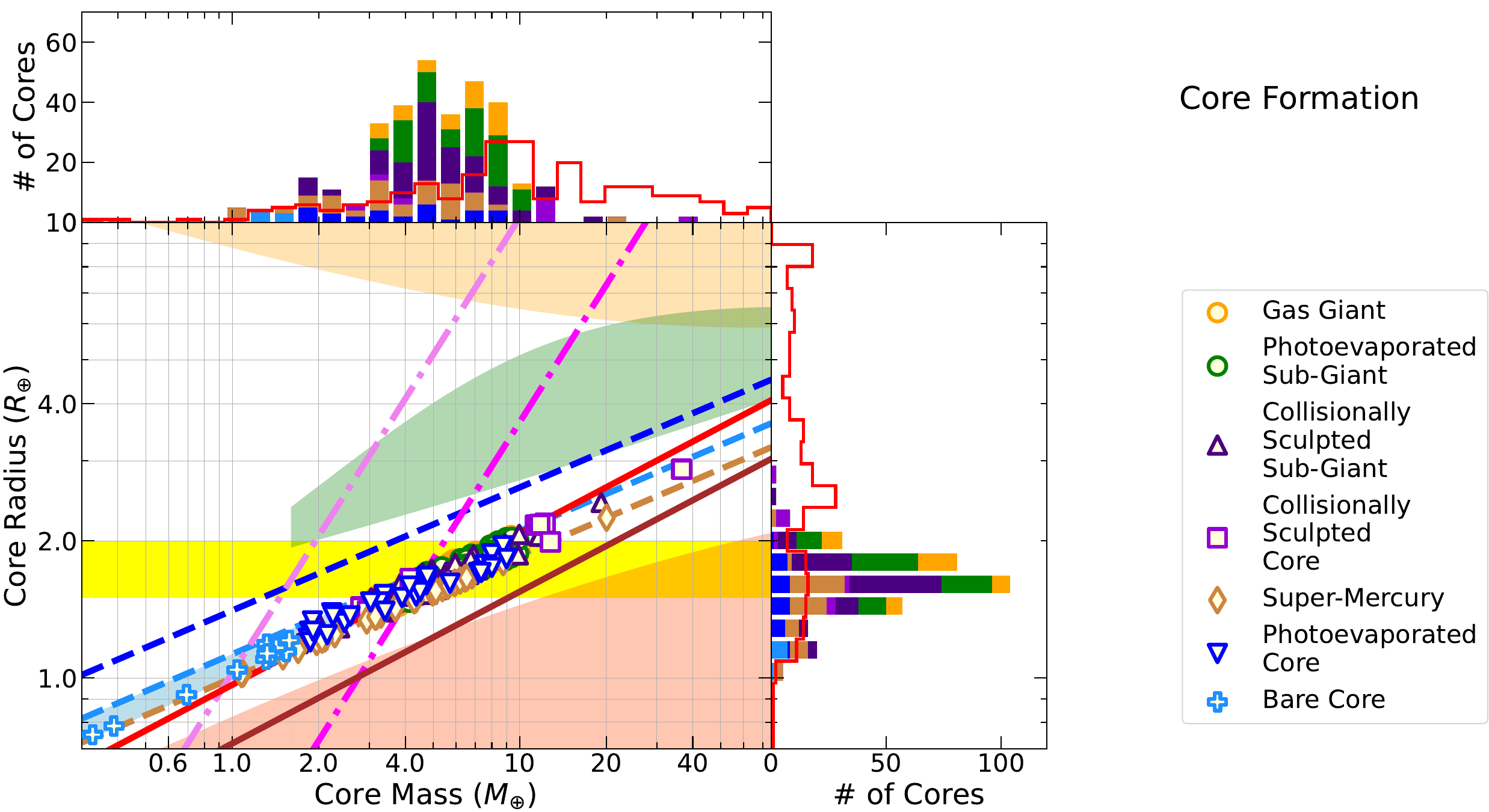}
\includegraphics[height=5.1cm]{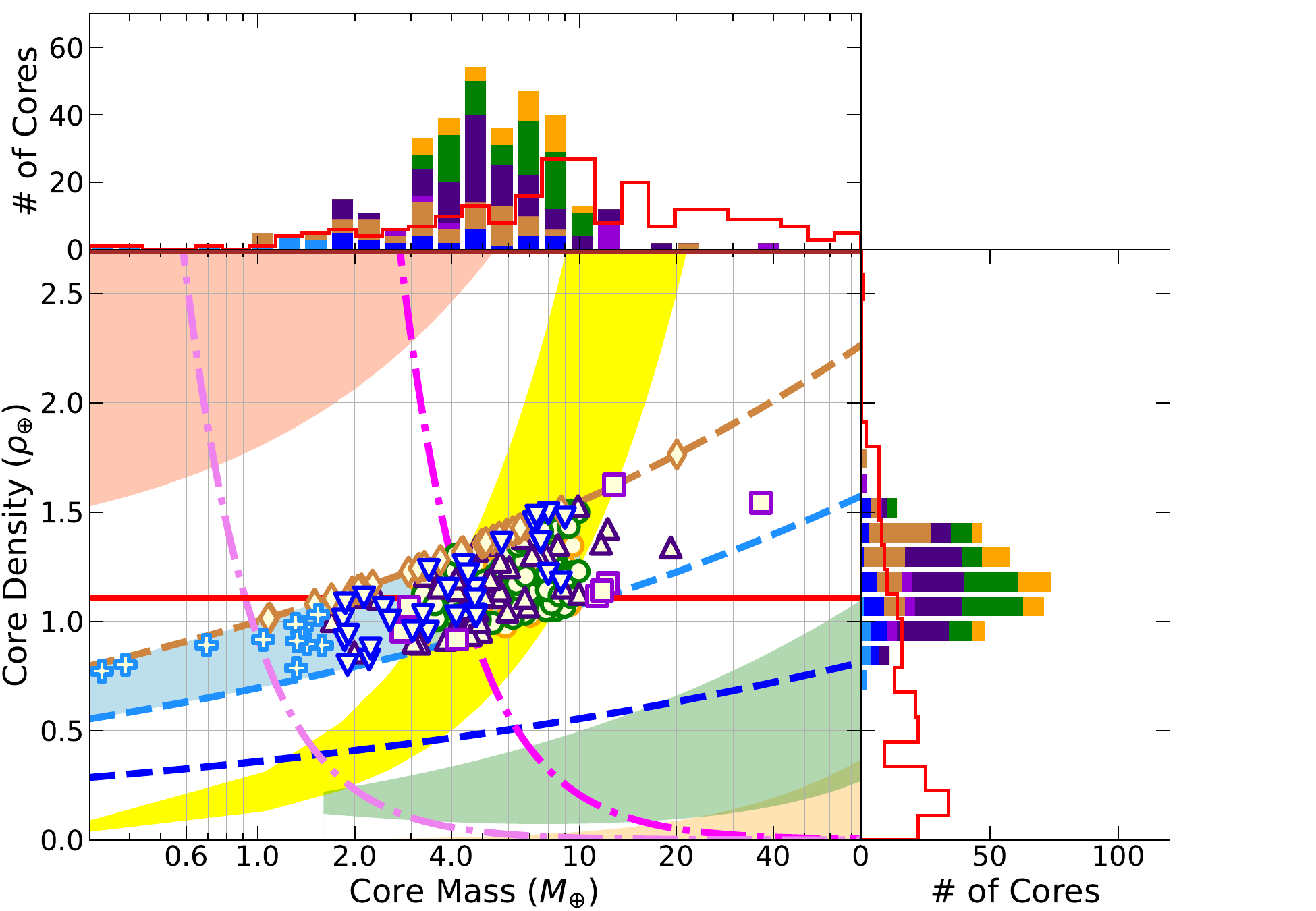}
\includegraphics[height=5.1cm]{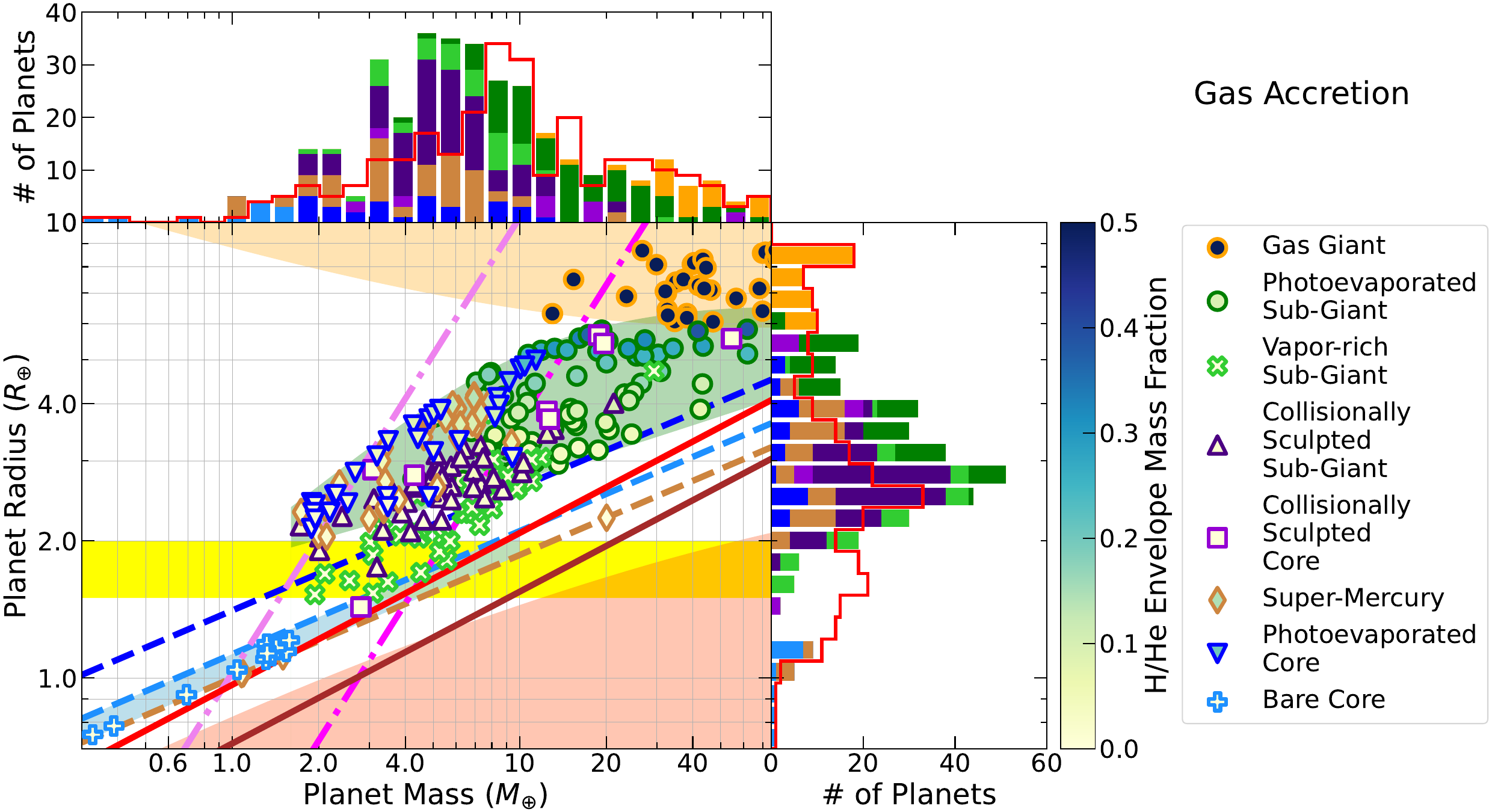}
\includegraphics[height=5.1cm]{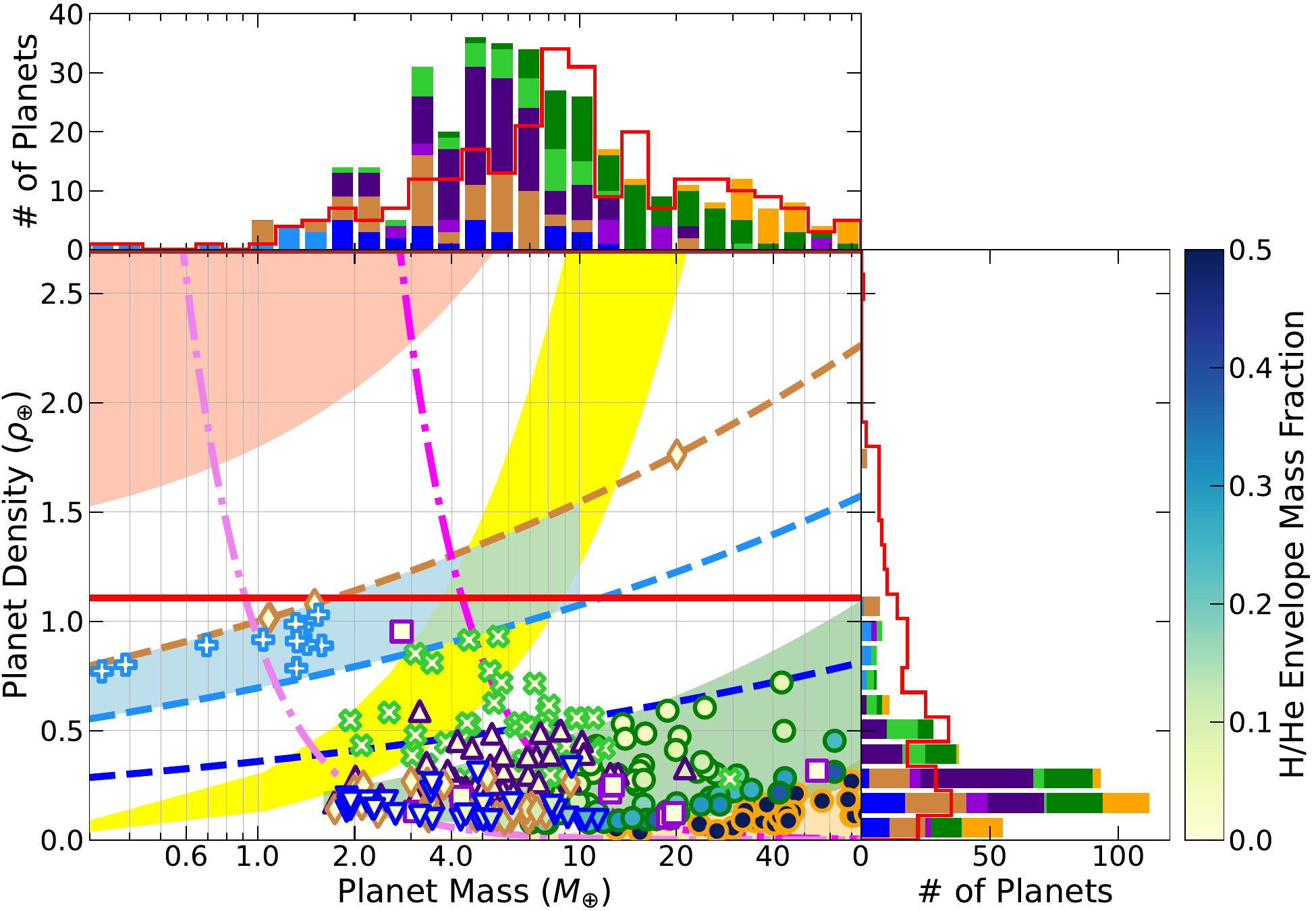}
\includegraphics[height=5.1cm]{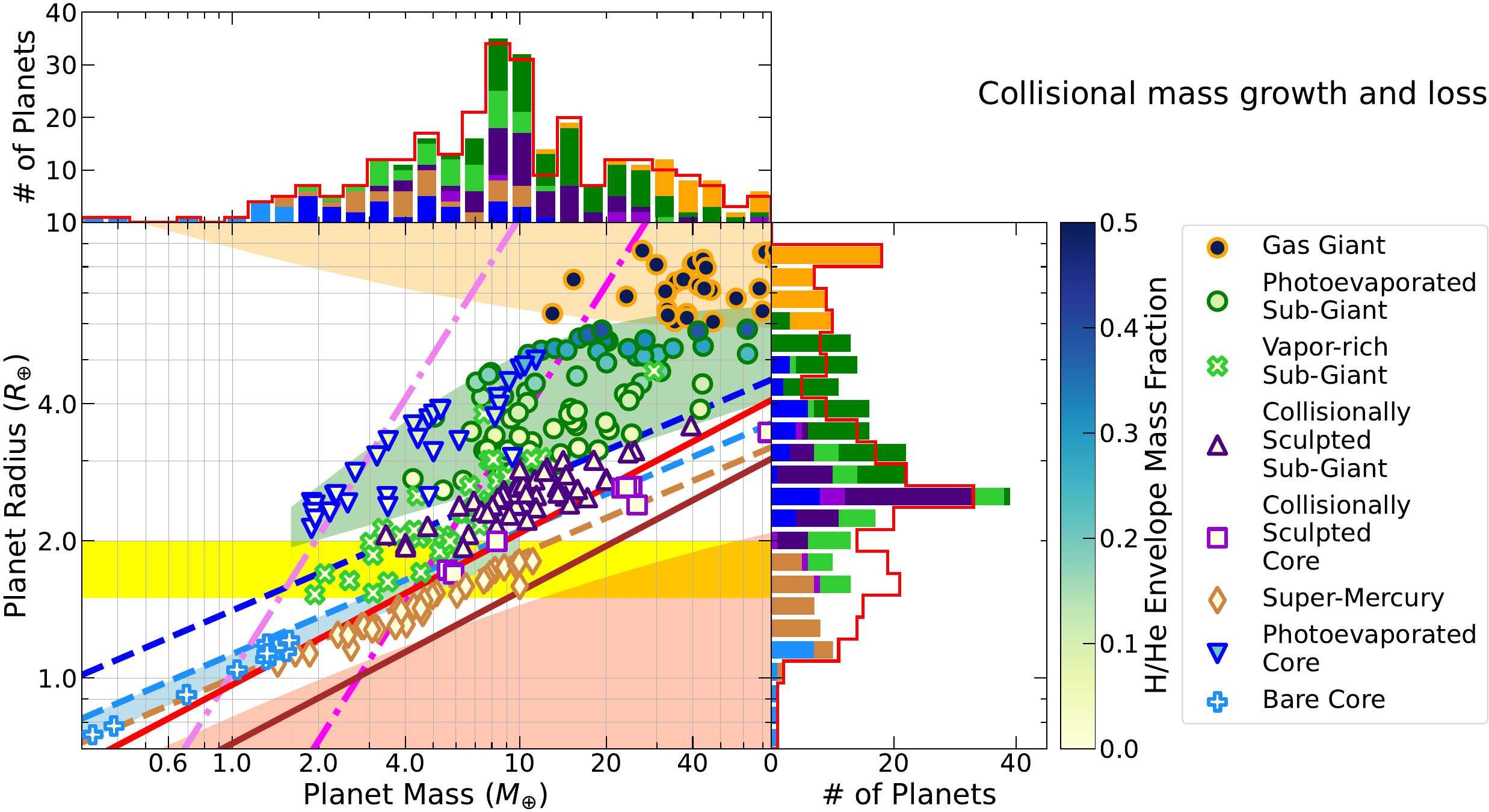}
\includegraphics[height=5.1cm]{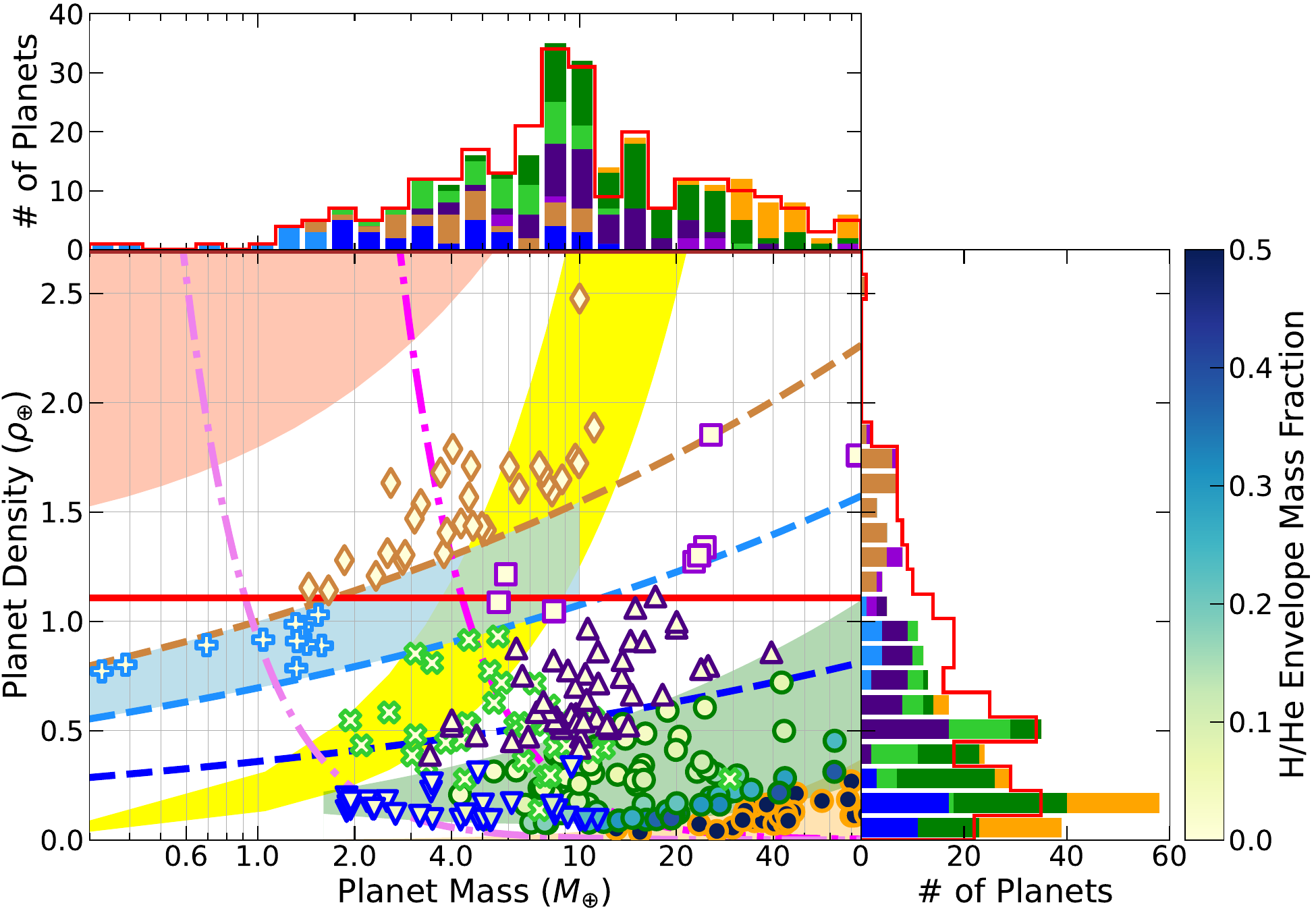}
\includegraphics[height=5.1cm]{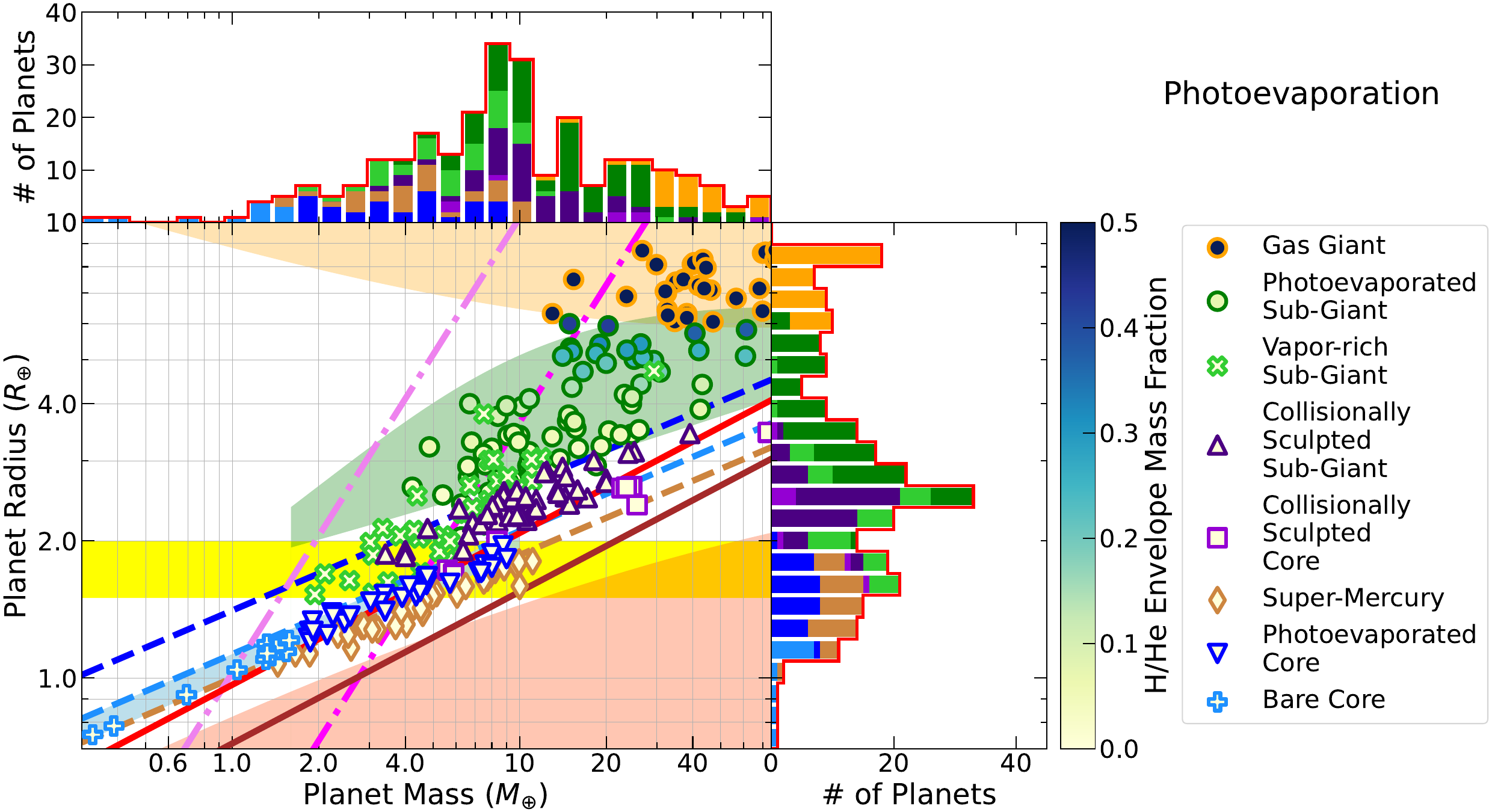}
\includegraphics[height=5.1cm]{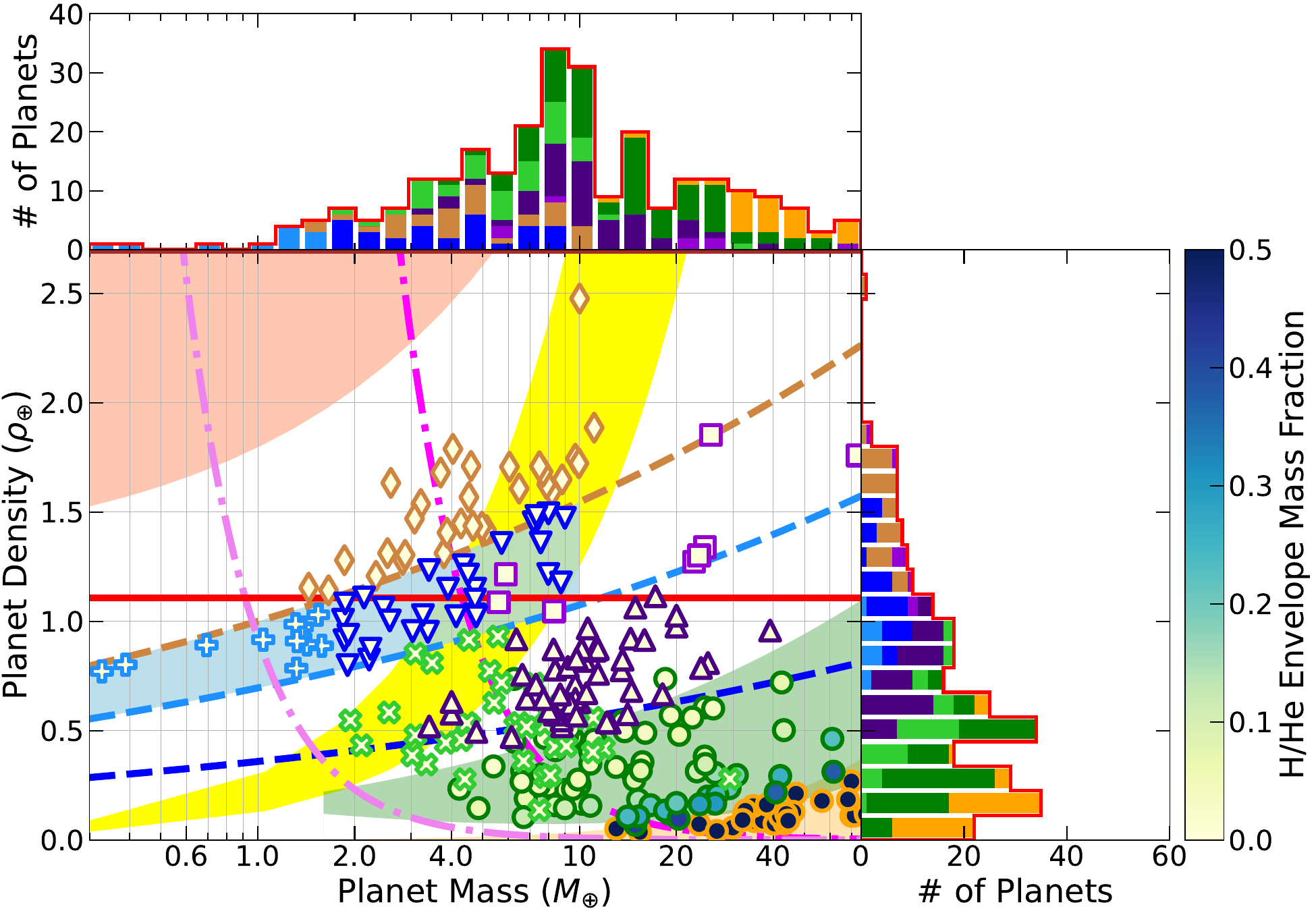}
\caption{The properties of planets at different formation and evolution stages in the mass-radius and mass-density diagrams on the left and right panels, respectively.
}
\label{fig14}
\end{center}
%\end{figure}
\end{minipage}
\end{figure*}

\addtocounter{figure}{-1}
\begin{figure*}
\begin{minipage}{17cm}
\caption{({\it Continued.}) Observed exoplanets treated in this work are categorized by the refined classification scheme (Figure \ref{fig3}).
The red line in two histograms represents the distribution of the current properties of exoplanets.
The color bar denotes the envelope mass fraction.
The top panels correspond to the core formation stage, and vapor-rich sub-giant planets are not included in these plots.
Most planets are distributed in the blue shaded region as expected.
The range of core composition is bounded by the Earth-like rock and water rich lines (Table \ref{table1}).
Such a range shrinks suddenly for planetary cores that are less massive than $\sim 1.3 M_{\oplus}$.
The second panels denote the gas accretion stage.
Planet radius expands due to gas accretion, and all of the planets now reside in the green and orange shaded regions, 
except for vapor-rich sub-giants and bare cores ($\lesssim 1.6 M_{\oplus}$) that did not undergo gas accretion as well as one outlier.
The third panels represent the stage of collisional mass growth and loss. 
This stage begins after gas disk dispersal.
Two new areas in these diagrams are populated by the formation of super-Mercuries and collisionally sculpted sub-giants.
The former fills out the radius valley, and the latter distributes in the parameter space, which otherwise requires the proposition of peculiar, water-dominated (i.e., nearly pure water) planets. 
The bottom panels show the final, photoevaporation stage.
Envelope mass loss by photoevaporation produces the population of photoevaporated cores,
which further fill out the radius valley if they are more massive than $\sim 4 M_{\oplus}$.}
\end{minipage}
\end{figure*}

\begin{figure*}
\begin{minipage}{17cm}
%\begin{figure}%[!ht]
\begin{center}
\includegraphics[height=12cm]{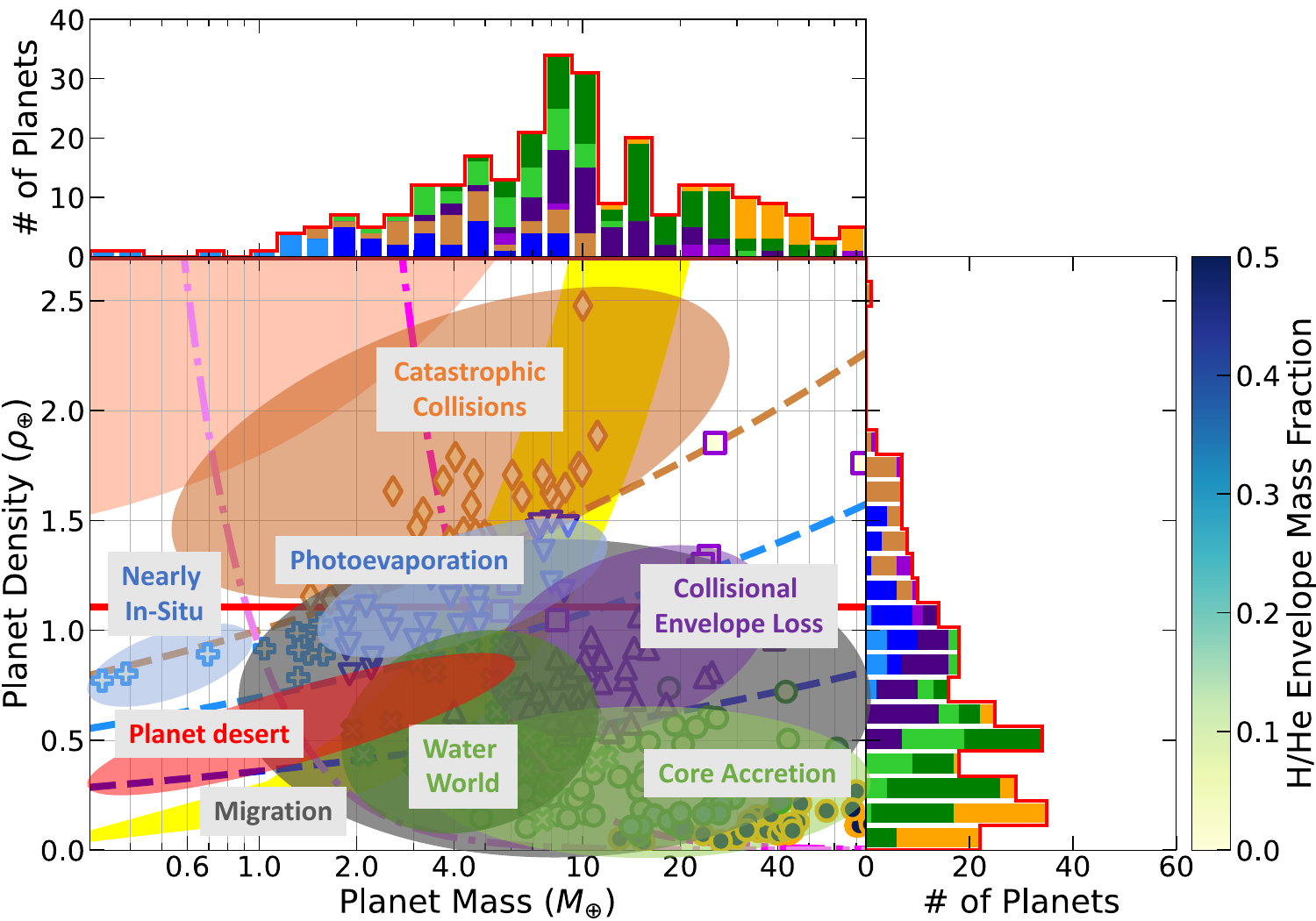}
\caption{Integration of  the formation and evolution pathways (as in Figure \ref{fig14}).
The shaded ellipses are added only for the purpose of visualization without quantitative estimates.
Identification of a planet desert makes it possible to constrain a possible maximum amount of water stored in planets.}
\label{fig15}
\end{center}
%\end{figure}
\end{minipage}
\end{figure*}

\textit{Core formation.} This is the first step of planet formation. 
Figure \ref{fig14} (the top panel) shows the behavior of currently observed exoplanets used in this work.
Detailed properties (e.g., core mass and envelope mass) of vapor-rich sub-giant planets are not constrained by the methodology used in this work, 
and hence they are not included in these plots.
Most planets distribute in the blue shaded region in the mass-radius and mass-density diagrams with some outliers.
This suggests that they likely formed out of solids with a range of the mixture between water and rock capped at $1/3$ at most, similar to solar system bodies (Table \ref{fig1}).
Since the planet mass after core formation is modest, the mass-density diagram is preferred to carefully examine the properties.
We find that the core composition is more or less Earth-like rock for cores that are less massive than 1.3 $M_{\oplus}$, 
and beyond that value, their composition becomes diverse within the blue shaded region.
This implies that delivery of such icy materials to the vicinity of the host star is likely to be conducted by gas-induced planetary migration;
such a sudden change is not expected if water is delivered by inward drift of dust particles.
In other words, large-scale migration becomes effective only for cores that are more massive than 1.3 $M_{\oplus}$, 
and less massive cores form within the water-snow line, so more or less in-situ like.
The number of data points currently available is small, and further examination of this discussion is surely needed.
We also notice efficient core formation in the mass range between $\sim 3 M_{\oplus}$ and $\sim 9 M_{\oplus}$ (the top histogram),
while the lower bound would be due to observational biases.

\textit{Gas accretion.} This is the second step of planet formation.
Figure \ref{fig14} (the second panel) shows the resulting properties of planets. 
For the envelope mass, the initial value (i.e., before collisions and photoevaporation) is used.
Due to gas accretion, the planet radius expands, and the mass-radius diagram is suitable for careful examination;
planets possessing gaseous envelopes move to the green and orange shaded regions.
The parameter space corresponding to the radius valley is populated by vapor-rich sub-giants, which is more visible in the mass-density diagram.
However, they are not abundant, suggesting that water-dominated planets are rare.

\textit{Collisional mass growth and loss.} This is the third step and occurs after gas disks are gone.
Collisions lead to mass growth and loss and eventually fill out two new parameter spaces in the mass-radius and mass-density diagrams.
Figure \ref{fig14} (the third panel) shows the results, and the mass-density diagram gives clearer illustration.
If collisions are energetic enough to remove silicate mantles (e.g., $v_{\rm imp} > v_{\rm esc}$),
then planets formed by the collisions have higher bulk densities than the Earth, 
and the parameter space above the Earth-like rock line (Table \ref{table1}) is populated.
On the other hand, if collisions are less energetic such that $v_{\rm imp} \simeq v_{\rm esc}$,
only gaseous envelopes are removed, and another new parameter space is occupied.
Importantly, the latter parameter space corresponds to the bulk density of water-dominated planets.
While the existence of such peculiar planets is not ruled out currently,
collisions provide a natural explanation for the origin of these planets.
Due to collisions, the parameter space identified by the radius valley is partially populated by super-Mercuries.

\textit{Photoevaporation.} This is the final step of evolution processes.
Figure \ref{fig14} (the bottom panel) shows the resulting distribution; while the refined classification (Figure \ref{fig3}) is explicitly labeled now,
the plots are identical to Figures \ref{fig1} and \ref{fig4}.
In essence, photoevaporation reduces the population of planets distributed in the green shaded region and increases that of planets distributed in the blue shaded region.
The latter parameter space overlaps with that defined by the radius valley, 
and hence planets are more populated in the valley due to photoevaporation if their core mass is larger than $\sim 4 M_{\oplus}$.
This implies that photoevaporation is a double-edged sword to produce the radius valley unless a certain distribution of core mass is achieved;
photoevaporation populates planets not only at the planet radius smaller than that of the valley, but also at the valley.
The diversity of massive bare cores and photoevaporated cores in bulk density suggests that they were likely to migrate to the current position by tidal torques arising from gas disks 
as discussed in the core formation stage.

Figure \ref{fig15} integrates the formation and evolution pathways discussed above.
In summary, the diversity of planets in the mass-radius and mass-density diagrams results from 
the combination of core accretion (i.e., core formation and gas accretion), collisional mass growth and loss, and photoevaporation.
The relative importance of each process is determined by the core mass that controls gas accretion and retention, 
the position of planets that regulates the occurrence and outcome of collisions and the efficiency of photoevaporation,
and the properties of planetary systems that dictate  the occurrence and outcome of collisions.
The efficiency of orbital evolution that is a function of planet mass determines the composition of planetary cores.
Verifying or falsifying the presence of a planet desert will constrain a possible maximum amount of water contained in planets.

\subsection{Planets in habitable zones}

The currently observed exoplanets are concentrated in the vicinity ($\lesssim$ 1 au) or at a distance ($\gtrsim$ 10 au) of the host star due to observational biases.
Upcoming new space telescopes such as Roman and HWO (Habitable World Observatory) will discover exoplanets in between, including planets in habitable zones.
We here discuss possible predictions to be deduced from our analysis that has been applied to close-in exoplanets in the above sections.

The properties of planets around habitable zones are almost unknown currently.
This is the case especially for planets around G-type stars.
Based on the exercise conduced in Section \ref{sec:post}, 
anticipated differences between close-in planets and habitable zone planets are
the efficiency of photoevaporation (e.g., equation (\ref{eq:massloss_PE})) and the occurrence of catastrophic collisions (e.g., equation (\ref{eq:rp_max})).
We therefore consider an extreme case that these processes will not operate for habitable zone planets.
It should be noted that in the following discussion,
the implicit assumption that the underlying population of habitable zone planets is similar to that of close-in planets observed currently is adopted.
Due to these limitations, great care is needed below.

\begin{figure*}
\begin{minipage}{17cm}
%\begin{figure}%[!ht]
\begin{center}
\includegraphics[height=5.1cm]{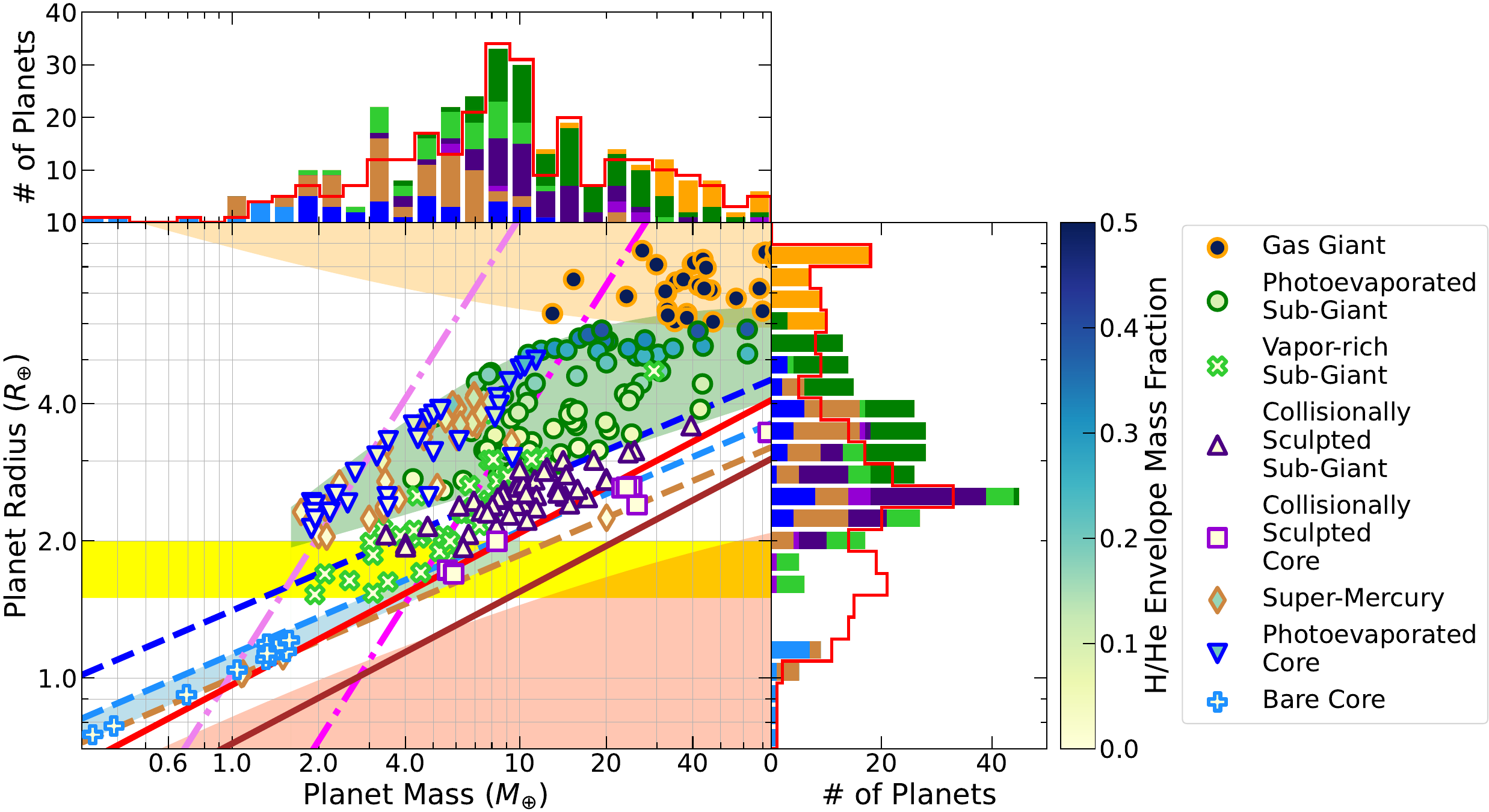}
\includegraphics[height=5.1cm]{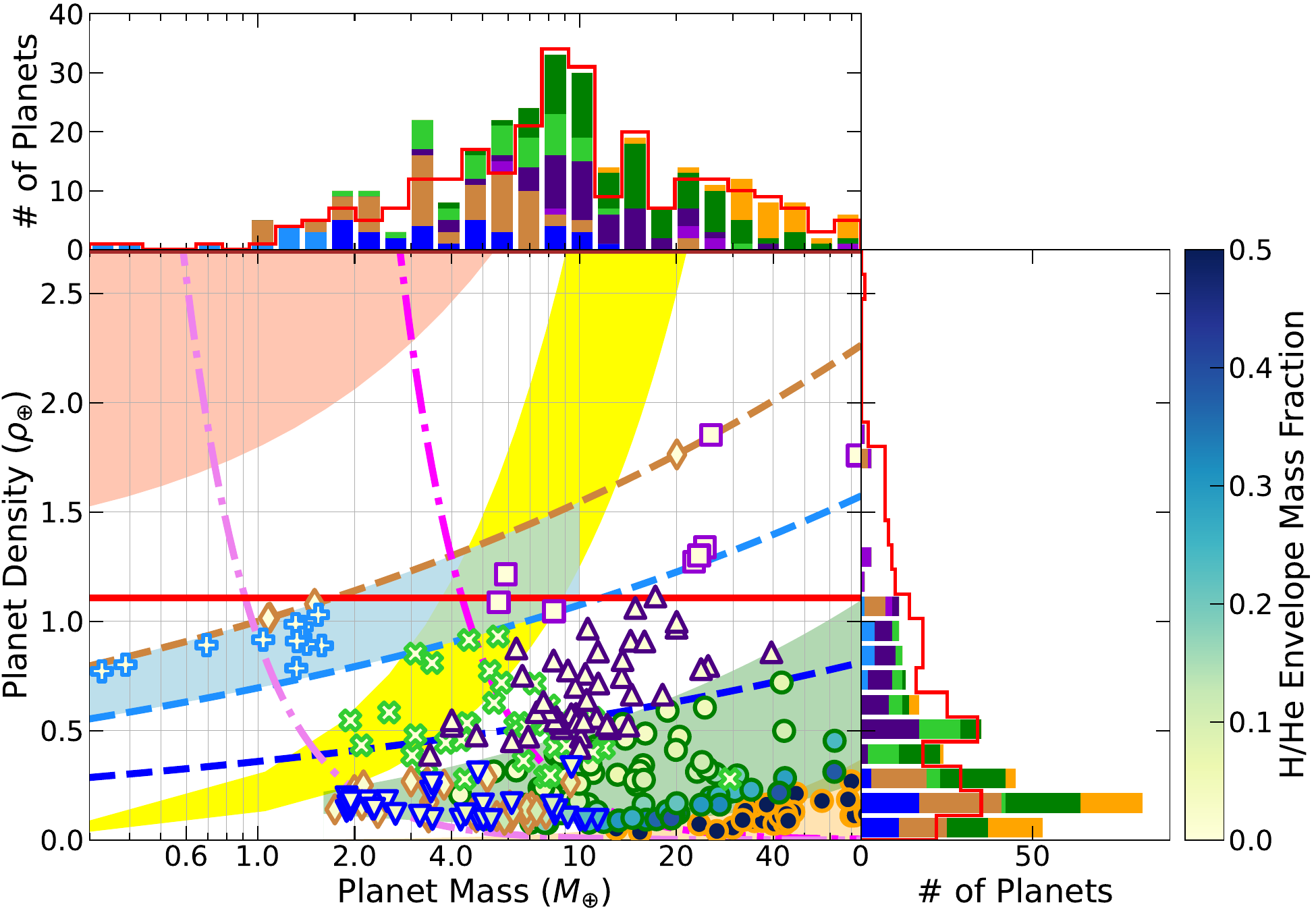}
\caption{Predictions about the properties of planets in habitable zones in the mass-radius and mass-density diagrams (as in Figure \ref{fig14}).
No photoevaporation and catastrophic collisions lead to the absence of photoevaporated cores and super-Mercuries, respectively, further clearing the radius valley;
vapor-rich sub-giant planets populate the valley predominantly.
Directly constraining the minimum value of the critical core mass becomes possible due to the lack of photoevaporated cores.
The composition of bare cores enables careful examination of water delivery processes (e.g., migration vs inward drift of dust particles).
Collisionally sculpted sub-giants are maintained due to less energetic collisions.
Precise mass and radius measurements are crucial for characterizing small-sized, solid planets including habitable eta-Earths,
while atmospheric characterization plays an important for constraining the composition and origins of planets that have gaseous envelopes.
}
\label{fig16}
\end{center}
%\end{figure}
\end{minipage}
\end{figure*}

Figure \ref{fig16} shows the resulting distribution of planets in the mass-radius and mass-density diagrams.
The absence of photoevaporation and catastrophic collisions do not move the population of photoevaporated cores and super-Mercuries to the current position in the diagrams.
As a result, the planet population in the radius valley decreases further, and the valley is filled out predominantly by vapor-rich sub-giant planets.
The disappearance of photoevaporated cores also makes it easier to tightly constrain the minimum value of the critical core mass.
The compositional diversity of bare cores provides a better hint of how water would be delivered to these solid planets.
The presence of collisionally sculpted sub-giants are maintained.
For planets distributed in the green shaded regions, 
differentiation between planets with H/He dominated envelopes and those with water (or other heavy volatiles) dominated envelopes becomes harder 
from precise measurements of planet mass and radius.
Instead, compositional determination of their atmospheres plays a critical role.

Thus, in the near-future era of exploring planets in habitable zones, 
precise measurements of both planet mass and radius become even crucial for characterizing solid planets including habitable eta-Earths,
while atmospheric characterization plays a dominant role for constraining the composition of planets with gaseous envelopes and their origins.

\section{Discussion} \label{sec:disc}

We have demonstrated in the above sections that 
our formulation leads to refined classification of observed exoplanets, identifies direct links with formation and evolution processes, and provides new insights into the origin of planets.
Our model is developed from the results of previous studies, where simplification and assumptions are used.
Accordingly, the findings of this work are valid within the framework of previous studies.
We here discuss uncertainties arising from such idealization and relevant caveats.

Our classification is based on theoretically derived mass-radius relations.
Uncertainties in adopted mass-radius relations therefore have direct influence on the results.
These uncertainties stem from those of EoS.
For Earth-like rock, EoS and the resulting mass-radius relation should reproduce the properties of the Earth,
and hence tight constraints are available at least for Earth-like planets.
For planets containing water, there is no such constraint.
Their EoS and mass-radius relation are one active research area \citep[e.g.,][]{2025ApJ...988..186A,2026arXiv260415304S},
and due to the uncertainty of EoS, we do not perform direct comparison of the results with different mass-radius relations.
Instead, we have adopted the one widely used in the literature \citep{2019PNAS..116.9723Z}.
If different mass-radius relations are utilized, quantitative differences may be seen.
However, we do not expect qualitative differences because recently derived mass-radius relations are more or less comparable to the one adopted in this work \citep{2026arXiv260415304S}.

The composition of envelopes provides another uncertainty.
As described above, previous studies assume H/He envelopes 
in order to compute planet radius and calibrate mass loss by photoevaporation \citep{2013ApJ...776....2L,2014ApJ...792....1L}.
This idealization has enabled identifying vapor-rich sub-giants (Section \ref{sec:post_photo}).
However, the presence of other molecules and dust is naturally expected,
which may become significant especially for planets experiencing collisions.
Collisions can dredge up interior materials, leading to metal enrichment of envelopes \citep[e.g.,][]{2019Natur.572..355L}.
If impact velocity is sufficiently high, vaporization of colliding materials occurs \citep[e.g.,][]{2024A&A...688A.103M}, 
and the composition of primordial envelopes may change.
It is ideal to explore such complexity.
However, expected parameter space is huge, and it may not be feasible to perform the corresponding study currently 
unless targets are specified and their properties are well constrained.

Finally, the results of this study also depend on the quality of observational data.
While we have selected exoplanets that have relatively small errors (Figure \ref{fig1}),
it does not necessarily mean that measurements of planet mass and radius are accurate.
In fact, discrepancies in measured values of planet mass and radius are found in the literarture occasionally.
Careful vetting of the exoplanet data is important especially for small-sized planets.
For instance, the planet mass and radius reported in the literature end up with planets that have higher densities than Iron \citep[e.g.,][]{2019ApJ...881..117S}.
Such vetting is imperative when the properties of individual planets are explored in detail.
This work focuses on statistical trends, 
and hence we expect that the overall finding of this work likely remains unchanged.

Thus, detailed examination of the finding of this work is necessary, and quantitative differences may arise.
However, we investigate statistical trends in this work, and qualitative differences are unlikely.

\section{Summary \& Conclusions} \label{sec:sum}

We have explored formation and evolution pathways of planets by synthesizing recent efforts made in the literature.
These include various mass-radius relations, gas accretion and retention recipes, and photoevaporative and collisional mass losses.
Our investigation has put the particular focus on the mass-radius and mass-density diagrams and 
constrained how these diagrams are populated by planets due to what process(es).
We have identified direct links between underlying processes and parameter space in the diagrams (Figures \ref{fig2}, \ref{fig7}, and \ref{fig10}), 
which are summarized in Figure \ref{fig15}.
Especially, the distribution of planets in the diagrams is diversified by two evolution processes: radiation-driven mass loss and collisional mass growth and loss.
We have further demonstrated this by tracing how the distribution of planets in the diagrams evolves with time.
In the diagrams, formation and evolution pathways of planets are divided into four stages:
Stage 1. Core formation (the top panel of Figure \ref{fig14} and Figure \ref{fig18}), 
stage 2. Gas accretion (the second panel of Figure \ref{fig14} and Figure \ref{fig19}),
stage 3. Collisional mass growth and loss (the third panel of Figure \ref{fig14} and Figure \ref{fig20}),
and stage 4: photoevaporation (the bottom panel of Figure \ref{fig14} and Figure \ref{fig21}).

We have developed the new classification scheme (Figure \ref{fig3}) and applied to exoplanets currently observed (Figure \ref{fig1}).
Below are the eight type of planets:

{\it Bare Cores.} These planets are less massive (i.e., $\sim 1.6 M_{\oplus}$) and did not undergo gas accretion.
Hence, they do not have primordial envelopes.
Sudden compositional diversity is identified for planets more massive than $\sim 1.3 M_{\oplus}$ (Figure \ref{fig15}).
This implies that large scale migration driven by disk gas was likely to be effective for such planets,
and they are potentially water rich.
On the other hand, less massive planets formed more or less in-situ, and their composition is Earth-like rock (Table \ref{table1}).
Detecting more exoplanets enables statistical examination of this argument.

{\it Photoevaporated cores.} These planets are more massive than $\sim 3 M_{\oplus}$ and surely experienced gas accretion during the gas disk stage;
equivalently, that planet mass represents a lower bound of the critical core mass (equation (\ref{eq:M_c_crit_range})). 
Due to high photoevaporation efficiencies, their primordial envelopes were lost completely (Figure \ref{fig6}).
Their compositional diversity is limited by the water rich case with the ice-to-rock mass ratio of 1/3 (Table \ref{table1}),
suggesting that the water abundance of solid bodies formed in extrasolar environments is comparable to that in the solar system.

{\it Collisionally sculpted cores.} These planets are more massive than $\sim 1.6 M_{\oplus}$, and some of massive planets could undergo gas accretion.
However, the presence of primordial envelopes is unlikely from their current properties, 
and the photoevaporation efficiency is also too low to remove the envelopes completely (labeled as Super-Earth in Figure \ref{fig6}).
This in turn indicates that their present mass is the outcome of collisions after gas disks are gone (Figure \ref{fig8}).
The properties of progenitors are consistent with the standard picture of core accretion.

{\it Super-Mercuries.} These planets have high iron mass fractions and certainly experienced mantle stripping due to high energetic events.
Their orbital distribution is comparable to other planets in the vicinity ($\lesssim$ 10 days) of the host star (Figure \ref{fig4}),
and hence the high energetic events should have occurred stochastically.
The required eccentricity is higher than that predicted from giant impacts (Figure \ref{fig10}).
These features point to the importance of long-term dynamical instabilities and the resulting catastrophic collisions as the origin of super-Mercuries.

{\it Collisionally sculpted sub-giants.} These planets have tenuous envelopes and definitely experienced gas accretion.
However, the current envelopes are too tenuous to be explained by gas accretion/retention, photoevaporation, or the combination of these processes (labeled as Sub-Neptune in Figure \ref{fig5}).
This invokes the operation of collisions after gas disks are gone as with the case of collisionally sculpted cores.
In fact, the computed properties of progenitors are accounted for by core accretion (Figure \ref{fig11}), 
and inclusion of collisions after core accretion naturally reproduces the current properties of these planets (Figure \ref{fig8}).
Present survival of gaseous envelopes against collisions suggests that they are of primordial origin and collisions occurred earlier than photoevaporation (Figure \ref{fig9}).

{\it Vapor-rich sub-giants.} These planets exhibit large radius and should possess gaseous envelopes.
On the other hand, the photoevaporation efficiency is high enough to remove the envelopes completely (Figure \ref{fig5}).
This (apparent) contradiction arises from the assumption used in previous work that envelopes consist purely of hydrogen and helium,
and the formulation employed in this work heavily relies on the outcome of such previous work.
Inclusion of heavy volatiles such as water in gaseous envelopes reduces photoevaporation efficiencies, and the contradiction becomes less problematic.
This is the reason of why these planets are referred to as vapor-rich sub-giants in this work.

{\it Photoevaporated sub-giants.} These planets keep primordial envelopes even under the action of photoevaporation.
This is partly because their cores are massive enough to weaken photoevaporation and partly because photoevaporation is indeed weak at their position (Figure \ref{fig5}).
Some of the planets are more massive than $10 M_{\oplus}$ that is widely quoted as the critical core mass beyond which gas accretion begins.
These planets avoid turning into gas giants possibly due to concurrent gas and solid accretion, 
which increase both core and envelope masses simultaneously (Figure \ref{fig13}).

{\it Gas giants.} These planets formed by standard core accretion. 
Their radius is larger than $6R_{\oplus}$ beyond which the envelope mass exceeds the core mass such that rapid gas accretion occurs (Figure \ref{fig4}).
While the onset of gas accretion is examined by the population of collisionally sculpted cores and collisionally sculpted sub-giants,
the resulting constraints are not tight enough currently (Figure \ref{fig12}).

We have finally discussed predictions about planets in habitable zones, 
by considering the extreme case, where both photoevaporation and catastrophic collisions are assumed to be negligible (Figure \ref{fig16}).
While great care is needed, the application of our framework developed from observed exoplanets suggests that 
precise measurements of planet mass and radius become crucial for tightly constraining the properties and origins of less massive planets including habitable eta-Earths.
For massive planets that possess gaseous envelopes, atmospheric characterization sheds light on their properties and origins.

In summary, this work constitutes an important and complementary step to reveal the origin of planets through integration of various efforts conducted in the literature.
Further examining the framework constructed in this work and the resulting classification scheme is necessary for carefully verifying the finding of this work (Section \ref{sec:disc}).
In the forthcoming paper, we will undertake it, focusing on the atmospheric metallicities of exoplanets constrained by HST and JWST.

\begin{acknowledgments}

The authors thank an anonymous referee for the useful comments on the manuscript and
Eiichiro Kokubo, Raissa Estrela, and Mark Swain for stimulating discussions.
This research was carried out at the Jet Propulsion Laboratory, California Institute of Technology, 
under a contract with the National Aeronautics and Space Administration (80NM0018D0004)
and funded through the internal Research and Technology Development program.
Y.H. thanks the hospitality of NAOJ, and part of this work was supported by the NAOJ Research Coordination Committee, NINS (NAOJ-RCC-2504-0101).

\end{acknowledgments}

\appendix

\section{Examining the validity of the power-law mass-radius relation for Earth-like rock} \label{sec:appned1}

We here examine briefly how the power-law mass-radius relation is valid for Earth-like rock and useful for the sample used in this work.

\begin{figure*}
\begin{minipage}{17cm}
%\begin{figure}%[!ht]
\begin{center}
\includegraphics[width=8.4cm]{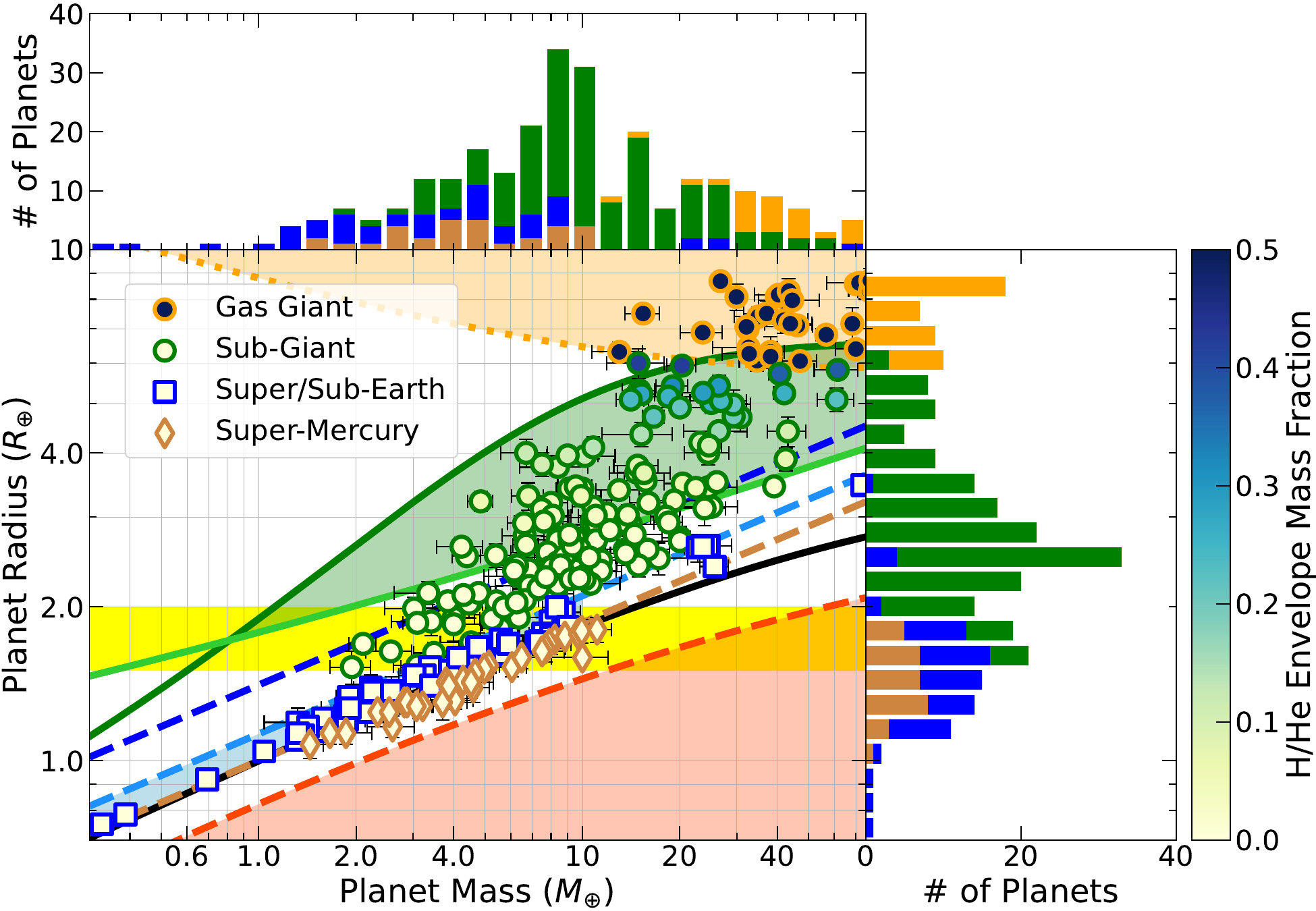}
\includegraphics[width=8.4cm]{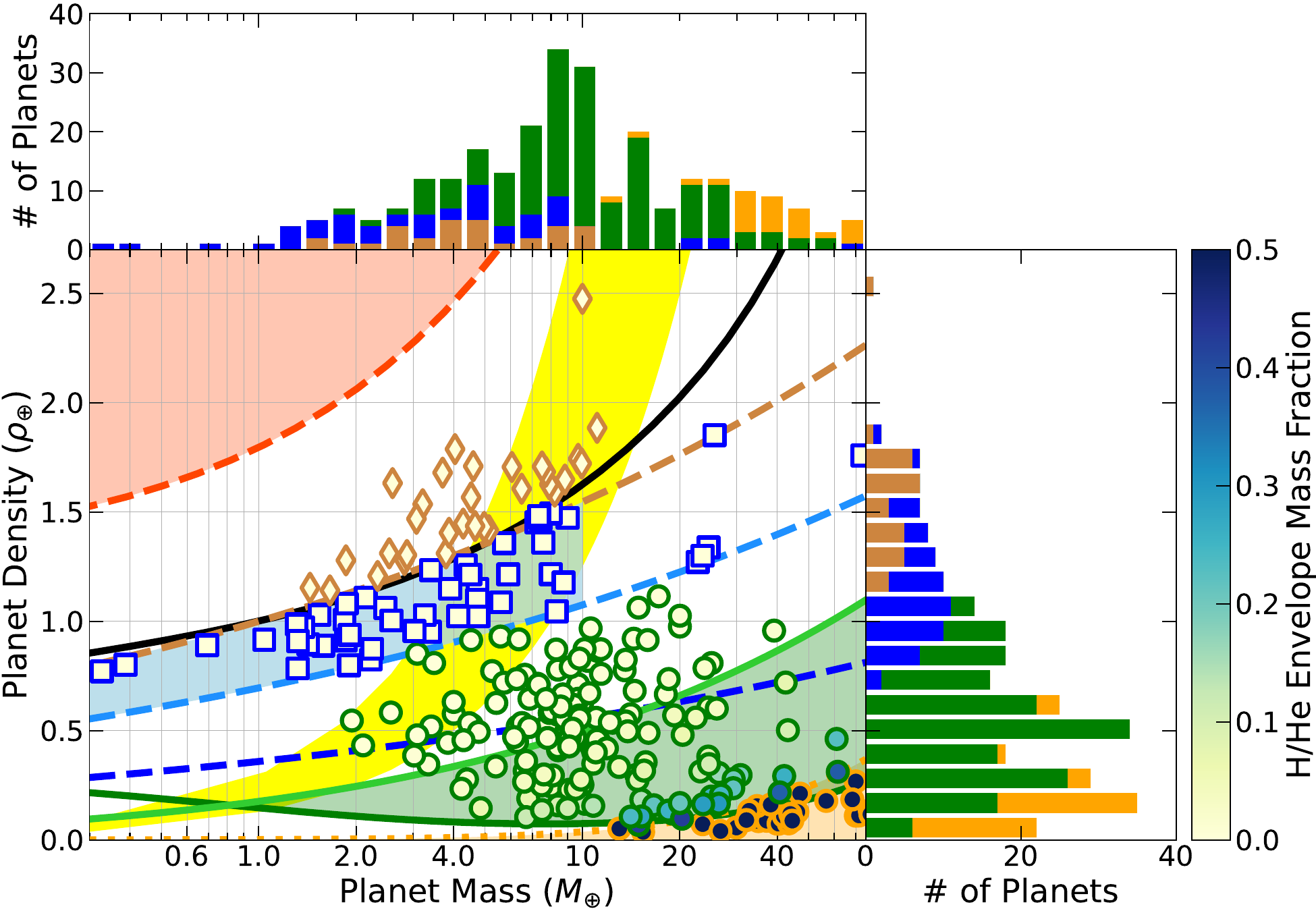}
\caption{Comparison between the power-law mass-radius relation and the detailed one (as in Figure \ref{fig4}).
The latter mass-radius relation is denoted by the black solid line.
Agreement between these two relations is excellent for less massive planets ($\la 10 M_{\oplus}$).
Deviation becomes non-negligible for more massive planets.
However, there is no planet affected by the deviation in the current samples.
}
\label{fig17}
\end{center}
%\end{figure}
\end{minipage}
\end{figure*}

Figure \ref{fig17} shows comparison between the power-law mass-radius relation and the one derived from detailed calculations \citep{2019PNAS..116.9723Z}.
It is clear that the power-law one agrees well with the detailed one for planets that are less massive than $\sim 10 M_{\oplus}$.
Deviation becomes large for more massive planets.
However, there is no such exoplanet in our samples.

In summary, adopting the power-law mass-radius relation is valid for this study and does not change the finding of this work.

\section{Detailed properties of planets at different formation stages} \label{sec:appned2}

\begin{figure*}
\begin{minipage}{17cm}
%\begin{figure}%[!ht]
\begin{center}
\includegraphics[height=5.1cm]{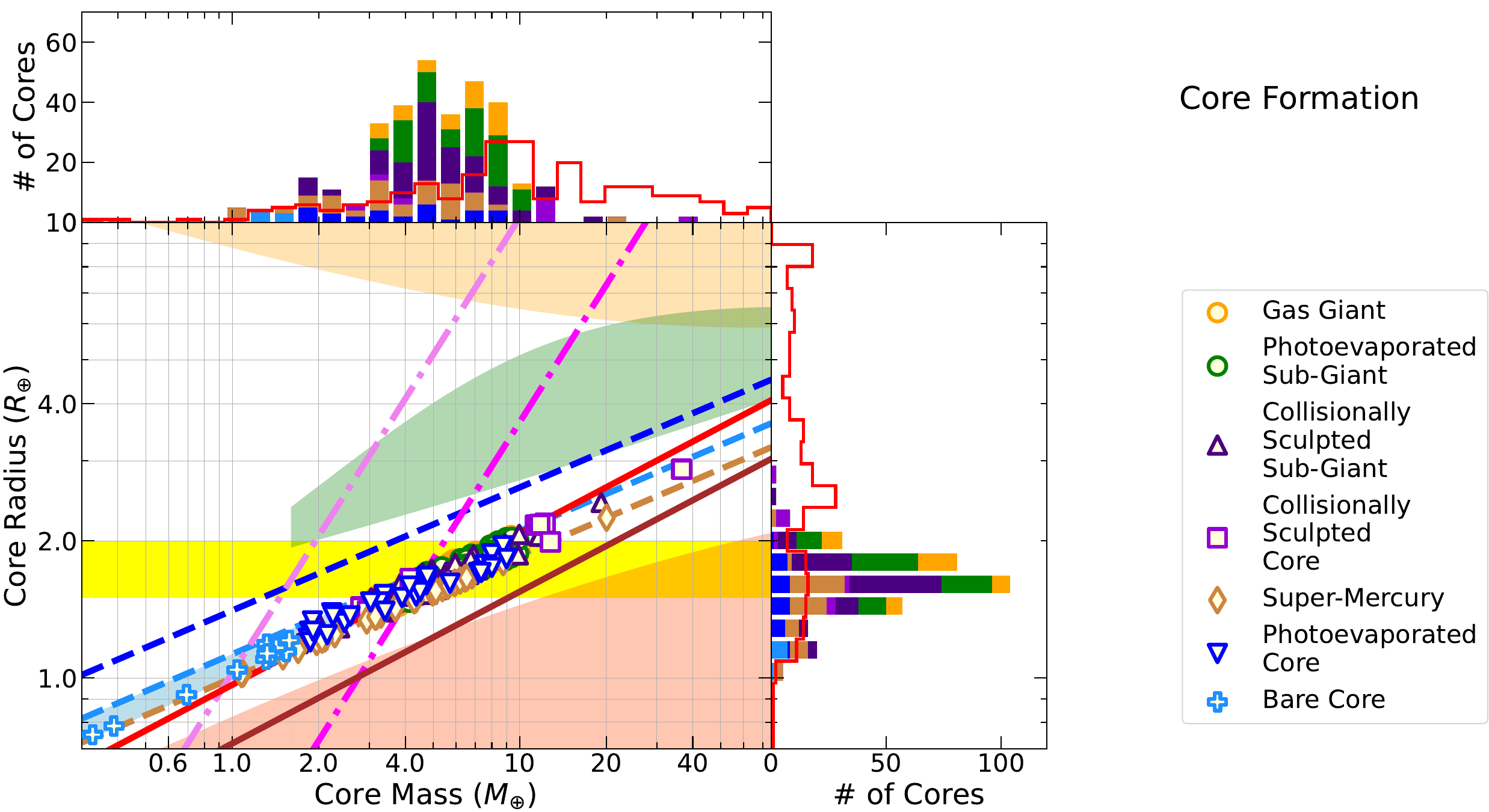}
\includegraphics[height=5.1cm]{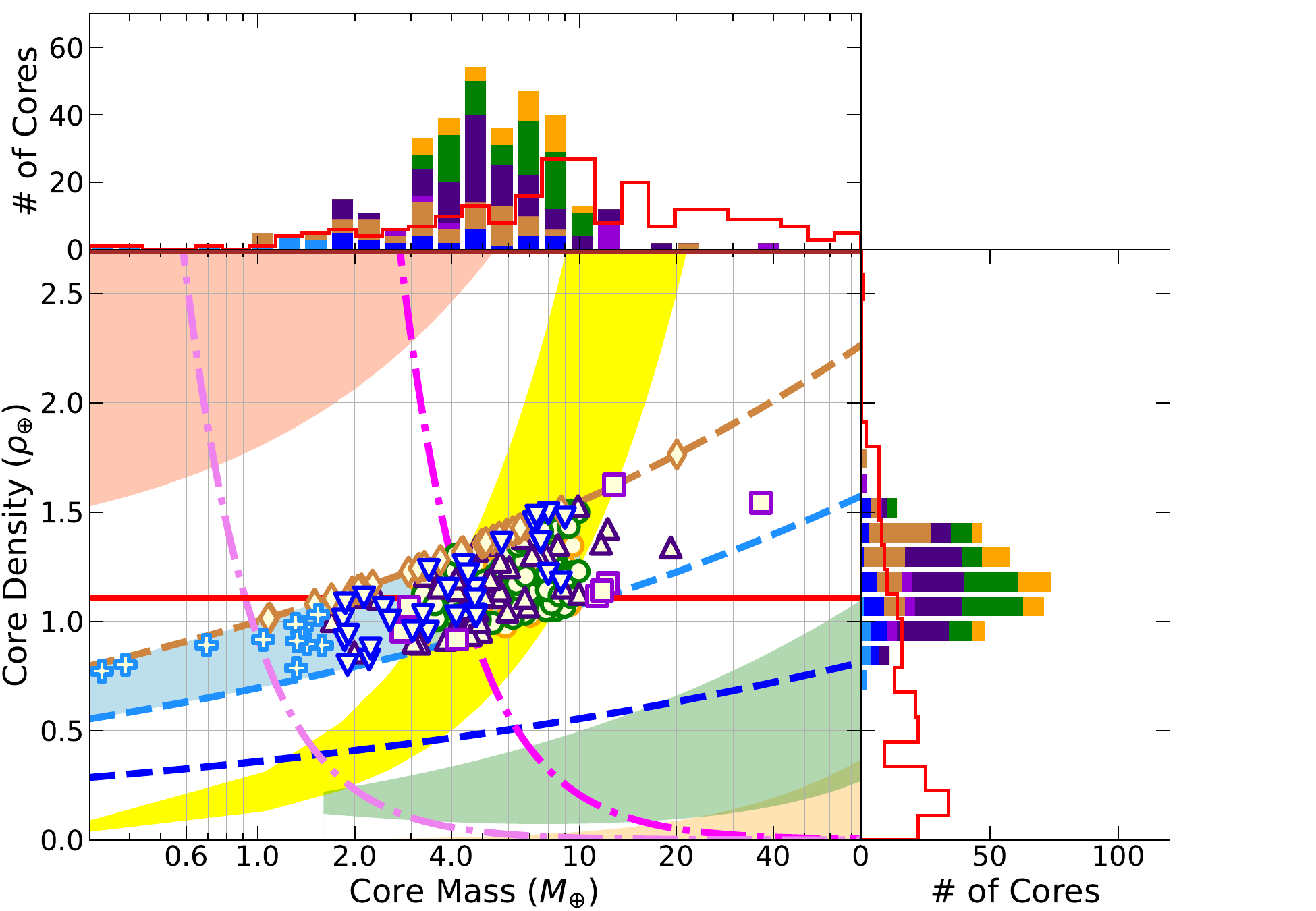}
\includegraphics[height=5.1cm]{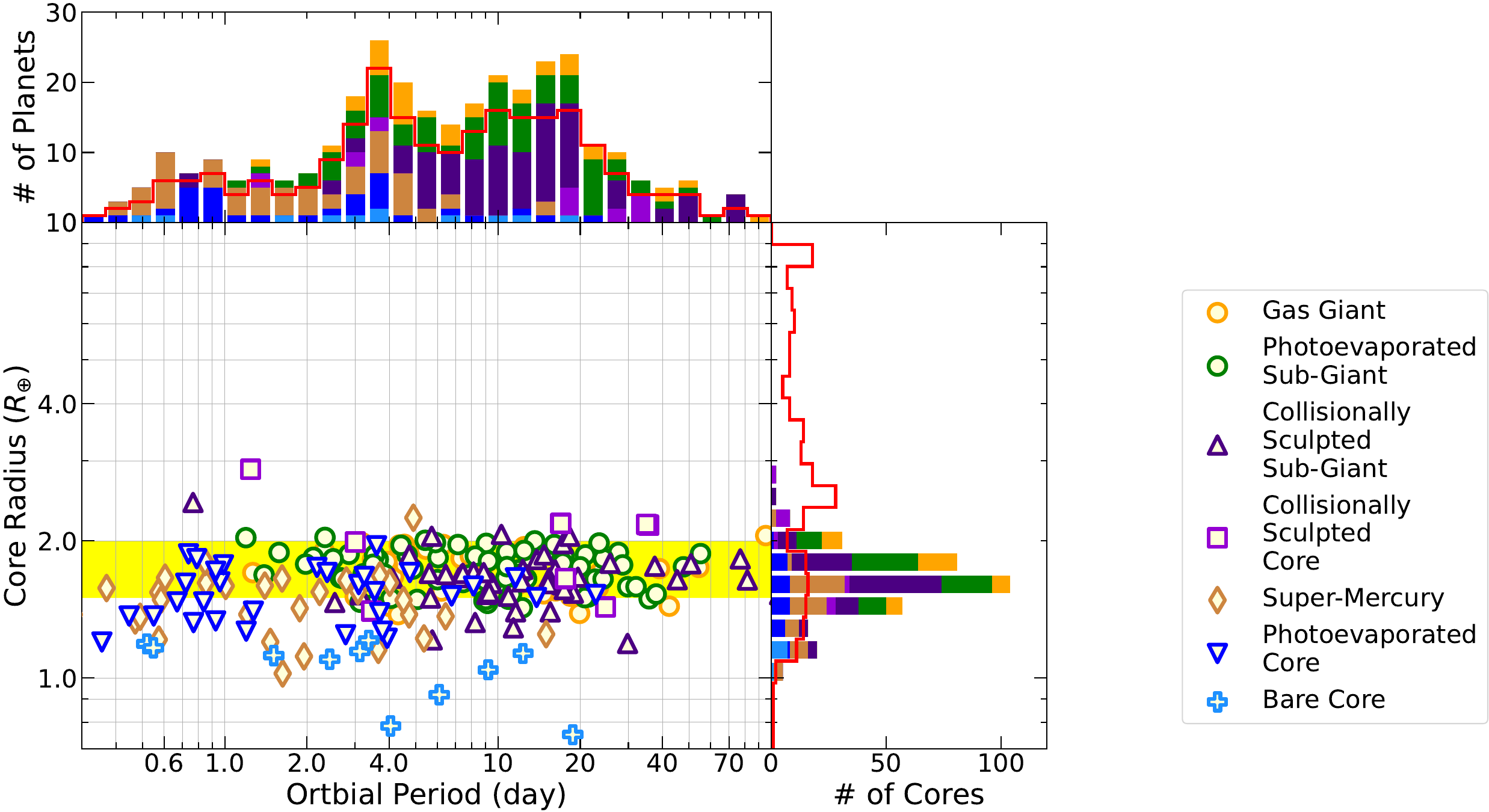}
\includegraphics[height=5.1cm]{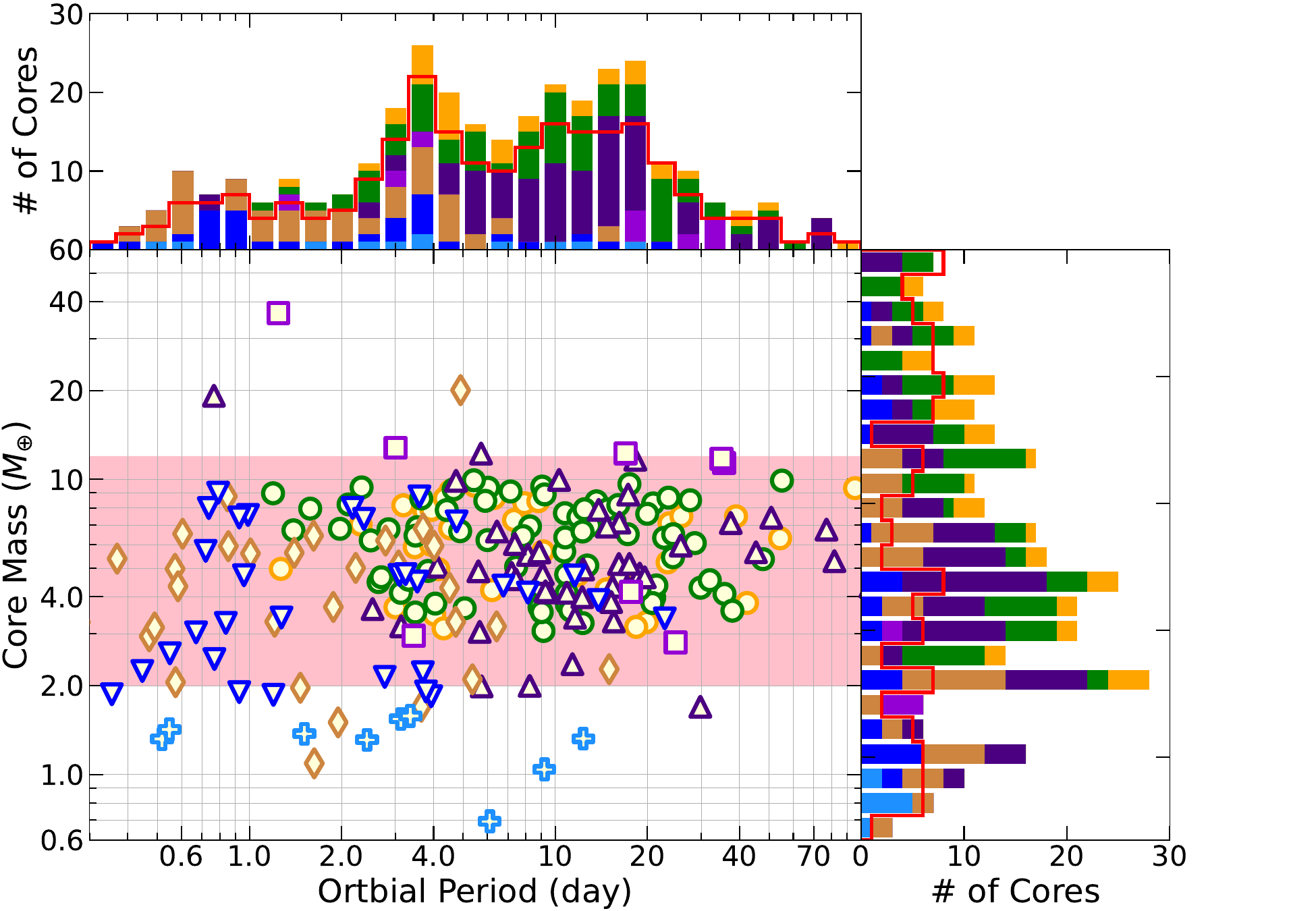}
\caption{The properties of planets after core formation (as in Figures \ref{fig4}, \ref{fig8}, \ref{fig11}, \ref{fig14}).
As expected, most planets are distributed in the blue shaded region in the mass-radius and mass-density diagrams on the upper left and right panels, respectively.
Their distributions in the orbital period-radius and orbital period-mass diagrams are shown in the lower left and right panels, respectively.
Vapor-rich sub-giant planets are not included in these plots.
The red line in two histograms represents the distribution of the current properties of exoplanets.
The diversity of core compositions is constrained by the Earth-like rock and water rich lines (Table \ref{table1}),
suggesting that the ice-to-rock ratio of exoplanetary materials is capped at 0.3, similar to solar system comets.
Also, such diversity becomes visible for planetary cores that are more massive than $\sim 1.3 M_{\oplus}$.
This sudden change implies the importance of planetary migration on these cores, 
and large-scale migration was not effective for less massive cores.}
\label{fig18}
\end{center}
%\end{figure}
\end{minipage}
\end{figure*}

\begin{figure*}
\begin{minipage}{17cm}
%\begin{figure}%[!ht]
\begin{center}
\includegraphics[height=5.1cm]{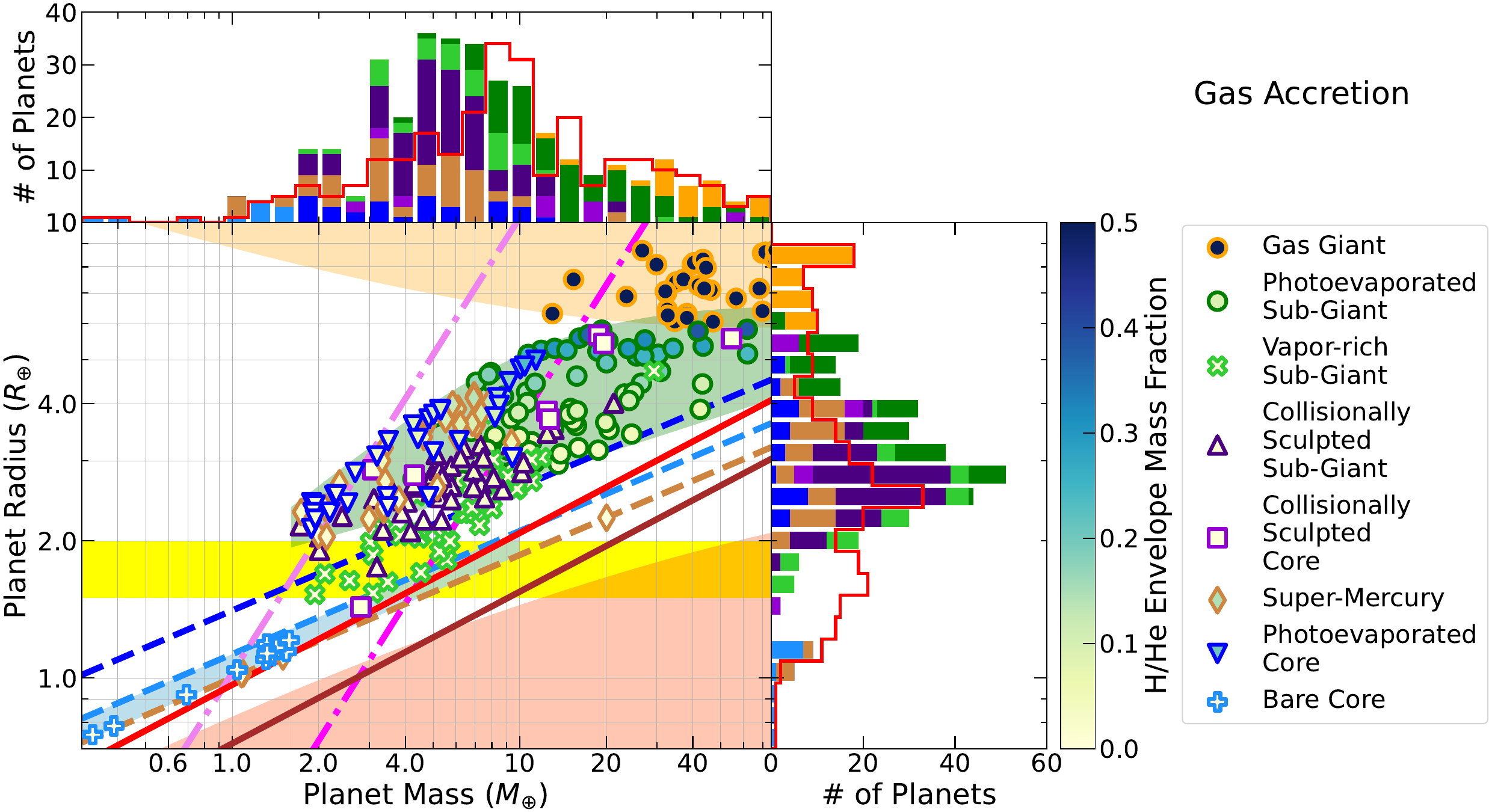}
\includegraphics[height=5.1cm]{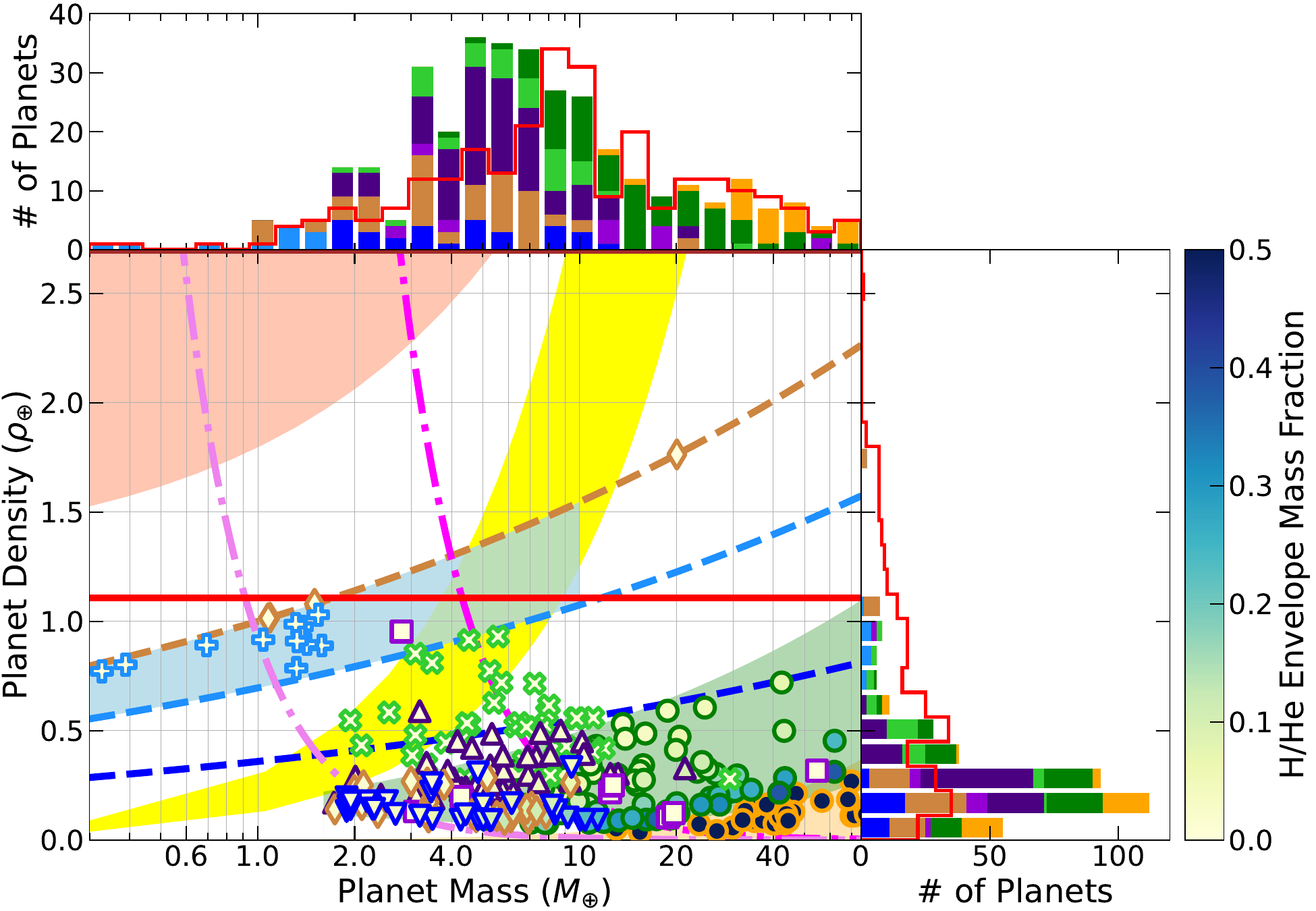}
\includegraphics[height=5.1cm]{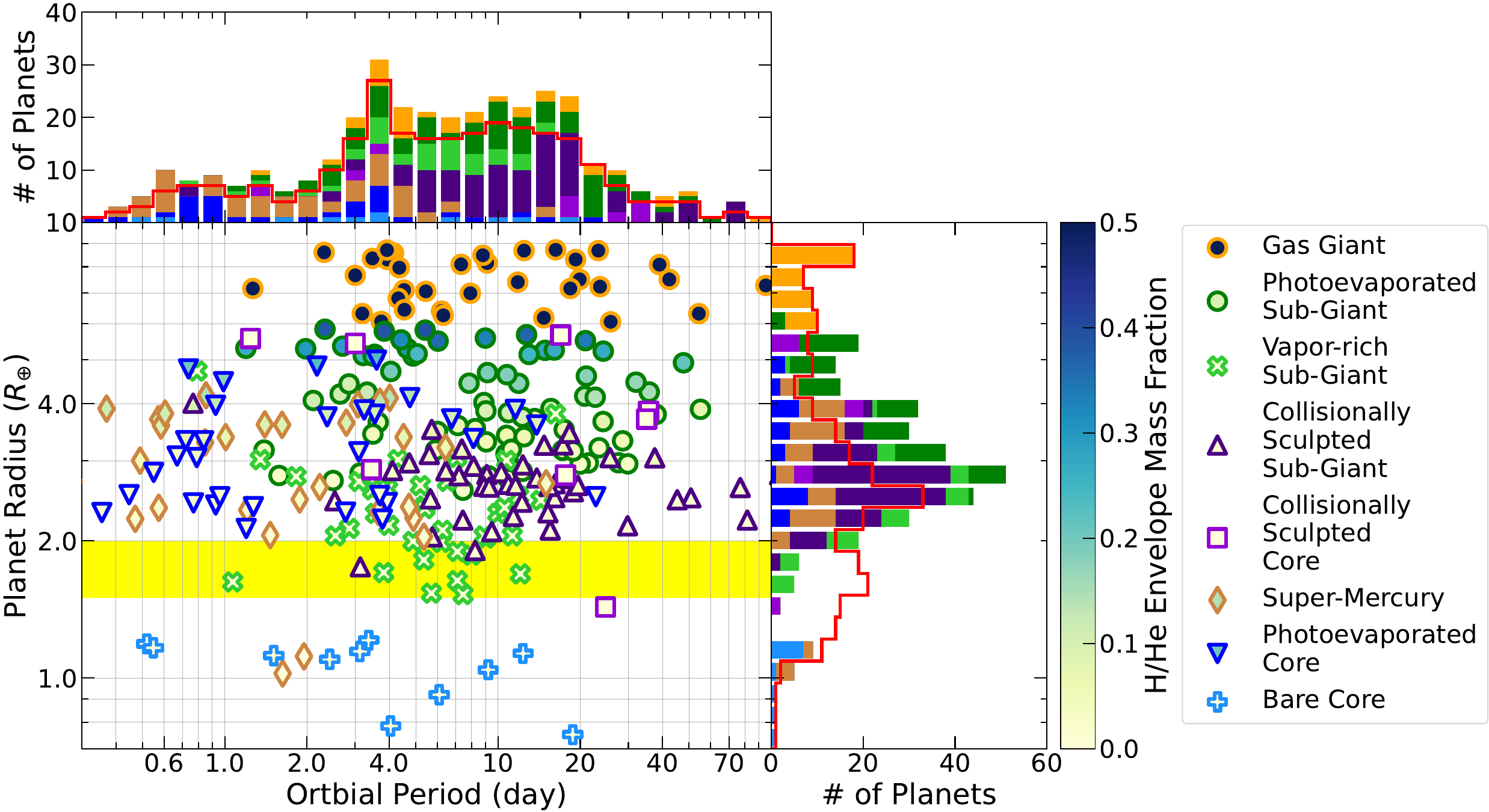}
\includegraphics[height=5.1cm]{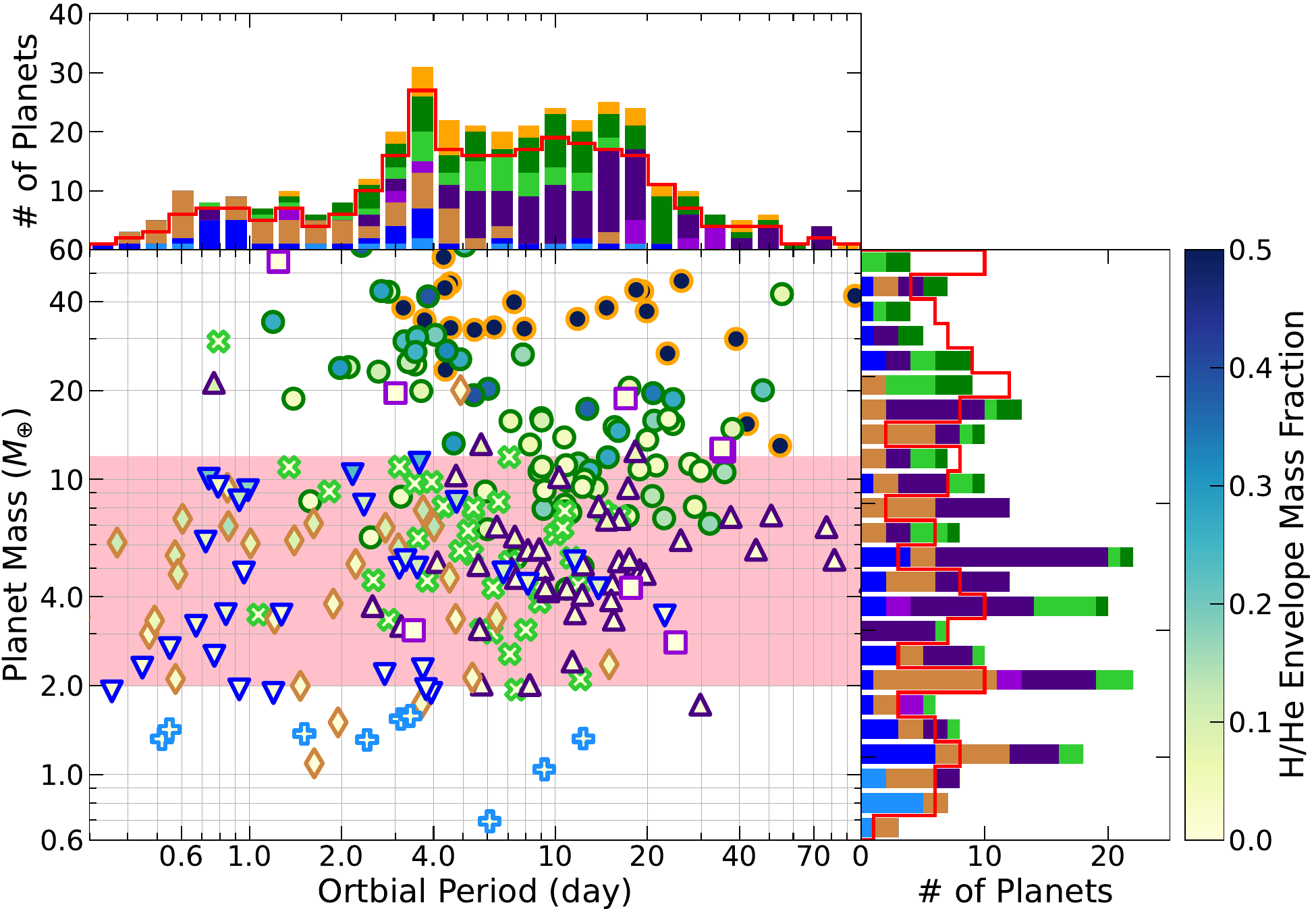}
\caption{The properties of planets after gas accretion (as in Figure \ref{fig18}).
Accretion of gaseous envelopes onto planetary cores expands the planet radius, and all of the planets are distributed in the green and orange shaded regions, 
except for vapor-rich sub-giants and bare cores ($\lesssim 1.6 M_{\oplus}$) that did not undergo gas accretion. 
One outlier is identified.
The color bar denotes the envelope mass fraction.}
\label{fig19}
\end{center}
%\end{figure}
\end{minipage}
\end{figure*}

\begin{figure*}
\begin{minipage}{17cm}
%\begin{figure}%[!ht]
\begin{center}
\includegraphics[height=5.1cm]{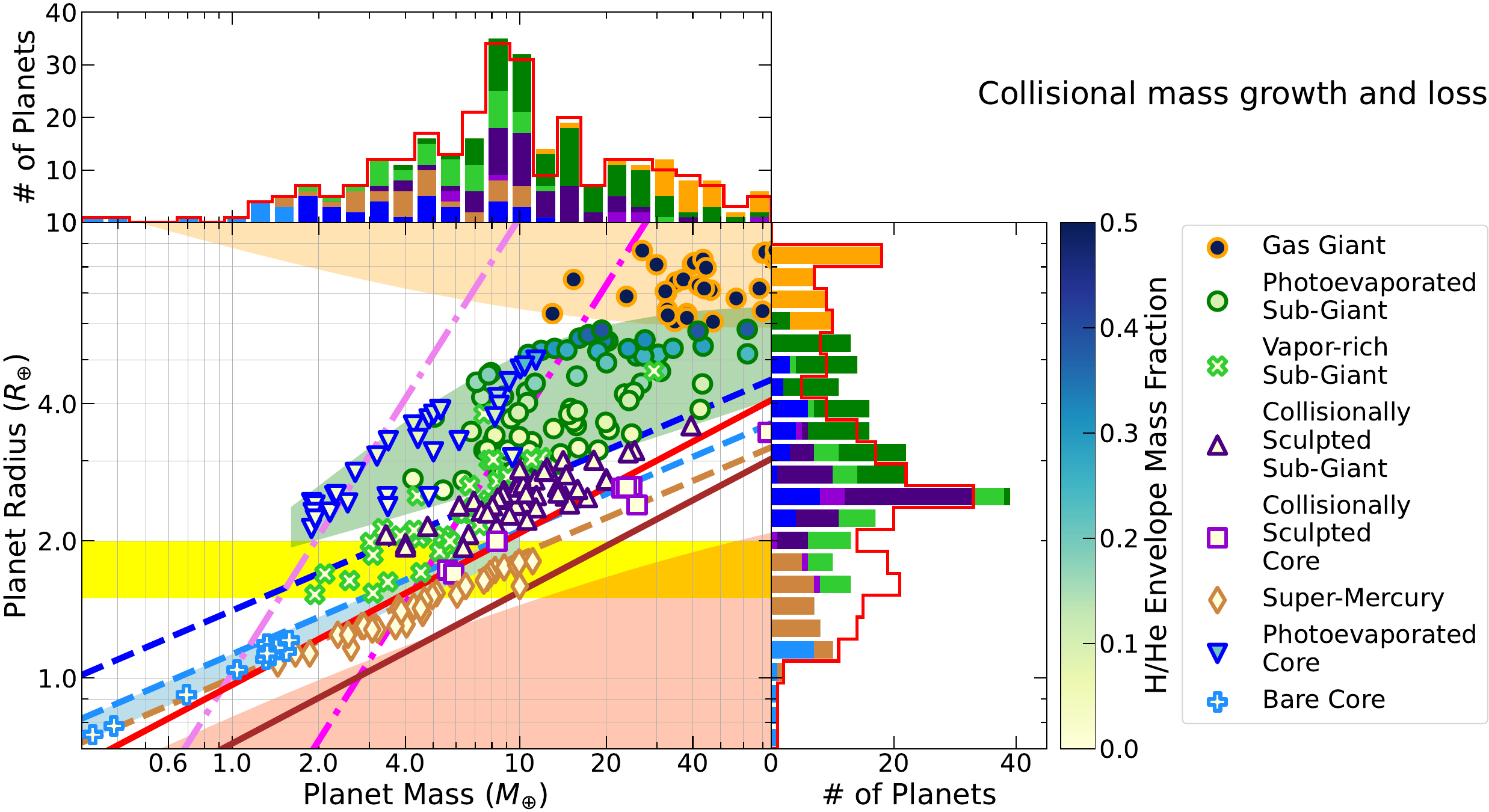}
\includegraphics[height=5.1cm]{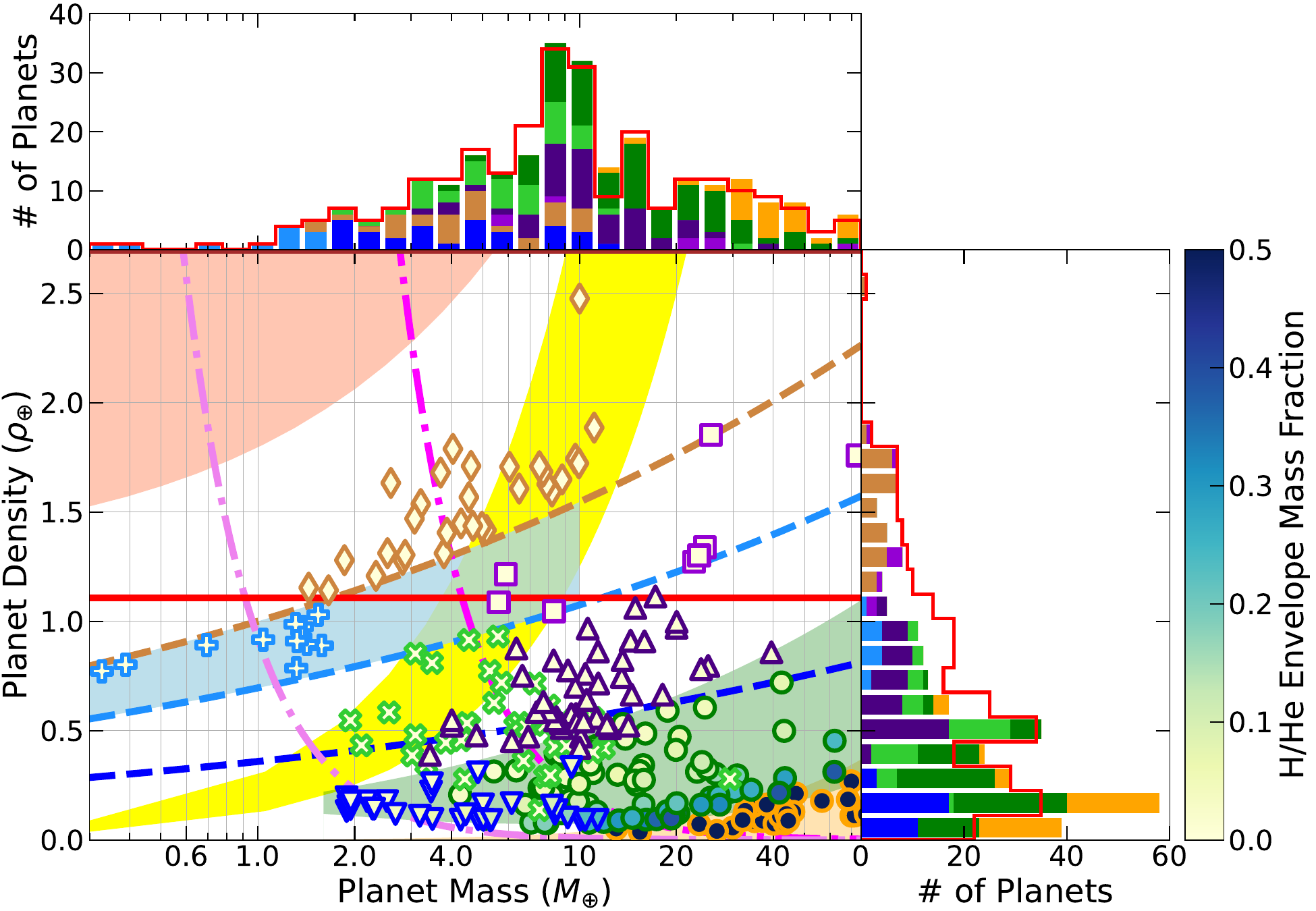}
\includegraphics[height=5.1cm]{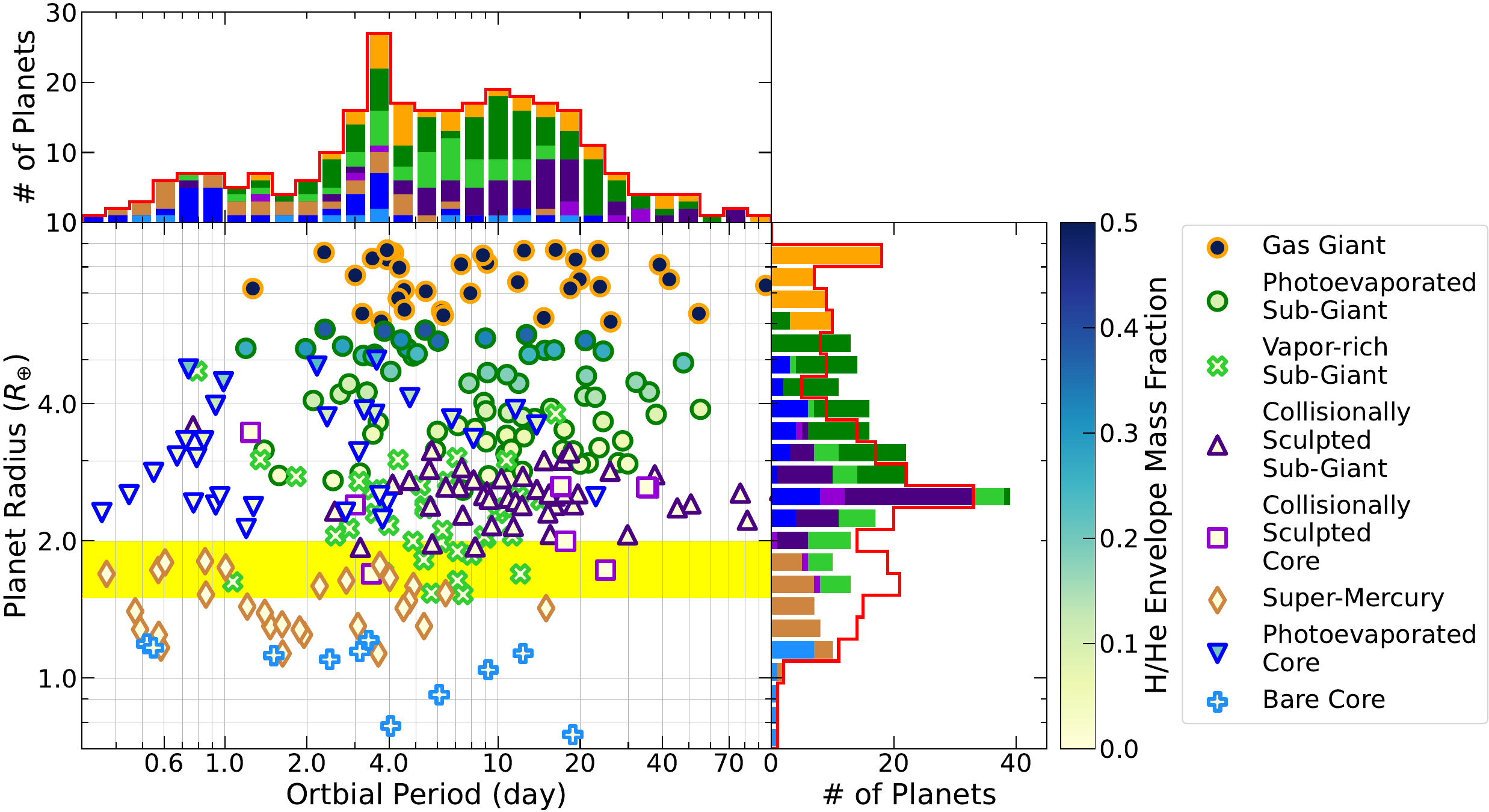}
\includegraphics[height=5.1cm]{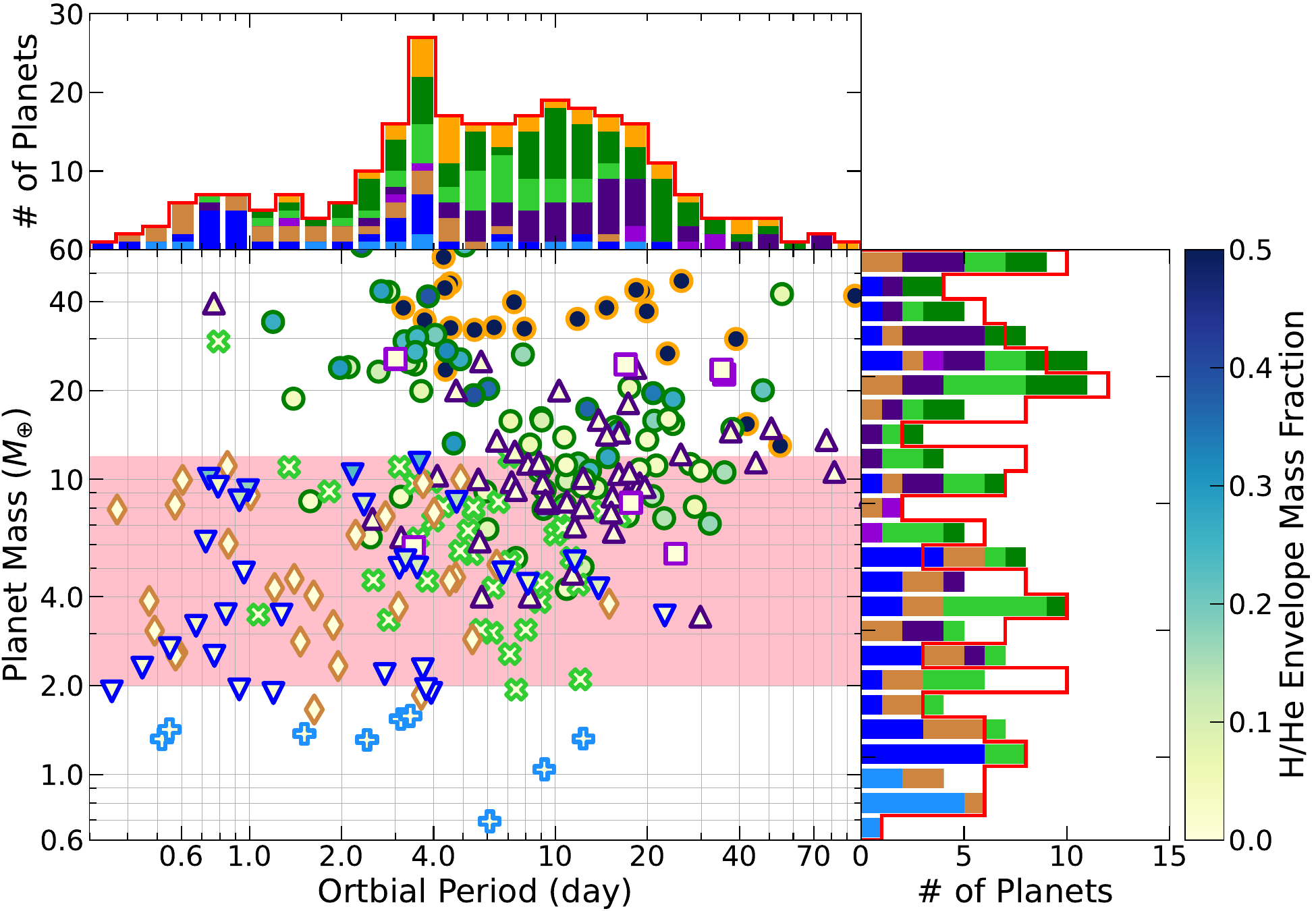}
\caption{The properties of planets after collisional mass growth and loss, which take place due to the dispersal of disk gas (as in Figure \ref{fig18}).
These evolution processes produce super-Mercuries and collisionally sculpted sub-giants and fill out two new areas in the mass-radius and mass-density diagrams.
Some of super-Mercuries populate the radius valley.
Most collisionally sculpted sub-giants distribute in the parameter space that is occupied by water-dominated planets if such peculiar planets would exist.} 
\label{fig20}
\end{center}
%\end{figure}
\end{minipage}
\end{figure*}

\begin{figure*}
\begin{minipage}{17cm}
%\begin{figure}%[!ht]
\begin{center}
\includegraphics[height=5.1cm]{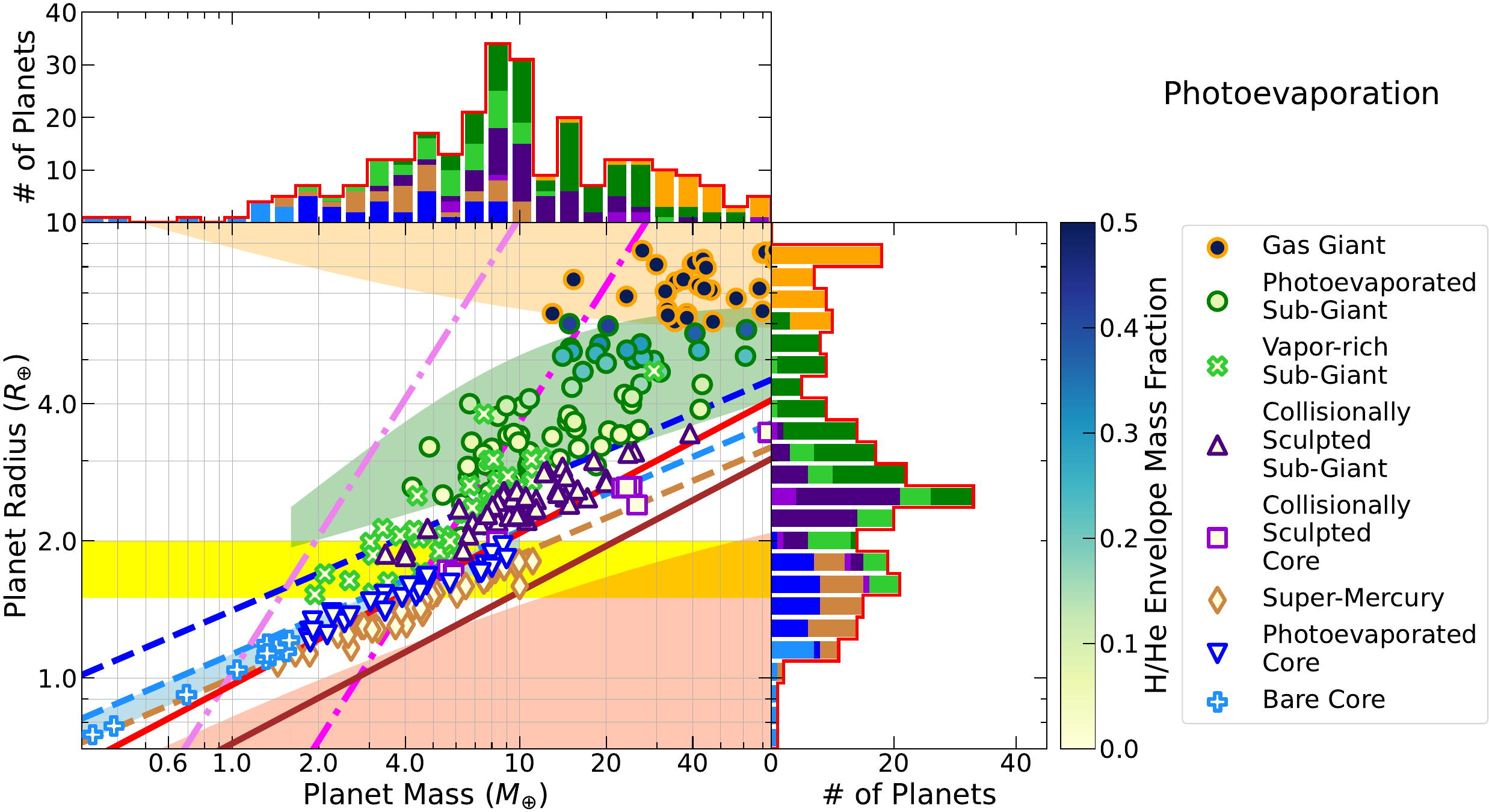}
\includegraphics[height=5.1cm]{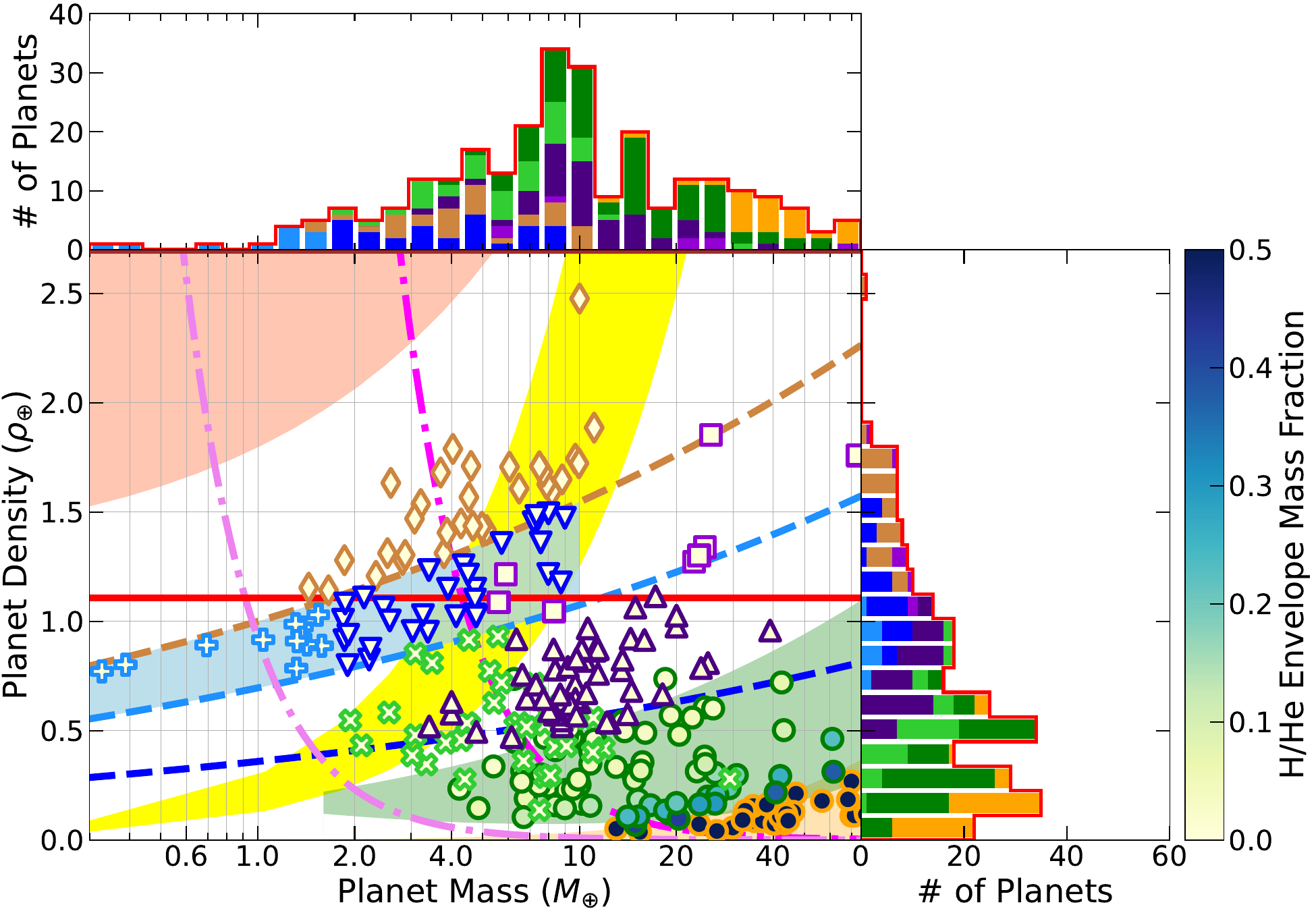}
\includegraphics[height=5.1cm]{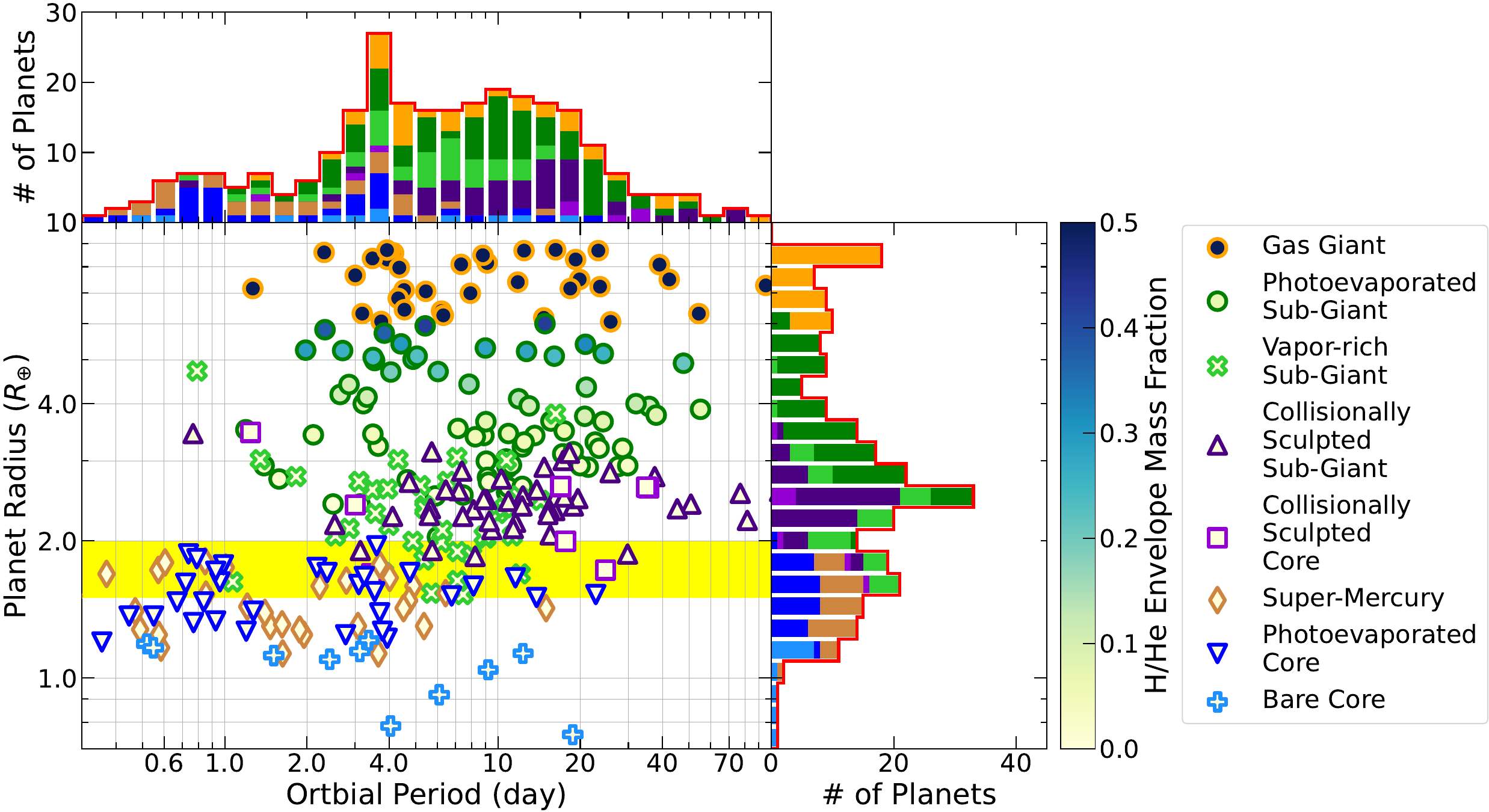}
\includegraphics[height=5.1cm]{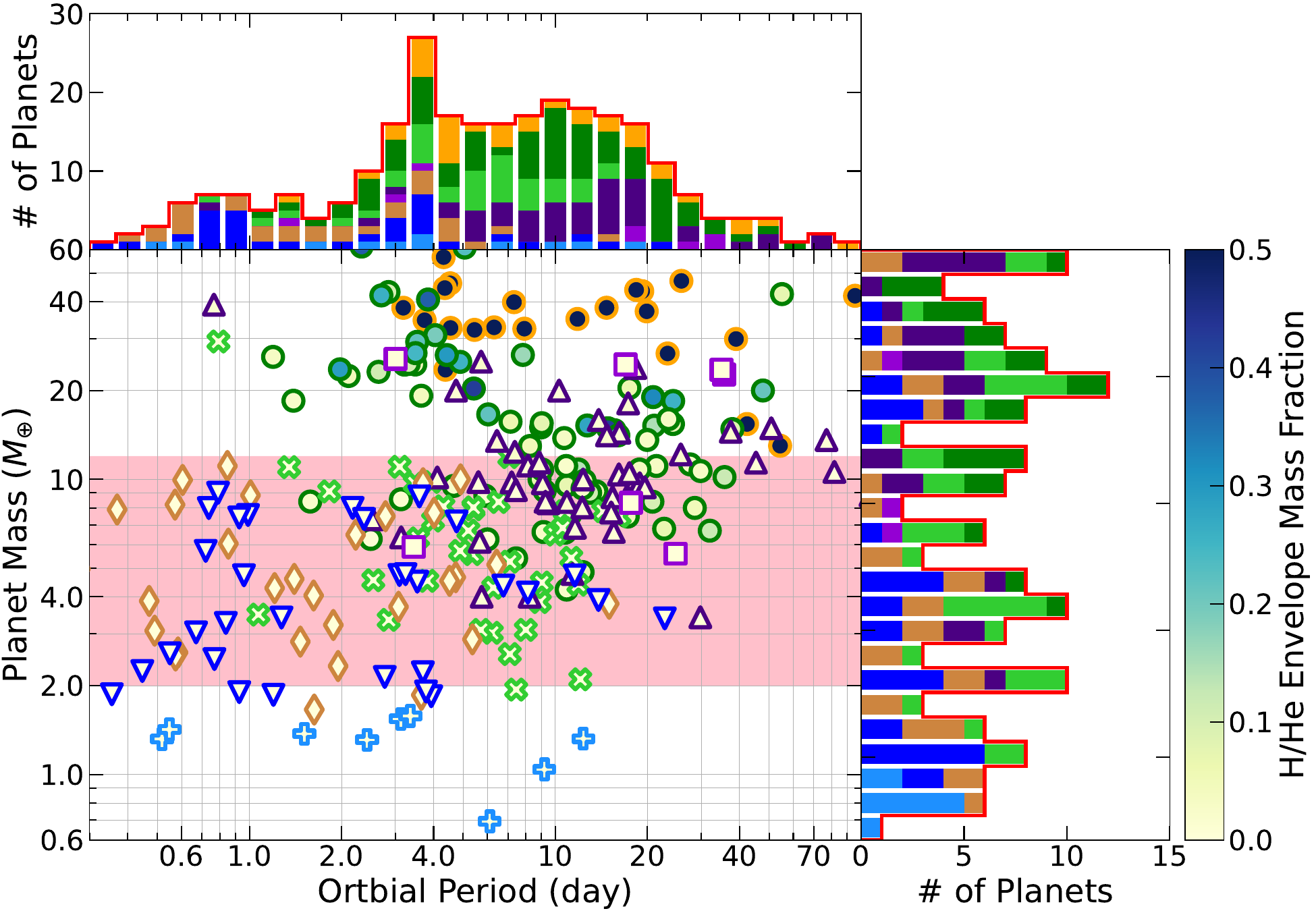}
\caption{The properties of planets after photoevaporation, which is the final process of planet formation and evolution (as in Figure \ref{fig18}).
Photoevaporation reduces the envelope mass of photoevaporated sub-giants and produces the population of photoevaporated cores.
The compositional diversity of photoevaporated cores supports the importance of migration, 
and its presence enhances the abundance of planets in the radius valley if they are more massive than $\sim 4 M_{\oplus}$.}
\label{fig21}
\end{center}
%\end{figure}
\end{minipage}
\end{figure*}

We summarize figures that show detailed properties of planets at different formation stages.
Figures \ref{fig18}, \ref{fig19}, \ref{fig20}, and \ref{fig21} correspond to the stage of core formation, gas accretion, collisional mass growth and loss, and photoevaporation, respectively.
To make comparison among various quantities readily, the top panels in each figure are reproduction of Figure \ref{fig14}.

\bibliographystyle{aasjournalv7}
\bibliography{adsbibliography}    %% includes the journal abbrevs



\end{document}